%% file: ms.tex
\documentclass[fleqn,usenatbib]{mnras}

\usepackage{newtxtext,newtxmath}

\usepackage[T1]{fontenc}

\usepackage[flushleft]{threeparttable}

\DeclareRobustCommand{\VAN}[3]{#2}
\let\VANthebibliography\thebibliography
\def\thebibliography{\DeclareRobustCommand{\VAN}[3]{##3}\VANthebibliography}

\usepackage{graphicx}	
\usepackage{amsmath}	
\usepackage{gensymb}
\usepackage{float}
\usepackage{placeins}
\usepackage{amsmath}
\usepackage{subcaption}
\usepackage{graphicx}	
\usepackage{booktabs}
\usepackage{xcolor}

\usepackage{comment} 

\newcommand{\kms}{\,\rm{km~s$^{-1}$}}
\newcommand{\msunyear}{$M_{\odot}\,\mathrm{yr}^{-1}$}

\newcommand{\msunpkpc}{$M_{\odot}\,\mathrm{kpc}^{-2}$}
\newcommand{\msunpkpcpyr}{$M_{\odot}\,\mathrm{kpc}^{-2}\,\mathrm{yr}^{-1}$}

\title[Spatially resolved gas flows around Milky Way]{Spatially Resolved Gas Flows Around the Milky Way}

\author[S. Clark et al.]{
Sean Clark,$^{1}$\thanks{E-mail: saclark7@ncsu.edu}
Rongmon Bordoloi,$^{1}$\thanks{E-mail: rbordol@ncsu.edu}
Andrew J. Fox,$^{2}$
\\
$^{1}$Department Of Physics, North Carolina State University, Raleigh, North Carolina, 27695\\
$^{3}$AURA for ESA, Space Telescope Science Institute, 3700 San Martin Drive, Baltimore, MD 21218, USA
}

\date{Accepted ---, Received ---; in original form ---}

\pubyear{2022}

\begin{document}
\label{firstpage}
\pagerange{\pageref{firstpage}--\pageref{lastpage}}
\maketitle

\begin{abstract}
We present spatially resolved measurements of cool gas flowing into and out of the Milky Way (MW), using archival ultraviolet spectra of background quasars from the \emph{Hubble Space Telescope}/Cosmic Origins Spectrograph. We co-add spectra of different background sources at close projected angular separation on the sky. This novel stacking technique dramatically increases the signal-to-noise ratio of the spectra, allowing detection of low column density gas (down to $EW$ > 2 m\AA). We identify absorption as inflowing or outflowing, by using blue/redshifted high velocity cloud (HVC) absorption components in the Galactocentric rest frame, respectively. The mass surface densities of inflowing and outflowing gas both vary by more than an order of magnitude across the sky, with mean values of 
$\langle \Sigma_{\rm in}\rangle \gtrsim 10^{4.6\pm0.1}$  \msunpkpc\ for inflowing gas and 
$\langle \Sigma_{\rm out}\rangle \gtrsim 10^{3.5\pm 0.1}$ \msunpkpc\ for outflowing gas, respectively. The mass flow rate surface densities (mass flow rates per unit area) also show large variation across the sky with 
$\langle \dot{\Sigma}(d)_{\rm in}\rangle \gtrsim (10^{-3.6\pm0.1})(d/12\,\mathrm{kpc})^{-1}$\,\msunpkpcpyr\ for inflowing and $\langle \dot{\Sigma}(d)_{\rm out}\rangle \gtrsim (10^{-4.8\pm0.1})(d/12\,\mathrm{kpc})^{-1}$\,\msunpkpcpyr\ for outflowing gas, respectively. The regions with highest surface mass density of inflowing gas are clustered at smaller angular scales ($\theta < 40^\circ$). This indicates that most of the mass in inflowing gas is confined to small, well-defined structures, whereas the distribution of outflowing gas is spread more uniformly throughout the sky. Our study confirms that the MW is predominantly accreting gas, but is also losing a non-negligible mass of gas via outflow. 
\end{abstract}




\section{Introduction}

Gas flows determine the evolution and fate of a galaxy. Ongoing star formation in galaxies requires intergalactic fueling and gas recycling to prevent quenching of the star formation \citep{Somerville2015}. The balance between inflows and outflows thus describes whether a galaxy will sustain star formation over long timescales \citep{Tumlinson2017}. The circumgalactic medium (CGM) acts as the moderator in this fuel exchange between the intergalactic medium and the host galaxy’s disk. Inflowing intergalactic material will flow into the CGM before accreting to the disk \citep{Ribaudo2011,Bouche2013,Kacprzak2012} while feedback driven by supernovae, stellar winds, and active galactic nuclei (AGN) push the material into the CGM where it can either be recycled, or be expelled from the dark matter potential well \citep{Bordoloi2011,Rubin2014,Bordoloi2014_outflow,Heckman2017}.

With the installation of the Cosmic Origin Spectrograph (COS) \citep{Green2012} on-board \textit{Hubble} space telescope, the multi-phase CGM has been traced through the wide range of far-UV metal absorption lines tracing out the cool [log($T$/K) > 4] \citep[e.g.][]{Tumlinson2013,Werk2014,Bordoloi2018,Liang2014,Chen2020}, warm [log($T$/K) >5] \citep[e.g.][]{Tripp2011,Tumlinson2011,Bordoloi_2014,Liang2014,Stocke2017,Bordoloi2017} and potentially hot [log($T$/K) > 6] phases of the CGM \citep{Qu2016}. However, spatially resolved mapping of the CGM has remained particularly challenging, simply owing to the paucity of multiple sightlines passing through the CGM of a high-$z$ galaxy. Recent efforts of such studies have relied on combining gravitationally lensed galaxies with moderate resolution ground based integral field spectrographs \citep{Lopez2018}. To conduct such spatially resolved (sub-kpc resolution) studies in high resolution spectroscopy remains the next big challenge in advancing our understanding of the CGM physics. Characterizing the CGM gas properties in these spatial scales is crucial as many of the sub-grid model assumptions in hydrodynamical simulations are also unresolved in these spatial scales.

Our unique vantage point inside the Milky Way (MW) makes the MW an ideal galaxy to study spatially resolved gas flows. Indeed, any quasi-stellar object (QSO) can be used as a background object to illuminate the foreground structures located within the MW CGM with high fidelity. Our vantage point also constrains us to use gas kinematics to identify such gas flows and differentiate these bulk flows from MW's thick and thin disks \citep[e.g.][]{Wakker1997}. Historically we call these bulk flows high velocity clouds (HVCs) and typically identify them with a characteristic local standard of rest velocity $|v_{LSR}| \gtrsim$ 100 \kms\ \citep[e.g.,][]{Richter_2017}. 

Our understanding of HVC properties comes from both radio 21 cm and UV metal-line absorption data obtained over last few decades \citep{Wakker1997,Putman2012,Collins2009,Sembach2003,Lehner2011,Fox_2014,Fox2015,Richter_2017,Shull2009}. HVCs show a range of metallicity ranging from 0.1-1.0 solar \citep[e.g.][]{Wakker1999,Richter2001,fox_2016} and are multi-phase in nature \citep{Putman2012,Sembach2003,Miller2016}.HVCs reside relatively close to us at distances $d <$ 15 kpc \citep{Wakker_2007,Thom2008,Lehner2012}. On average, cool-ionized gas is inflowing onto the MW galaxy at a rate of $\gtrsim$ 0.53 \msunyear and outflowing at a rate of $\gtrsim$ 0.16 \msunyear, respectively \citep{Fox_2019}, and a significant fraction of this mass may also be hidden at lower velocities \citep{Zheng_2019,Qu2020}.

These HVCs are not uniformly distributed over the sky and very distinct local structures are seen both in absorption as well as in 21 cm emission \citep{Putman2012,fox_2016,Bordoloi_2017,Richteretal2017}. Well known gaseous structures such as Complex C represent infalling gas \citep{Wakker1999}, whereas the Smith Cloud might be part of a galactic fountain \citep{Fraternali2015,fox_2016}, where enriched gas is being recycled back onto the disk of the Galaxy. Similarly, kinematics of entrained gas inside the Fermi Bubbles \citep{su_2010} suggest that a nuclear wind is being driven by the nuclear wind of the MW \citep{Fox2015,Bordoloi_2017, Ashley2020}.

These UV absorption-line studies of individual sightlines rely on high signal to noise ratio (SNR, $\geq$ 10--15) and moderate to high resolution individual spectra of background objects \citep{Lehner2012, Richteretal2017,Fox_2019}. However, to-date most \emph{HST}/COS observations of AGN have a SNR $<$ 10. This means that majority ($>$70\%) of archival \emph{HST}/COS spectra are not suitable to study the gas flows around the MW in this traditional method \citep{Peeples2017}. 

A novel way to tap into these otherwise unused low SNR spectra and study the gas flows in the halo of the MW in a spatially resolved manner is to co-add these spectra at different parts of the sky. In this work, we perform spectral stacking of these \emph{HST}/COS spectra from different parts of the sky and create a high-fidelity spatially resolved map of the gas flowing into and out of the MW. Employing a co-addition process gives three primary advantages over the traditional single line of sight approach. First, co-addition process allows us to use a larger sample of available \emph{HST}/COS QSO spectra, which immediately results in better sampling of MW halo. Second, the stacking method offers an increase in the sensitivity of the observation, with detection of absorption lines possible down to EW $\geq$ 2 m\AA, whereas most individual line analysis are restricted to sensitivities of EW $\geq$ 30-50 m\AA . Finally, since the co-added spectra will result in an average spectra in a given spatial region, the potential impact of intervening high-$z$ absorbers as contaminant is significantly reduced \citep{Bordoloi2011}.

In this work, we restrict our study to HVCs, since they can be identified directly from a large sample of spectra. We use these HVCs to quantify the rate at which cool ionized gas is flowing into or out of the MW far away from the galactic plane but within the halo of the MW. However, one caveat to analyzing only HVC components is that gas structures moving with slower radial velocities within the halo ($|v_{\rm LSR}|<100$\kms) are not accounted for \citep{Zheng_2019}. Another caveat is that the HVC observations only cover cool and warm gas phases, but not the hot phase. For both these reasons, the presented flow rates represent a lower limit on the total flow rates.

This paper is organized as follows. Section \ref{Section 2} describes how quasar spectra are selected. Section \ref{Section 3} describes how localized sky regions are selected for co-addition and the processing, co-added, analysis methods used in this work. Section \ref{Section 4} presents the results showing the spatial distribution of inflowing and outflowing gas column densities, velocities, surface mass densities and regional mass flow rates per unit area respectively. Section \ref{Section 5} summarizes the findings of this work.  

\section{Description of Observations}\label{Section 2}

Our analysis utilizes archival medium-resolution (G130M or G160M grating) UV spectra of AGN observed with the \emph{Hubble Space Telescope (HST)} Cosmic Origins Spectrograph (COS) \citep{Green2012}. We retrieve all the archival COS AGN spectra from the Hubble Spectroscopic Legacy Archive (HSLA)  Data Release 2  \citep{Peeples2017} resulting in 775 unique AGN spectra and 602 unique M-grating spectra. We apply the following two criteria to select the final set of spectra that are used in this work. First we identify all sightlines (with M-grating COS spectra) with an average  signal to noise ratio (SNR) per pixel $\langle SNR \rangle$ >2 near the transition of interest, which is necessary to ensure a reasonable continuum fit for that local slice, where we compute the $\langle SNR \rangle$ around $\pm$5\,\AA\ of each transition of interest. Second, we filter out any heavily contaminated spectra, where the spectral slice have $\geq 50\%$ of the pixels registering zero flux. This results in a final sample of 440 total sightlines for the analysis. Our sample contains 170 additional sightlines compared to \cite{Richteretal2017}, the largest individual sightline HVC survey to date. Due to variation in SNR and wavelength coverage of the grating, the available number of sightlines vary slightly between different ionic transitions. We primarily focus on the \ion{Si} {II} 1193/\ion{Si} {III}  1206/\ion{Si} {IV} 1393 transitions, for which there are 391/388/412 available sightlines, respectively. Over all species, we utilize 73\% of all the targets available in the HSLA catalog with G130M or G160M grating observations. Table \ref{tab:Transitions} presents all atomic transitions analyzed in this work.

\begin{table}
	\centering
	\caption{Atomic transitions and oscillator strengths used in this work. Atomic data obtained from \citet{2003ApJS..149..205M}.}
    	\label{tab:Transitions}
        	\begin{tabular}{lcc} 
        		\hline
        		Ion & $\lambda_0$ (\AA) & $f$\\
        		\hline
        		\ion{Si} {II} & 1190.416 & 0.292\\
        		\ion{Si} {II} & 1193.290 & 0.582\\
        		\ion{Si} {II} & 1260.422 & 1.180\\
        		\ion{Si} {II} & 1526.707 & 0.133\\
        		\ion{Si} {III} & 1206.500 & 1.630\\
        		\ion{Si} {IV} &  1393.760 & 0.513\\
        		\ion{Si} {IV} &  1402.773 & 0.254\\
        		\ion{C} {II} & 1334.532 & 0.128\\
        		\hline
            \end{tabular}
\end{table}

Table \ref{tab:Mean SNR} shows the number of spectra in different SNR bin around each individual transition. Typical absorption line spectroscopy relies on higher SNR data to accurately measure individual absorption lines \citep[e.g.][]{Tumlinson2013,Bordoloi_2014,Richter_2017,Danforth2014}. Assuming an optimistic scenario where a spectrum with SNR >5 allows us to detect individual ions, we can compute how many additional spectra are available for this analysis that will be unsuitable for a traditional individual sightline analysis. We separate Table \ref{tab:Mean SNR} into three categories: rejected spectra (SNR $\leq$ 2), spectra that traditional absorption line analysis typically reject (2 $\leq$ SNR $\leq 5$), and spectra with sufficient SNR  ($S/N > 5$). Our approach  enables the use of an additional 225--240 spectra to investigate the spatial variation of mass flow rates around the MW.

\begin{table}
	\caption{Number of \emph{HST}/COS quasar spectra at different SNR bins around individual transitions.}
	\label{tab:Mean SNR}
	\begin{tabular}{lccc}
		\hline
	    SNR & \ion{Si}{IV} $\lambda 1393$ (\AA) & \ion{Si}{III} $\lambda 1206$ (\AA)& \ion{Si}{II} $\lambda 1193$ (\AA) \\
	    \hline
	    SNR $\leq$2 & 148 & 182 & 179\\
		2$<$SNR $\leq$5 & 240 & 225 & 231\\
		SNR  >5 & 181 & 163 & 160\\
		\hline
	\end{tabular}
\end{table}

\begin{figure*}
    \includegraphics[width=16cm,height=8cm]{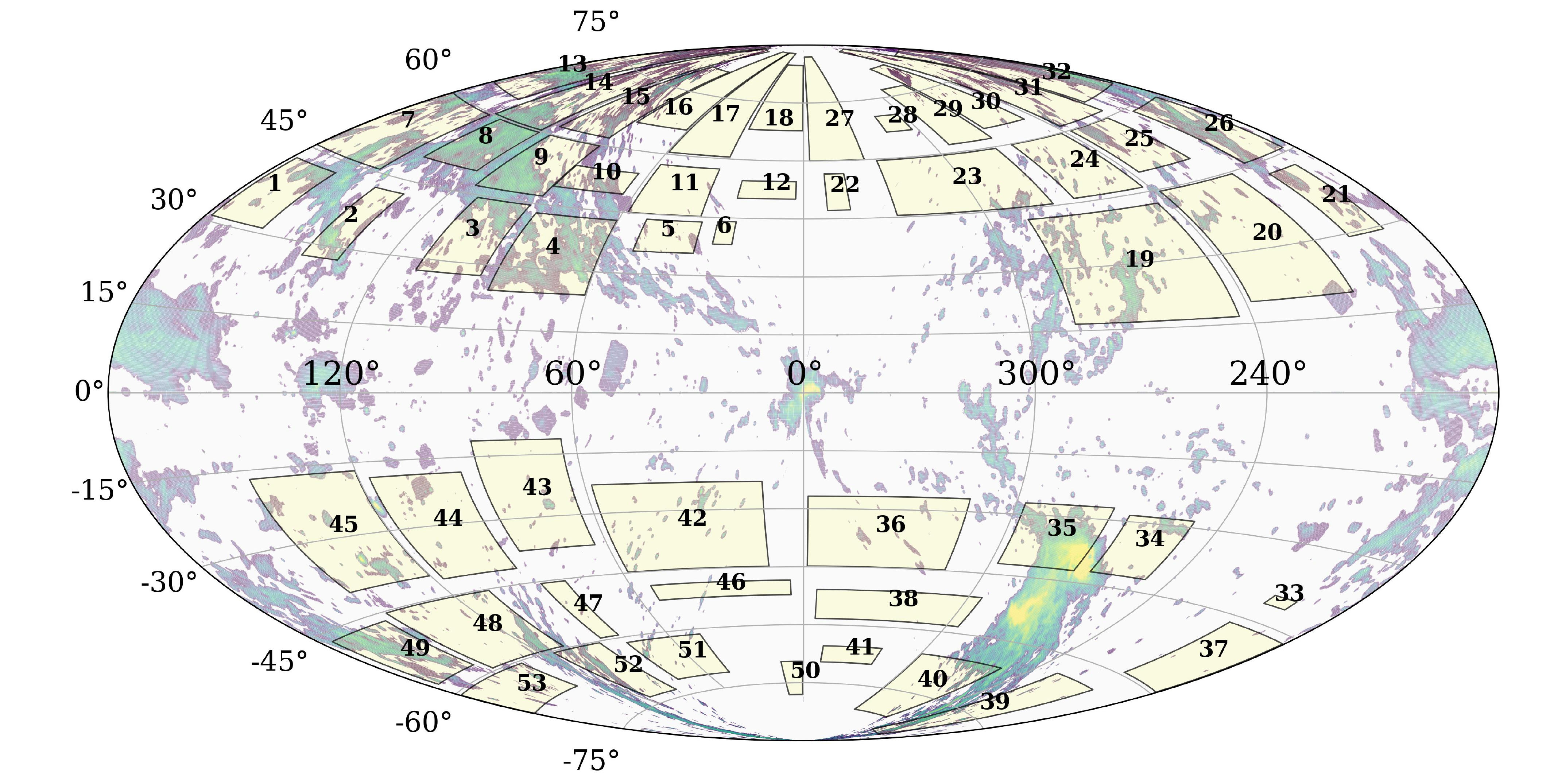}
    \caption{All-sky map in Galactic coordinates showing the location of our target regions. Each shaded region represents the solid angle subtended by the sightlines used to perform the co-addition of the AGN spectra. The solid angles are determined by the maximum difference in latitude and longitude between the lines of sight within that region. The region identifiers in the middle of each shaded region are as given in Table \ref{tab:SolidAngProperties}.  The background colored map shows the location of major 21 cm emission HVCs on the sky adopted from \citet{10.1093/mnras/stx2757} and reoriented to display the Galactic center at the center of the display.}
    \label{fig:SolidAng}
\end{figure*}

\section{Methodology}\label{Section 3}

In the following section, we lay out the methods used to divide the sky into distinct local regions, co-add the  spectra, identify the HVCs, filter out contaminants, and quantify the average gas column densities and kinematics towards different regions of the sky.

\subsection{Selecting local regions on the sky}

Our aim is to study the spatial variation of gas flows into and out of the MW. To achieve that, we initially divide the sky into 53 independent distinct regions with $\delta l=30\degree$ and $\delta b=15\degree$.
This region size was defined to ensure that each initial region would include a minimum of 3 QSOs. As there are smaller number of UV bright QSOs near the Galactic disk, the regions near the Galactic disk are selected with an initial $\delta b=45\degree$. 

We select all the AGN within the footprint of each region that satisfy the criteria described in section 2. We  compute the projected area on the sky subtended by the AGN in each region. For each individual region, the associated area on the sky is the product of the largest separation in latitude and longitude between the individual targets in that region, and we convert them to the solid angle subtended on that region of the sky as:

\begin{equation}
    \Omega = (\delta l) (\delta b) \left(\frac{\pi}{180}\right)^2 , 
\end{equation}

The size of each region in Figure~\ref{fig:SolidAng} corresponds to the solid angle associated with that region. We present all the individual regions along with their general properties (i.e. region ID, location, QSO count, and solid angle) in Table~\ref{tab:SolidAngProperties}.

\subsection{Processing and co-adding the spectra}

We bin each individual 1D quasar spectrum obtained from HSLA to Nyquist sampling with two bins per resolution element to maximize their SNR without losing any information \citep{Bordoloi_2014}. For the rest of the analysis we use these re-sampled spectra. All the 1-D spectra in the HSLA database have wavelength solutions in the heliocentric frame of reference. We transform the wavelength solution to the local standard of rest (LSR) using the following equations:

\begin{eqnarray}
    v_{helio} = \frac{\lambda- \lambda_{0}}{\lambda_{0}}c, \\
     v_{LSR} = v_{helio} + 9\,\cos{l} \cos{b} + 12\,\sin{l} \cos{b} + 7\sin{b}, 
\end{eqnarray}
here $v_{helio}$ is the heliocentric velocity centered on an individual transition at a rest frame wavelength $\lambda_{0}$, $\lambda$ is the observed wavelength and $c$ is the speed of light. $ v_{LSR}$ is the LSR velocity in \kms, $l$ and $b$ are galactic latitudes and longitudes, respectively.

For individual lines of sight, we select a $\pm5$\AA\ region around the rest-frame wavelength of each transition. We perform a local continuum fit on these sliced spectra, in the LSR frame. We perform an iterative sigma-clipping process to isolate the continuum flux from any absorption and emission components and then fit a  multi-order polynomial. We divide the observed flux by this continuum fit to derive the continuum-normalized flux for the spectral slice. We visually inspect each continuum fit to ensure quality. For any bad continuum fit, we interactively perform a local continuum fit with spline function using the
\textit{rbcodes} package\footnote{https://github.com/rongmon/rbcodes} \citep{rbcodes}. 
Each QSO's continuum fit for all transitions analyzed can be found in the repository for this analysis \footnote{https://github.com/saclark7/Investigating-Spatially-Resolved-Gas-Flows--MW}.

All individual spectral slices, now in the LSR frame and continuum normalized, are ready to be co-added. The co-addition process combines the individually processed spectra within their respective, predefined regions of the sky. Co-addition increases the SNR of the resultant spectra, enabling detection of diffuse, low-column-density gas in the MW halo, 
and negating the need to filter out contamination from higher-$z$ intergalactic absorption lines.

For each transition, we select all continuum-normalized spectral slices within each region on the sky. We co-add these spectra by computing their inverse-variance-weighted mean flux at each pixel. We also attempt to co-add the spectra using sigma-clipped mean and median values and find that such methods are less sensitive to the weakest absorption features hidden in the noise. This is primarily due to fainter (noisier) objects dominating the resultant spectra. We find that a co-add computed through the inverse-variance-weighting best captures the HVC absorption components of all individual sightlines contained within the resultant spectra. Additionally, each QSO spectrum included in the HSLA database uses the most recent CalCOS wavelength calibration available at the time of release \citep{Peeples2017}, because the HSLA spectra are formed by co-adding the standard CalCOS reduction of the data. The velocity zeropoint error in CalCOS is 7.5 km/s \citep[STScI Instrument Science Report COS 2018-24]{Plesha2018}. This uncertainty is small enough to not affect the science conclusions presented in the paper, because it should not introduce a significant contribution in the shape or characteristics of the stacked spectra. Once a co-added spectra has been obtained, average coordinates from the individual sight lines comprising the co-add are assigned to this region and the weighted average flux values are used for the remainder of the analysis. As an example, a region analyzed for the Si transitions (and \ion{C}{II} for comparison to the \ion{Si}{III} transition) is presented in Figure \ref{fig:Patch31}. All the co-added spectra used in this work are presented in Appendix~\ref{Coadds}.

\begin{figure}
    \centering
    \includegraphics[width=\columnwidth,height = 15cm]{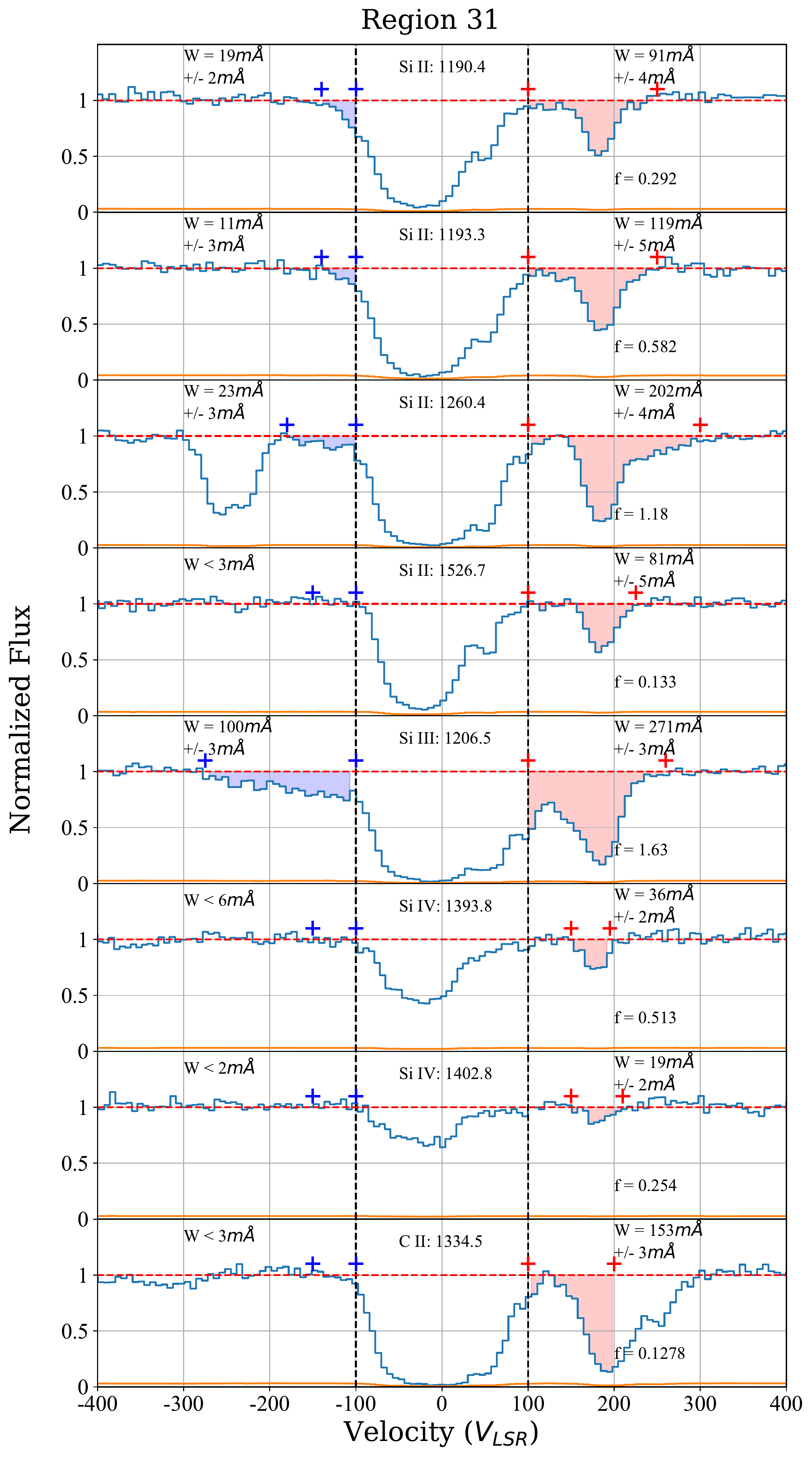}
    \caption{Example stacked \emph{HST}/COS  QSO spectrum (light blue line) towards region 31 showing the \ion{Si} {II}, \ion{Si} {III}, \ion{Si} {IV}, and \ion{C} {II} transitions, respectively. The blue shaded regions represent blueshifted HVCs detected in the $v_{\rm LSR}$ frame, whereas red shaded regions depict redshifted HVCs. The blue and red crosses indicate the integration limits used to compute absorption strength and kinematics for each co-added spectrum. The measured rest frame equivalent widths are in m\AA. For any non-detection, we report the 2$\sigma$ upper limit of non-detection in rest frame equivalent width. The error spectrum associated with the co-add (pale red line) is presented in the bottom of each panel. The absorption component located at $-$215 \kms\ in \ion{Si}{II} 1260 is \ion{S}{II} 1259, and we exclude that in the analysis. The absorption component at 263 \kms\ in the \ion{C}{II} 1334 transition is the \ion{C}{II}* 1335 line, and we exclude that in the analysis.}
    \label{fig:Patch31}
\end{figure}

\subsection{Contamination Filtering} \label{subsec:contamination}

This study focuses on measuring the properties of inflowing and outflowing gas across the entire MW halo. We exclude contributions from two well known structures, the Fermi Bubbles and the Magellanic Stream, which may unduly influence the results \citep[see ][]{Fox_2019}. We treat any HVC contribution towards these structures as contamination and remove them before further analysis as follows.

The Fermi Bubbles are two giant lobes of plasma seen towards the Galactic Center \citep{su_2010}. They host entrained cool and warm gas that trace the kinematics of the MW nuclear outflow \citep{Fox2015,Bordoloi_2017,Karim_2018,Ashley2020}. We first roughly identify the spatial location of the Fermi Bubbles as two 27 degree radius circles centered at ($l,b$) = (0,27$\degree$) and (0,$-$27$\degree$) for the Northern and Southern hemispheres, respectively. The regions where the lines of sight are completely encompassed within these circles are removed from the analysis. However, region 22 and region 38 are borderline regions that have a mixture of individual QSOs falling both within and outside the identified Fermi Bubble contours. For these two regions, we cross match the coordinates of individual QSOs to reject any sightlines that are definitely within the Fermi Bubble boundaries. We compare the QSO coordinates  to the catalogs from  \citet{Bordoloi_2017} and \citet{Karim_2018}  for the northern and the southern Fermi Bubbles, respectively.

We also identify the HVC contribution from the Magellanic Stream \citep{donghia2016} and filter it out as follows. We first cross-match the individual sightlines with the targets in the \citet{Fox_2014} survey of the Magellanic Stream. For each sightline in that survey, any velocity range identified in \citet{Fox_2014} as Magellanic stream is flagged and not used for the rest of the analysis. If we cannot cross-match a target, its coordinates are matched with the closest target from the \cite{Fox_2014} catalogue and the velocity range associated with Magellanic Stream is removed from the rest of the analysis. If all sightlines within a region contain a contribution from the Magellanic Stream, sightlines with contamination beginning at lower absolute velocities are masked and then a regional bound on the velocity integration limits is  placed on the resultant co-add. For example, region 41 consists of five sightlines. Targets HE0226-4110 and RBS144 have MS contamination beginning at 
$v_{\rm LSR}\approx100$\kms, whereas the remaining three sightlines exhibit contamination starting at $\gtrsim 155$\kms. The two sightlines with contamination near 100\kms\ have their individual spectra masked within the contaminated regions, while a general integration-bound restriction is placed at $v_{\rm LSR}$ = 155\kms\ for the co-added spectra. Finally, the majority of targets within the Leading Arm of the Magellanic Stream contain contamination that begins below 100\kms, and therefore the contamination regions of individual targets cannot be simply masked. Therefore, the Leading Arm (Region 19) is completely removed from the analysis. 

\subsection{HVC Detection and Measurement}

For a given spatial region, once all the spectra are co-added, we visually inspect the absorption profiles of each transition of a given ion to search for contamination and to set the velocity range for column density calculation. For each identified HVC absorption component, we compute the apparent optical depth (AOD) column densities \citep{1991ApJ...379..245S, doi:10.1146/annurev.astro.34.1.279} and AOD-weighted line centroids as follows:

\begin{equation}
    N_{\lambda}(v) = \frac{m_{e}c} {\pi e^{2}} \frac{-{\rm ln}[F_{n}(v)]} {f\lambda_{0}} 
    = 3.768\times10^{14} \frac{-{\rm ln}[F_{n}(v)]} {f\lambda_{0}},
    \label{eq: Tau_a} 
\end{equation}
\begin{equation}
    N_{\lambda} = \int_{v-}^{v+} N_{\lambda}(v)dv,
\end{equation}

where $m_{e}$ is the mass of an electron, $f$ is the oscillator strength of a given transition, $F_{n}$ is the co-added normalized flux, and the integration limits ($v-,v+$) represent the velocity domain of the identified HVC absorption region. We define the line centroid velocities as the AOD-weighted mean velocity of the transition.  Figure \ref{fig:Patch31} shows the co-added spectra around region 31 of the sky. There are 11 QSO spectra being used for this co-add. Here we detect both blue and redshifted HVC absortion (marked as red and blue shaded regions) in \ion{Si} {II/III/IV} and \ion{C}{II} ions. Note that the co-added spectrum has very high SNR and the mean 3$\sigma$ detection threshold in this co-added spectrum is $\gtrsim$8.5 {m\AA}.  For any non-detection, we present the rest frame equivalent width 2$\sigma$ upper limit of non-detection.

The detection thresholds in the co-added spectra are sensitive to two factors: the number of sightlines available to be coadded within a region, and the SNR of the individual spectra used in the co-addition. We quantify the 3$\sigma$ detection threshold in all the co-added spectra and compute the mean and standard deviation of all 3$\sigma$ detection thresholds for \ion{Si} {II/III/IV} used in this work.  The mean 3$\sigma$ detection threshold for $\log N_\ion{Si}{III}$ is 11.66$\pm$ 0.36. Similarly, for $\log N_\ion{Si}{II}$ and $\log N_\ion{Si}{IV}$, the mean detection thresholds are 12.07$\pm$0.34 and 11.84$\pm$0.34, respectively. The small spread in detection thresholds between different regions suggests there is no large bias in detection sensitivity across the sky. To investigate if there is local bias within the detection thresholds, we divide the sky into four quadrants and group the 3$\sigma$ detection threshold values for the regions within each quadrant.  A two sample K-S test shows that these detection thresholds are consistent with being drawn from the same parent distribution across the four quadrants. This shows that there is no systematic bias in detection sensitivity of this survey across the sky.  Furthermore, the presence of individual regions with higher detection thresholds where weak lines may not be detected will not change the total mass budget calculation, because the highest column density regions dominate the contribution to the overall mass budget.

To check for contamination from high-$z$ intervening absorption line systems, for each blue or red shifted absorption system, we compute the rest frame equivalent widths of all associated transitions. For the \ion{Si} {II} absorption, we use the 1190, 1193, 1260 and 1526 transitions. We first compute the rest frame equivalent width ratios between the \ion{Si} {II} 1190 and \ion{Si} {II} 1193 transitions and compare them to their corresponding oscillator strengths \citep[$f$-values;][]{2003ApJS..149..205M}.  If the equivalent width ratios (within their measurement uncertainties) are between 1 and the ratio of their corresponding oscillator strengths, then the line ratios are considered to be physical and without contamination \citep{Bordoloi_2014} and the \ion{Si} {II} 1193 measurements are used in the study. In our analysis the majority (78\%) of the  \ion{Si} {II} 1190/ Si II 1193 line ratios are found to be acceptable (Figure Appendix \ref{fig:EW_Ratio}). For the equivalent width ratios that do not meet this criteria, the regions are flagged and the process is repeated between the  \ion{Si} {II} 1193 and  \ion{Si} {II}  1260 transitions. Ten out of the twelve absorbers are within acceptable equivalent width ratios; and in these cases we adopt their \ion{Si} {II} 1193 measurements for the rest of the analysis.  For the two remaining regions (regions 18 and 40), this process is repeated for the \ion{Si} {II} 1190 and \ion{Si} {II} 1260 transitions. These line ratios are in agreement with their corresponding oscillator strength ratios and so we adopted their \ion{Si} {II} 1260 line measurements. The HVC components of these two \ion{Si} {II} 1260 lines do not appear to be saturated and so the adopted AOD column densities and velocities are robust. All the line ratios and the corresponding acceptable equivalent width ratio ranges are presented in Figure Appendix \ref{fig:EW_Ratio}. For the \ion{Si}{IV} 1193 and \ion{Si}{IV} 1402 transitions, we perform the same line ratio test as described above and find that all the \ion{Si}{IV} transitions have a physically acceptable doublet ratios.

\ion{Si}{III} 1206 is a singlet, and we do not have any additional lines to check for contamination. However, as \ion{Si}{III} 1206 is blueward of the observed frame 1215 {\AA}, any intervening (IGM/CGM) Ly-$\alpha$ line will never contaminate it. Another possibility is contamination by (z $\sim$0.176) Ly-$\beta$ line and other higher-$z$ metal lines. This has low probability as most of the HST QSOs are not at very high redshifts. \cite{Bordoloi_2017} studied a sample of 47 QSO sightlines around the Fermi Bubbles region to quantify the Fermi Bubbles kinematics. A detailed line identification with high SNR individual spectra revealed that only 1 \ion{Si}{III} 1206 HVC had contamination from a high-$z$ Ly-$\beta$ absorber, corresponding to a contamination rate of $\sim$ 2\%. Low contamination rates in high SNR literature samples and the fact that several sightlines are co-added in each stack results in the co-adds washing out any spurious contribution from intervening IGM absorbers. Hence on average \ion{Si}{III} 1206 measurements are quite robust.

After we identify and measure the kinematics and absorption strengths of individual HVC regions in the co-added spectra, we proceed to segregate the HVCs as inflowing or outflowing material. We follow the method described in \cite{Fox_2019} by transforming all measured HVC velocity centroids to the Galactic Standard of Rest (GSR) frame using:

\begin{equation}
    v_{GSR} = v_{LSR} + 240\; {\rm km s}^{-1}\,\sin{l} \cos {b},
\end{equation}

where $l$ and $b$ are the mean coordinates of the local regions under study, $v_{LSR}$ and $v_{GSR}$ are the velocity centroids of the identified HVCs in the LSR frame and GSR frame, respectively. We adopt a galactic rotation velocity of 240 \kms\ \citep{Reid_2014} for this transformation. Following \citet{Fox_2019}, we define the inflowing gas as any HVC with $v_{\rm GSR}<0$\kms\ and outflow gas as any HVC with $v_{\rm GSR}>0$\kms. 
This distinction is generally valid for distant clouds, but breaks down at low distance. Since the mean distance to HVCs is
12 kpc \citep[][which we adopt in our calculations,]{Lehner2011} the clouds are distant enough that they would not be expected to have a significant rotational component associated with the disk, and the use of the GSR frame
is justified \citep{1991A&A...250..499W}.

\section{Results}\label{Section 4}
In this section, we present the variation in column density and velocity of inflowing and outflowing gas across the sky. We then discuss how mass inflow and outflow rates vary across different parts of the sky and how the total mass of inflowing and outflowing gas varies across the sky.  

\subsection{Spatial Variations of Column Densities and Velocities}
\begin{figure*}
    \centering
    \includegraphics[width = 2\columnwidth]{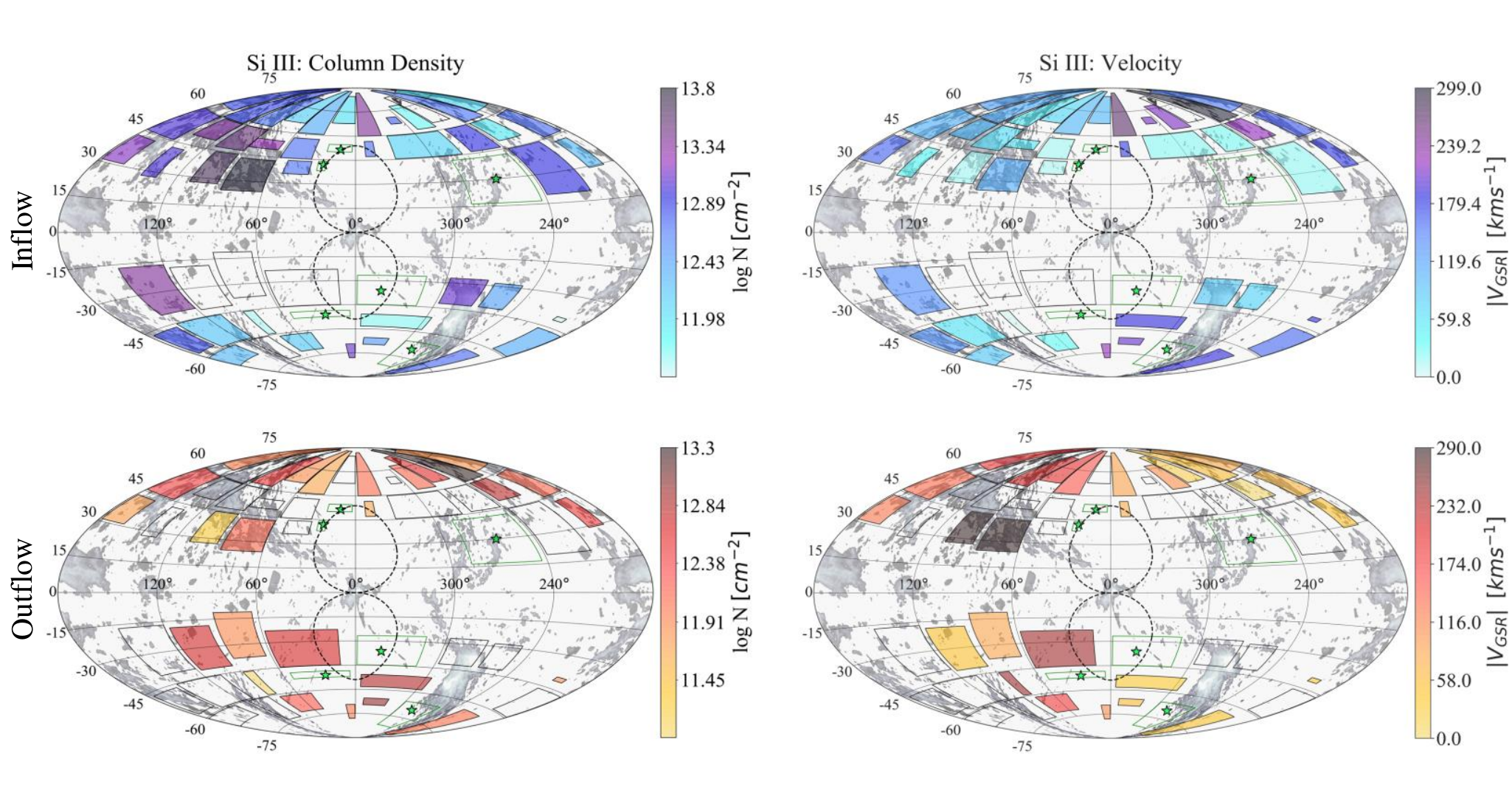}
    \caption{All-sky maps showing the spatial distribution of inflow and outflow in \ion{Si} {III} absorption. The top row presents the characteristics for inflowing material ($v_{\rm GSR}<0$\kms), while the bottom row shows the distribution for outflowing material ($v_{\rm GSR}>0$\kms). The left column shows the variation in column density across the sky, while the right column shows the mean absolute centroid velocity in the GSR frame of detected HVCs. The H I 21cm maps from \citet{10.1093/mnras/stx2757} are overlaid to provide context for the significant variation. The highest detected column density regions are associated with the well known Complex C. Additionally, non-detections are denoted by empty regions and completely filtered regions are denoted by a green border and green stars. The marker points represesnt the mean galactic coordinates, but are still representative of the spatial regions as defined in Figure \ref{fig:SolidAng}.}
    \label{fig:SiIII_Characteristics}
\end{figure*}

We first present the variation of column densities and velocities of individual transitions across the sky. We first focus on \ion{Si} {III} 1206 transition, as it is the strongest UV transition line used in this work.  Figure \ref{fig:SiIII_Characteristics} shows the observed variation in column density and  velocity ($v_{\rm GSR}$) of both inflowing and outflowing material, for the \ion{Si} {III} transition. The individual regions are color coded to reflect their corresponding column densities (left panels) and velocities (right panels), respectively. The open patches mark regions with no statistically significant detection. The green stars denote regions that are filtered out for contamination as described in Section~\ref{subsec:contamination}.

\ion{Si} {III} ion is detected almost evenly throughout the sky at $|b|>0\degr$, in  both inflowing and outflowing material, respectively. However, the incidence of detection of inflowing \ion{Si} {III} absorption is 34\% higher than that of outflowing \ion{Si} {III} absorption. Across all regions studied, the incidence of detection of \ion{Si} {III} absorption for inflowing gas is 91.5$\pm $4.1\% (43/47), and that for outflowing gas is 57.4$\pm$7.1\% (27/47), respectively.

This difference in incidence of detection between inflowing and outflowing gas is even more stark when we look at \ion{Si}{II} and \ion{Si}{IV} transitions (Appendix \ref{Regional Properties}), respectively. For \ion{Si}{II}, the fraction of regions showing inflow and outflow are 72.3$\pm$6.4\% (34/47) and 25.5$\pm$6.3\% (12/47), and for \ion{Si}{IV} the same numbers are 51.1$\pm$7.2\% (24/47) and 25.5$\pm$6.3\% (12/47), respectively. It is noteworthy that for both the the \ion{Si}{II} and \ion{Si}{IV} outflows are primarily detected in the regions around the galactic center with nearly all detection being confined within $\pm 120 \degree$ of the Galactic Center (Figures \ref{fig:SiII_Characteristics}, \ref{fig:SiIV_Characteristics}). This suggests that the inflowing and outflowing gas are not spatially correlated, particularly in these ionization states.

We now investigate if there is spatial variation in the properties of inflowing/outflowing gas detected in different parts of the sky. For that purpose we divide the whole sky into two cases. Case 1: we divide the sky into two hemispheres, North ($b>$0$\degr$) and South ($b<$0$\degr$) hemispheres. Case 2: we divide the sky between East ($360\degr \leq l \leq 180 \degr$) and West ($180\degr \leq l \leq 0 \degr$), hemispheres.
 
We first investigate if the incidence of detection of inflowing/outflowing gas is the same in different parts of the sky. We first focus on the difference between the northern and southern hemispheres. 

Case 1: The incidence of detection for inflowing gas on the Northern hemisphere is 83$\%\pm6.9\%$ (24/29), 97$\%\pm3.7\%$ (28/29), and 69$\%\pm8.4\%$ (20/29) for the \ion{Si} {II/III/IV} transitions respectively. The same for the Southern hemisphere is $56\%\pm11.3\%$ (10/18), 78$\%\pm9.6\%$ (14/18), and 22$\%\pm9.6\%$ (4/18) respectively. These measurements clearly suggest that for the \ion{Si}{IV} transition there is a clear North-South asymmetry in terms of incidence of detection of inflowing gas. We perform the adjusted Chi-squared test with Yate’s correction for continuity on the incidence of inflowing \ion{Si} {II/III/IV} transitions. The P-values for these transitions are 0.09, 0.12 and 0.004, respectively. These P-values suggest that particularly for inflowing \ion{Si}{IV} gas, we can rule out the null hypothesis that the incidence of detection between Northern and Southern hemisphere are the same at more than 99.5$\%$ confidence level. This asymmetry is also present with marginal significance for the detection rate of inflowing \ion{Si}{II} gas (P=0.09). These findings suggest that if we exclude the Magellanic Stream, most of the inflowing gas around the MW is located in the Northern hemisphere ($b>$0$\degr$).
 
 Similarly for outflowing gas, the \ion{Si} {II/III/IV} detection rates for the Northern hemisphere are 17$\%\pm6.9\%$ (5/29), $59\%\pm8.9\%$ (17/29), and $28\%\pm8.1\%$ (8/29), and that for the southern Hemisphere are $39\%\pm11.1\%$ (7/18), $56\%\pm11.3\%$ (10/18), and $22\%\pm9.6\%$ (4/18), respectively. Here we do not find any statistically significant North-South asymmetry in detection rates of any ion. The P-values for the adjusted Chi-Squared test for outflowing \ion{Si} {II/III/IV} transitions are 0.19, 0.92, and 0.95, respectively. These results show that there is no North-South asymmetry in the incidence rate of outflowing gas.
 
 Case 2: Here we investigate any variation in incidence of detection between the Eastern and Western hemispheres. The incidence of detection for inflowing gas in the Eastern hemisphere is $74\%\pm8.3\%$ (20/27), $85\%\pm6.8\% (23/27)$, and $52\%\pm9.4\%$ (14/27) for \ion{Si} {II/III/IV} transitions respectively. The same for the Western hemisphere is 70$\%\pm10.0\%$ (14/20), $95\%\pm5.2\%$ (19/20), and $50\%\pm10.9\%$ (10/20) respectively. Here, the incidence of detections do not clearly show any East-West asymmetry. The P-values from an adjusted Chi-Squared test for these transitions are 0.98, 0.55, and 0.87 for \ion{Si} {II/III/IV}, respectively.
 
 For the outflowing gas, the incidence of detection in the Eastern hemisphere are 19$\%\pm7.4\%$ (5/27), 52$\%\pm9.4\%$ (14/27), and $26\%\pm8.3\%$ (7/27), respectively. While for the Western hemisphere the detection rates are $35\%\pm10.4\%$ (7/20), $65\%\pm10.4\%$ (13/20), and $25\%\pm9.5\%$ (10/20), respectively. Like the outflowing results for case 1, there are no statistically significant asymmetries apparent from detection rates of outflowing gas in the eastern and Western hemispheres with P-values of 0.35, 0.55, and 0.79.
 
This exercise shows that the incidence of detection for outflowing gas do not show any East-West or North-South asymmetry. However, they are more confined towards the Galactic Center in \ion{Si}{IV} (Figure \ref{fig:SiIV_Characteristics}). For inflowing gas, once we exclude the Magellanic Stream \citep[which will eventually merge into the MW, ][]{Fox_2014}, bulk of the inflowing gas is detected in the Northern hemisphere.

As seen in Figures \ref{fig:SiIII_Characteristics}, \ref{fig:SiII_Characteristics}, and \ref{fig:SiIV_Characteristics}, there is significant variation in the column densities across the sky for both inflows and outflows. Specifically, for inflowing metals, the variation in column density across regions is observed to be on the order of 1.64, 2.34 and 2.68 orders of magnitude for \ion{Si}{IV}, \ion{Si}{III}, and \ion{Si}{II}, respectively. For outflowing metals, the variation is 1.19, 2.31 and 1.8 orders of magnitude for \ion{Si}{IV}, \ion{Si}{III}, and \ion{Si}{II}, respectively. 

 We compute the column density ratios between the \ion{Si}{II/III/IV} ions and check if the column density ratios vary across the sky. These measurements are shown in Figure \ref{fig:column_density_ratios} (Appendix  \ref{Regional Properties}). For inflowing gas, in both \ion{Si}{II}/\ion{Si}{IV} and \ion{Si}{III}/\ion{Si}{IV} ratios, the regions with relatively higher \ion{Si}{IV} column densities are typically found towards higher latitudes $(|b|>60)$, and typically towards the galactic center (l $\sim\pm90 \degree$). For outflowing absorbers, regions with lower \ion{Si}{II}/\ion{Si}{IV} column density ratios  are restricted to the Northern hemisphere around the regions associated with Complex C and M. However, these primarily consisted of lower limits (i.e. only detection of \ion{Si}{IV}), and are not conclusive.

\subsection{Regional Mass and Gas Flow Rate} \label{Mass and flow rates}

We now compute the total mass, surface mass density and mass flow rates per unit area of inflowing and outflowing gas in each individual region on the sky, and describe their variation across the sky. The mass estimates described below are derived from the three primary tracers \ion{Si}{II}, \ion{Si}{III}, and \ion{Si}{IV}.

\subsubsection{Gas Mass and Surface Mass Density}\label{Gas Mass and Total Mass}

We follow \cite{Fox_2019} and assume that all inflowing and outflowing gas are distributed in a thin shell at a fixed distance in each individual regions. High-velocity clouds are measured to exist within a range of $\sim$ 5--15 kpc. This range is measured using statistical analysis of HVC distances \citet{Lehner2011} as well as individual cloud distances from absorption-line studies towards stars at known distance 
\citep{Wakker_2007,Wakker_2008,Thom2008,Smoker2011}. For simplicity we assume a constant HVC distance $d$= 12 kpc throughout this work, but our measurements can easily be scaled for any distance using the parametric equations as follows. The total (inflowing/outflowing) gas mass within each spatial region of solid angle $\Omega$ is defined as:
\begin{align}
        M_{\rm in/out}(\Omega) &= 1.4m_{\rm H}\langle N_{\rm H} \rangle_{\rm in/out}\Omega d^{2} \\
         \nonumber &\approx 10^{6}M_{\odot}\left(\frac{\langle N_{\rm H} \rangle_{\rm in/out}}{10^{18.8} \mathrm{cm}^{-2}}\right)\left(\frac{\Omega}{0.1\,\mathrm{sr}}\right)\left(\frac{d}{12\,\rm{kpc}}\right)^{2},
\end{align}
\label{eq: Mass_eq}
where $\Omega$ is the solid angle subtended by the individual region studied, $d$ is the distance to the inflowing/outflowing gas, $m_{\rm H}$ is the mass of the hydrogen atom, the factor 1.4 accounts for the mass of helium and metals, and the average column density of hydrogen $N_{\rm H}$ can be written as:

\begin{equation}
    \langle N_{\rm H} \rangle_{\rm in/out}  = \frac{\langle N_{\rm Si}\rangle_{\rm in/out}}{(Z/Z_{\odot}) 10^{-4.49} 10^{-0.26}},
\end{equation}

where $10^{-4.49}$ is the solar silicon abundance \citep{doi:10.1146/annurev.astro.46.060407.145222}, $10^{-0.26}$ accounts for dust depletion in the MW halo \citep{sembach1996gas}, and we adopt a metallicity relative to solar ($Z/Z_\odot$) of 20${\%}$ for inflowing clouds \citep[as measured in Complex C;][]{shull_2011}
and 50${\%}$ for outflowing clouds \citep[as measured in the Smith Cloud;][]{fox_2016}, and:

\begin{equation}
    \langle N_{\rm Si}\rangle =\langle N({\rm Si\,II}) \rangle + \langle N({\rm Si\,III}) \rangle + \langle N({\rm Si\,IV}) \rangle. 
    \label{eq: Navg}
\end{equation}

We assume that the contribution from higher ionization states of silicon is negligible, so that all the sillicon mass in confined within the \ion{Si} {II/III/IV} phase. This conservative approximation ensures that the calculated $\langle N_{\rm Si}\rangle$ value is the minimum Si column density and therefore the total regional gas mass can be considered lower limits.

\begin{figure}
    \includegraphics[width = \columnwidth]{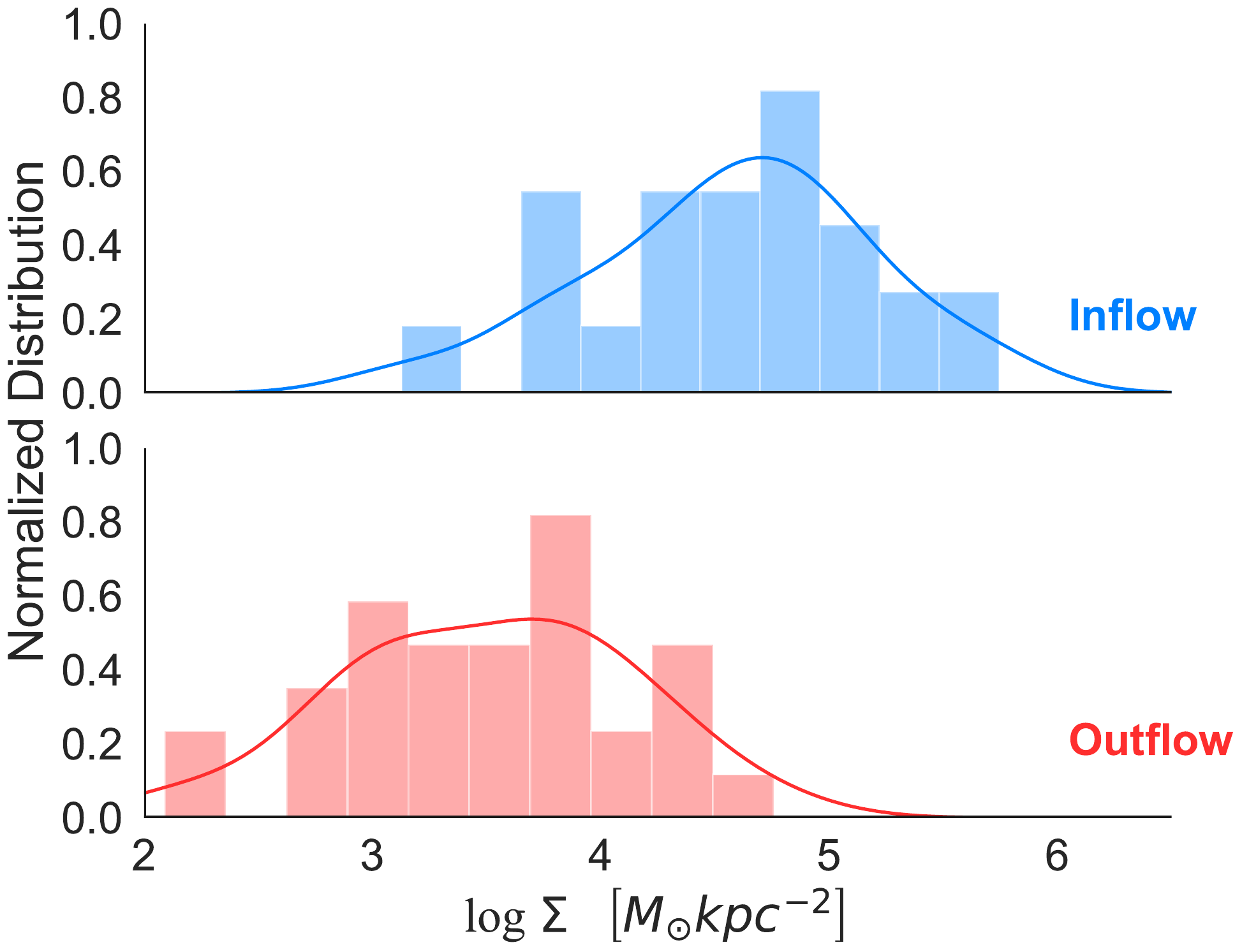} 
    \caption{Distribution of inflowing (top panel) and outflowing (bottom panel) gas surface mass densities. Both the surface mass distributions are normalized for presentation. The coloured lines represent the probability density estimation for inflowing (blue line) and outflowing (red line) gas, respectively.}
    \label{fig:Mass Distribution}
\end{figure}

\begin{figure*}
     \includegraphics[width =2\columnwidth]{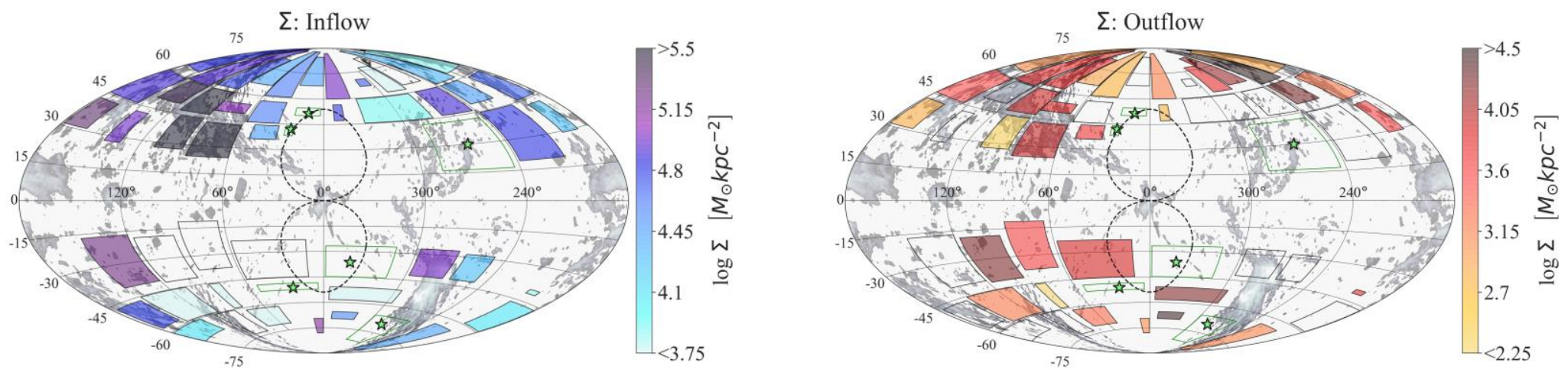}
     \caption{All-sky maps of the regional surface mass density for inflowing (left panel) and outflowing (right panel) material, respectively. The surface mass densities are obtained by summing  each region's average ionic column densities (from \ion{Si} {II}, \ion{Si} {III}, and \ion{Si} {IV} contributors)  as presented in equation 10. Green stars represent completely filtered regions, and open regions denote non-detections.}
     \label{fig: SMD_allsky}
 \end{figure*}

The co-added metal column densities are averages over individual regions. We can infer the regional surface mass density of inflowing or outflowing material within each region according to: 
\begin{equation}
    \Sigma_{\rm in,out} = 7.07 \times 10^{4} M_{\odot} \; \mathrm{kpc}^{-2}\,\left(\frac{\langle N_{\rm H} \rangle_{\rm in/out}}{10^{18.8}\,\mathrm{cm}^{-2}}\right).
\label{eq: SMD}
\end{equation}
These surface mass densities are distance-independent and proportional to the mean hydrogen column density across the region. They are
presented in Figures \ref{fig:Mass Distribution} and \ref{fig: SMD_allsky} and should be considered conservative lower limits similar to Equation \ref{eq: Navg}, because they only account for observed ionization states.

Figure \ref{fig:Mass Distribution} presents the $\Sigma$ distribution of inflowing (top panel) and outflowing (bottom panel) gas around different regions of the MW. The surface mass density of inflowing gas is on average significantly higher than outflowing gas showing that the MW is dominated by inflowing gas. The mean $\log \Sigma$ of inflowing gas is 4.57$\pm$0.09 $M_{\odot}\,\mathrm{kpc}^{-2}$ while individual regions show a variation of 1.21 dex within a 95\% confidence interval. For outflowing material, the mean $\log \Sigma$ is 3.51$\pm$0.11 $M_{\odot}\,\mathrm{kpc}^{-2}$ with a regional variation of 1.23 dex within a 95\% confidence interval. Clearly, the mean inflowing surface mass density is an order of magnitude higher than the mean outflowing surface mass density. We perform a two sample KS test between the $\Sigma$ distribution of inflowing and outflowing gas and rule out the null hypothesis that these two are drawn from the same parent distribution at $>$5$\sigma$ confidence level.

Figure \ref{fig: SMD_allsky} shows the measured values of $\Sigma$ across the whole sky for inflowing (left panel) and outflowing (right panel) gas, respectively. This clearly demonstrates that the inflowing and outflowing gas surface mass densities are significantly different over the whole MW galaxy. The highest surface mass density of inflowing gas is detected towards the spatial regions that cover Complex C, a large HVC in the northern hemisphere. We identify regions 3, 4, 8, 9, 10 to approximately cover the entire Complex C structure and compute the total cool ionized gas mass in those regions to be $1.84\times10^{7} M_{\odot}$ ($d/10$ kpc). This is in excellent agreement with the upper estimates of Complex C mass from \citet{Wakker_2007} of $1.4\times 10^{7} M_{\odot}$ ($d/11.2$ kpc) and \citet{Thom2008} of $1.28\times 10^{7} M_{\odot}$ ($d/10$ kpc). We note that our selected regions were selected blindly, without any prior knowledge of spatial location of the Complex C structure. Therefore our regions contain additional sightlines which were not used in the \citet{Wakker_2007} and \citet{Thom2008} analyses of this structure. The broad agreement of total mass estimates of these works with our study provides further evidence of robustness of our approach.

\begin{table}
	\centering
	\caption{Average surface mass density of inflowing and outflowing gas}
    	\label{tab:SMD_Quadrants}
    
        	\begin{tabular}{lcc} 
        		\hline & $\langle \log (\Sigma/M_{\odot}\,\mathrm{kpc}^{-2}) \rangle$ & $\langle \log (\Sigma/M_{\odot}\,\mathrm{kpc}^{-2})\rangle$\\
        		Quadrant & Inflow & Outflow \\
        		\hline
        		Q1 ($l > 180$, $b \geq 0$) & 4.60 $\pm$ 0.09 & 3.98 $\pm$ 0.25 \\
        		Q2 ($0 \geq l \geq 180$, $b \geq 0$)& 5.16 $\pm$ 0.12 & 3.60 $\pm$ 0.10 \\
        		Q3 ($0 \geq l \geq 180$, $b <0$)& 4.71 $\pm$ 0.25 & 3.65 $\pm$ 0.21 \\
        		Q4 ($l > 180$, $b < 0$)& 4.42 $\pm$ 0.19 & 4.10 $\pm$ 0.19 \\
        		\hline
            \end{tabular}
\end{table}

We further highlight the regional differences in $\Sigma$ variation of inflowing and outflowing gas by dividing the sky into four quadrants. These are tabulated in Table \ref{tab:SMD_Quadrants}. In each quadrant, $\langle \log \Sigma \rangle$ of inflowing gas is much higher than that of outflowing gas. Q2 exhibits the highest $\langle \log \Sigma \rangle$ for inflowing gas and it also hosts Complex C. Q4 shows the highest  $\langle \log \Sigma \rangle$ for outflowing gas, the dominating contribution coming from the two regions immediately above the Fermi Bubbles. These regions will likely have contribution from the nuclear outflow of the MW.

\begin{figure*}
     \includegraphics[width = 8cm]{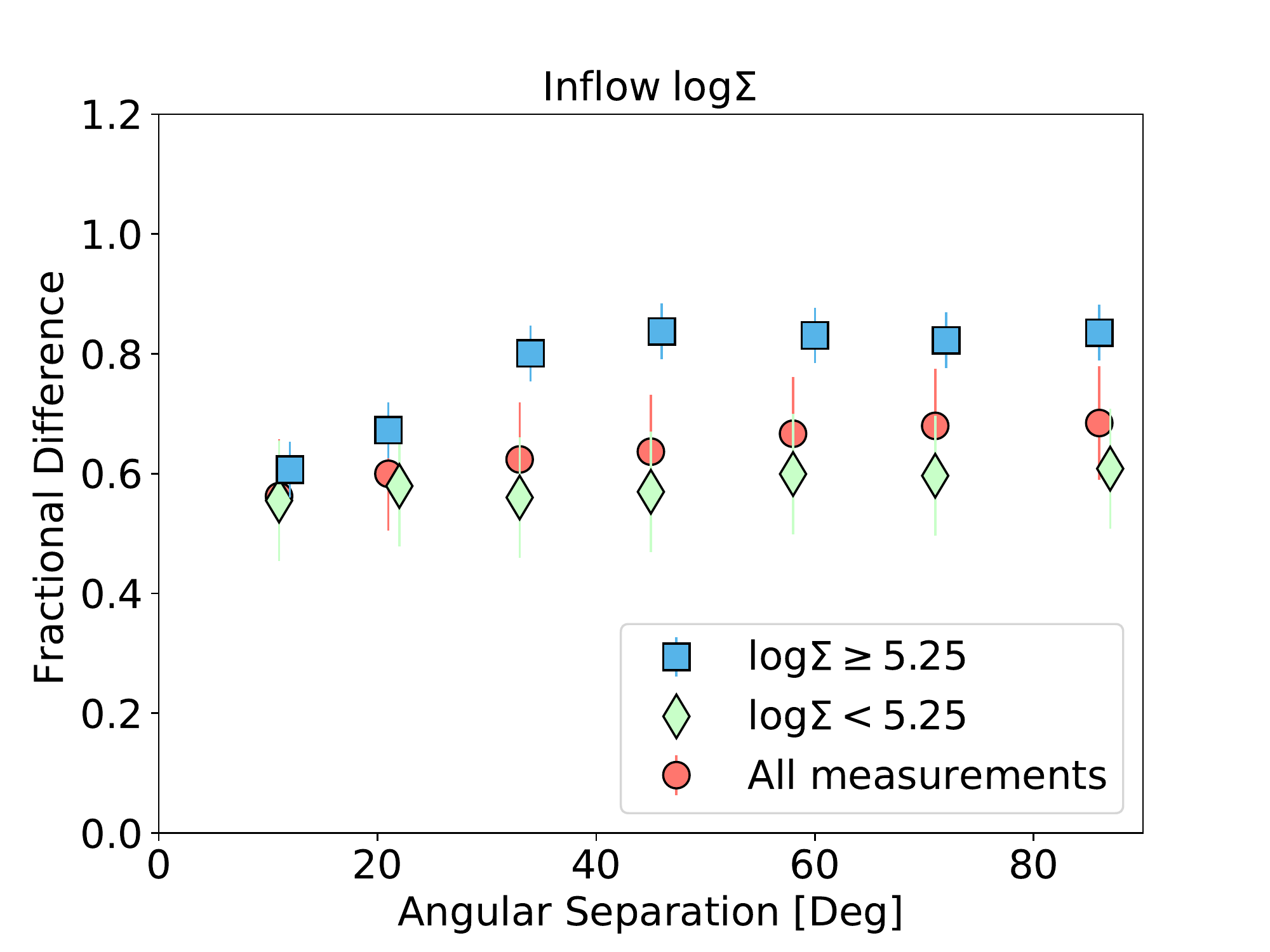}
     \includegraphics[width = 8cm]{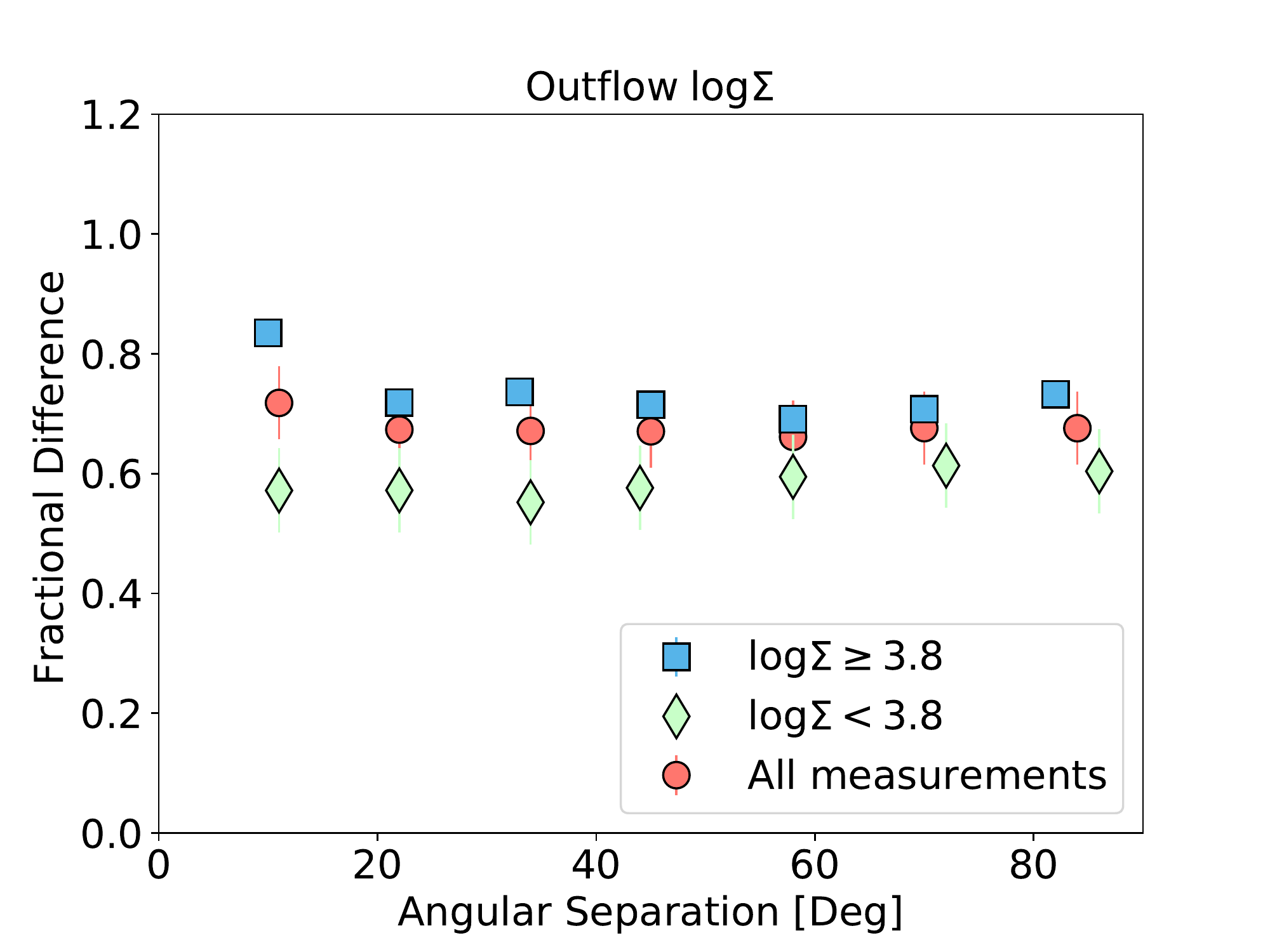}
     \caption{Pairwise fractional differences in surface mass density of inflowing (left panel) and outflowing (right panel) gas as a function of angular separation on the sky. Fractional difference of zero indicates that the spatial regions have identical mass densities, and fractional difference of 1 indicates that they are 100\% different. \textit{Left Panel:} Fractional difference is estimated for the full sample (red circles) and in two $\Sigma$ bins, high surface mass density regions (blue squares, $\log \Sigma \geq 5.25$), and low surface mass density regions (green diamonds, $\log \Sigma < 5.25$). The high mass inflowing regions are more clustered at smaller angular scales ($\theta < 40^\circ$), where as lower $\Sigma $ regions are more consistent with being randomly distributed. \textit{Right panel:} Same as left panel, now for outflowing gas. Outflowing gas fractional difference shows no evidence of angular dependence in all surface mass density ranges as compared to the inflowing gas. $\Sigma$ is in units of \msunpkpc.}
     \label{fig: Frac_diff}
 \end{figure*}

We further explore the variation in spatial distribution of inflowing and outflowing gas by comparing the pairwise fractional difference of surface mass densities between different regions on the sky as a function of their angular separation. The fractional difference ($FD$) in surface mass density of two regions on the sky is defined as:

\begin{equation}
     FD (\theta) = \frac{\Sigma_i - \Sigma_j }{\mathrm{max}(\Sigma_i  , \Sigma_j)},
\end{equation}

here $\theta$ is the angular separation between the $i^{\rm th}$ and $j^{\rm th}$ regions, $\Sigma_i$, $\Sigma_j$ are surface mass density of those two regions. This implies that if $FD $ = 0, the surface mass density between two regions are identical, and if $FD $ =1, the surface mass densities are 100\% different. Using this definition we compute the $FD$ for both inflowing and outflowing gas, and compute the mean $FD$ in angular bins. By dividing the sky into two bins, we are able to see prominent clustering effects for small-scale high-surface-mass-density inflowing gas. However, if the full range of the sample is considered, this effect would be masked out. Figure \ref{fig: Frac_diff} presents the mean angular fractional differences of inflowing (left panel) and outflowing (right panel) gas in different bins of surface mass densities. Both for inflowing and outflowing gas the mean fractional differences increase as we probe higher surface mass density gas. However, for outflowing gas, in all the two surface mass density bins the fractional differences are more or less flat as a function of angular separation. This suggests that the outflows are typically more or less randomly distributed across the sky. Whereas, for the highest surface mass density bins of inflows (blue squares), at the smallest angular bins ($\theta <$ 40$^\circ$), the fractional differences are 20\% smaller than those at higher angular bins. In the full sample of surface mass density range (red circles), this trend cannot be seen as it is washed away by the rest of the sample. This suggests that at those smaller scales, typical inflowing gas structures are statistically more similar than those at higher angular scales, or the most massive inflowing structures are smaller and more concentrated on the sky as compared to the outflowing structures. This is consistent with the canonical understanding at higher redshifts that typically inflowing gas has a smaller sky covering fraction than outflowing gas \citep{Melso2019}. 

\subsubsection{Mass flow rate surface density}

To quantify the rate at which gas is flowing into or out of the MW, we convert the inferred surface mass densities to mass flow rates per unit area by multiplying them by $v/d$. We define the mass flow rate surface density ($d \Sigma/dt$) for each individual region as:

\begin{align}
    & \frac{d \Sigma(d)_{\rm in/out}} {dt} = \dot{\Sigma}(d)_{\rm in, out}\gtrsim  \\ 
    \nonumber & 0.1 M_{\odot}\,\mathrm{kpc^{-2}\,yr^{-1}}
    \left( \frac{\Sigma}{10^{7}\,M_{\odot}\,\mathrm{kpc^{-2}}} \right) \left( \frac{d}{12\,\mathrm{kpc}} \right)^{-1} \left( \frac{|\langle v_{GSR} \rangle|}{120\,\mathrm{kms}^{-1}} \right),
    \label{eq: Mdot}
\end{align}

\begin{figure*}
    \includegraphics[width = 2\columnwidth]{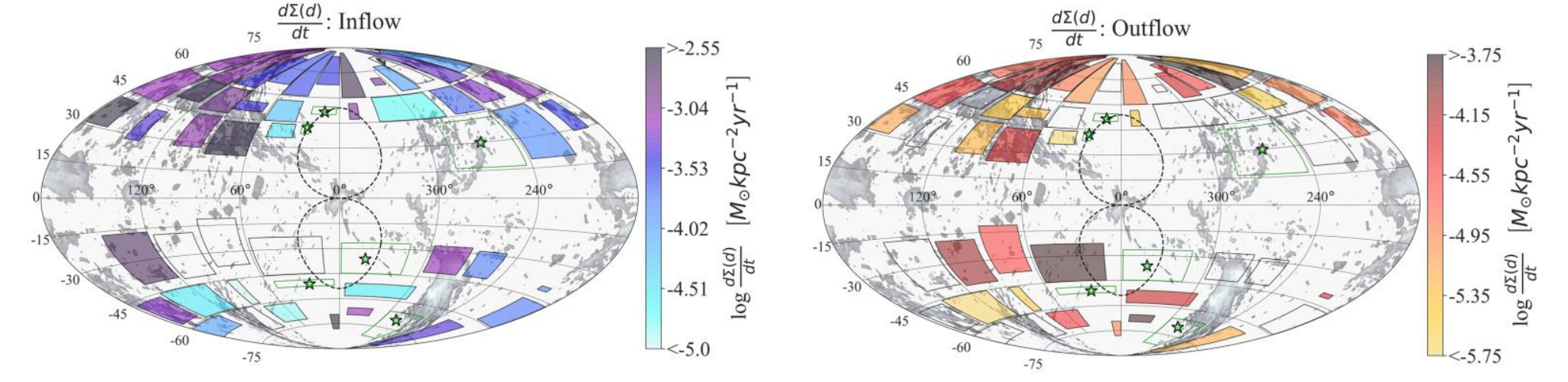} 
    \caption{All-sky maps of the mass flow rate surface density (mass flow rate per unit area) for inflowing and outflowing material, which is obtained by scaling our regional measurements by its associated surface area at a shell distance of 12\,kpc. Each colored circle represents the regional flow rate density estimates from all contributions of \ion{Si} {II}, \ion{Si} {III}, and \ion{Si} {IV}. Green stars represent completely filtered regions, and open regions  denote non-detections.}
    \label{fig:MassFlow}
\end{figure*}

where $|\langle v_{GSR} \rangle|$ is the mean absolute GSR gas velocity over all the detected phases of \ion{Si} {II/III/IV}. Figure \ref{fig:MassFlow} shows the mass flow rate surface densities of inflowing (left panel) and outflowing (right panel) gas, respectively. For inflowing gas, we measure a large variation across the sky for regional $\log \dot{\Sigma}(d)$, with a mean $\log \dot{\Sigma}(d) \gtrsim$ -3.55 $\pm$ 0.08 $M_{\odot}\,\mathrm{kpc}^{-2}\, \mathrm{yr}^{-1}$ and more than two dex of variation across the sky. After excluding the Magellanic Stream, the highest mass inflow rate surface densities are observed towards region 1 with $\log \dot{\Sigma}(d)$ $\approx$ $-$2.55$\pm$0.27 $M_{\odot}\,\mathrm{kpc^{-2}\,yr^{-1}}$. This also happens to be one of the regions with highest surface mass density of inflowing gas.  

For outflowing gas, the typical mass outflow rate surface densities are almost an order of magnitude smaller than that for inflowing gas with a  mean $\log \dot{\Sigma}(d) \gtrsim$ -4.79 $\pm$ 0.14 $M_{\odot}\,\mathrm{kpc^{-2}\,yr^{-1}}$ and 1.2 dex of variation between different outflowing regions. The region with the highest mass outflow rate surface density is region 31 with $\log \dot{\Sigma}(d)$ $\approx$ $-$3.38 $\pm$ 0.27 $M_{\odot}\,\mathrm{kpc^{-2}\,yr^{-1}}$. Regions with the highest mass flow rate surface densities are not the same regions as those with the highest gas velocities (Figure \ref{fig:SiIII_Characteristics}). Regions with higher mass inflow rates surface densities are clustered together more closely on the sky than regions with lower values, or outflowing regions. This again shows that gas inflows are more concentrated across the sky, whereas outflows are more uniformly dispersed. 

\subsubsection{Error Propagation}

The error in the column densities is the statistical error associated with the observed flux computed using the AOD method as given in equation \ref{eq: Tau_a}. To account for dust depletion, a 38\% error associated with the dispersion of [Si/Zn] in halo clouds \citep{sembach1996gas} is used. For distance, a 33\% error is adopted from the mean distance of 12kpc $\pm$ 4kpc as reported by \citet{Lehner2011}. A 20\%  error is adopted for the metallicity from \cite{Fox_2014}, which mapped all metallicity measurements from Complex C onto a solar abundance scale. Additionally, the uncertainty on the line velocities is measured as the error on the mean velocity of the line centroids. The errors associated with the summed contributions of ionization states are added in absolutes (equation \ref{eq: Navg}, while all other calculations have  their errors summed in quadrature. The surface mass density and mass flux density values associated with each region along with their relative errors are given in Table \ref{tab:SMD}. 

\section{Results and Summary}\label{Section 5}

In this study we have used a novel co-addition (stacking) technique to observe the spatial variation of MW inflows and outflows across the sky. Stacking spectra significantly improves the SNR of the co-added spectrum, allowing for detection of small variations in metal column densities across the sky. This technique also enables the use of a larger sample of spectra from the HSLA QSO catalogue than is available from individual sightline analyses, and allows for characteristics of the CGM to be defined and analyzed over their different regions of the sky. The results of this analysis are summarized below.

\begin{enumerate}
    \item We use the largest sample of archival medium-resolution \emph{HST}/COS spectra of background AGN ($N$=440) to date to perform a spatially resolved study of gas flows into and out of the MW galaxy. We divide the sky into 53 individual regions and co-add the spectra that are in each of these regions. By stacking these individual spectra within different spatial regions of the sky, we detect absorption systems with absorption strengths low as $EW\gtrapprox{2}$\,m\AA, allowing for diffuse and previously-undetected gas to be included in the characterization of the Galactic CGM. The typical 3$\sigma$ detection threshold for the co-added spectra is $\log (N/\mathrm{cm}^{-2}) \approx$ 11.75 for \ion{Si}{III} 1206.
    
    \item For both inflowing and outflowing gas, \ion{Si}{III} is the most easily detected transition. For the whole sky, the frequency of detection of inflowing \ion{Si} {III} absorption is 34\% higher than that of outflowing \ion{Si} {III} absorption. Of all the regions studied, the incidence of detection of \ion{Si} {III} absorption is 91.5$\pm $4.1\% (43/47) for inflowing gas and 57.4$\pm$7.1\% (27/47) for outflowing gas, respectively. Between the different spatial regions, we find significant variation in column densities of individual metals greater than 2.5 orders of magnitude for inflowing material and over 2.3 orders of magnitude in outflowing material, highlighting the dynamic range in absorption strengths of HVCs around the MW. 

    \item We search for differences between incidence of detection of different ions by dividing the sky between different hemispheres: North ($b>$0$\degr$), South ($b<$0$\degr$), East ($360\degr \leq l \leq 180 \degr$) and West ($180\degr \leq l \leq 0 \degr$), respectively. The incidence of detection of inflowing \ion{Si}{II/IV} absorption is significantly different between northern and southern hemisphere of the MW. Once we exclude the Magellanic Stream, the bulk of the inflowing gas is confined towards $b > 0^\circ$. In particular, for \ion{Si}{IV} absorption, an adjusted Chi-squared test shows that the difference in detection rates between northern and southern hemispheres are different at $>$99.5\% confidence.
    
    \item For outflowing gas, we find no asymmetry in detection rates between northern and southern hemispheres. However, particularly for the \ion{Si}{II/IV} transitions, the outflowing absorption is mostly detected within the region
    $120^\circ \leq l \leq 240^\circ$. 
    
    \item We compute the surface mass densities of both inflowing and outflowing gas in individual regions. These mass densities are distance-independent and are proportional to the mean HVC hydrogen column density in each region. For inflowing gas, the mean surface mass density is $\langle \Sigma_{\rm in}\rangle \gtrsim 10^{4.57\pm 0.09}$\msunpkpc, while individual regions show a variation of 1.21 dex  within 95\% confidence interval. For outflowing gas, the mean surface mass density is $\langle \Sigma_{\rm out}\rangle \gtrsim 10^{3.51\pm 0.11}$ \msunpkpc, with a regional variation of 1.23 dex within 95\% confidence interval. The mean outflowing surface mass density is an order of magnitude lower than the mean inflowing surface mass density.
    
    \item A pairwise angular fractional difference analysis shows that the inflowing regions with highest surface mass densities are more clustered at smaller angular scales ($\theta < 40^\circ$). This suggests that most of the mass in inflowing gas are confined in well-defined structures like Complex C, whereas the distribution of outflowing gas is more diffuse and uniform throughout the sky.
    
    \item We compute the mass flow rate surface densities of both inflowing and outflowing gas and and observe more than two dex of variation within each. We find that the average mass flow rate surface density for inflowing gas is $\langle \dot{\Sigma}(d)_{\rm in}\rangle \gtrsim (10^{-3.55\pm0.08})(d/12\,\mathrm{kpc})^{-1}$ \msunpkpcpyr, and that for outflowing gas is $\langle \dot{\Sigma}(d)_{\rm out}\rangle \gtrsim (10^{-4.79\pm0.14})(d/12\,\mathrm{kpc})^{-1}$ \msunpkpcpyr, over an order of magnitude lower. 
   
    \item The regional ionization fractions (see Appendix \ref{Regional Properties}) show that for inflowing material, highly ionized regions are constrained to higher latitudes $(|b|>60\degr)$ and typically confined to the Galactic Center ($|l|\lesssim 90\degr$).

\end{enumerate}

These results demonstrate that the stacking technique performs well relative to individual sightline analyses, while adding the complementary function of glimpsing the localized characteristics of very diffuse metals in the MW CGM. These localized characteristics are useful for tuning the sub-grid parameters of hydrodynamical simulations. Additionally, the findings we report further support the fact that the MW is currently in a phase dominated by inflow over outflow. This net influx of will lead to an increased star formation rate over next few 100s of Myr, enabling the MW to give birth to a new generation of stars.

\section*{Acknowledgements}
This research made use of Astropy\footnote{http://www.astropy.org}, a community-developed core Python package for Astronomy \citep{astropy:2013, astropy:2018}. This research has made use of the HSLA database, developed and maintained at STScI, Baltimore, USA \citep{Peeples2017}.

\section*{Data Availability}
The data underlying this article are available in \emph{HST} Spectroscopic Legacy Archive (HSLA). The processed data and the analysis codes used for this project are available in https://github.com/saclark7/Investigating-Spatially-Resolved-Gas-Flows--MW. The absorption line measurement and interactive continuua fitting codes are available in https://github.com/rongmon/rbcodes.




\bibliographystyle{mnras}
\bibliography{Bibliography.bib} 

\newpage
\appendix

\section{Regional Properties}\label{Regional Properties}
In this section we present the properties of each individual regions and the measured average column densities and velocities for each region, and their corresponding surface mass densities and mass flow rate surface densities in Tables \ref{tab:SolidAngProperties}, \ref{tab:regional_characteristics}, and \ref{tab:SMD}. We further present the spatial variation of \ion{Si}{II} and \ion{Si}{IV} column densities and velocities and the column density ratios between \ion{Si}{II}, \ion{Si}{III} and \ion{Si}{IV} ions across the sky.

We further investigate if the choice of the boundaries of the local regions themselves bias our measurement of gas flow rates. We isolated two regions (region 44 and region 45) that hold a high number (23) of QSOs within them. This domain is randomly sub-divided into two new spatial regions where we maintain the criteria that each region has a minimum of 3 QSOs, all of which have a SNR  greater than 2. We co-add the spectra in these two new spatial regions and measure the line column densities as described in section 3. This process is repeated 10 times while we randomly sub-divide these regions into two new spatial regions, following the selection criteria described above. These measurements are shown in Appendix A, Figure \ref{fig:N Distribution}. For each new sub-division we get slight variations in column densities, but all measurements are consistent within 0.28 dex in column density of the initial measurements. This consistency is very clearly seen for stronger Si III and Si II absorption. For Si IV column densities, the absorption is very close to the detection threshold and therefore more noisy. Nonetheless, we get consistent results across different realizations and different species. This exercise shows that the spatially resolved outflow/inflow column density measurements are broadly insensitive to the choice of the selected local regions.

\input{Patch_Identity_table.tex}

\input{PatchProperty_Final.tex}
\input{SurfaceDensityCharacteristics.tex}

\begin{figure*}
    \centering
    \includegraphics[width=0.95\textwidth]{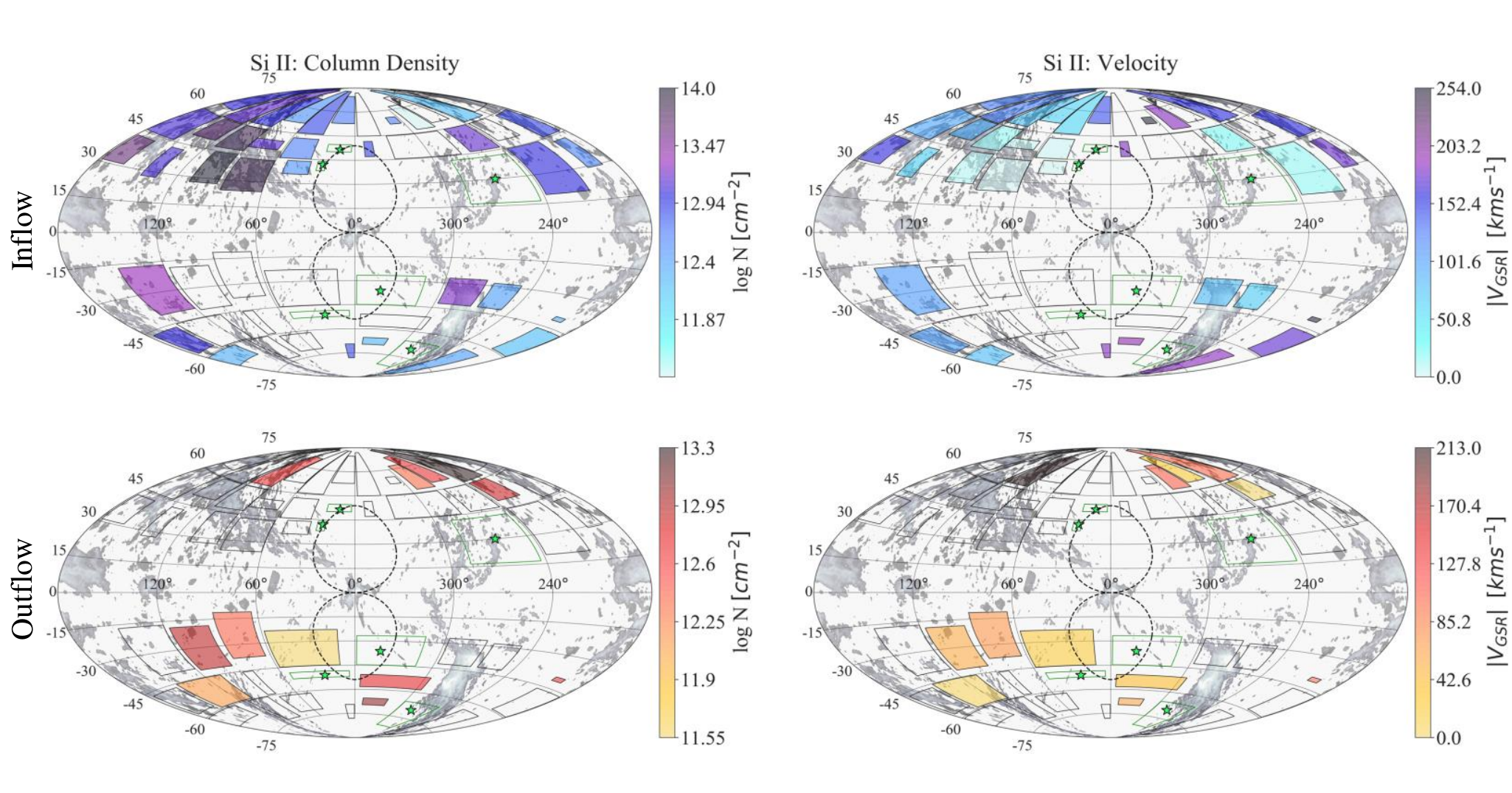}
    \caption{All-sky maps showing regional variation in column density (left panels) and absolute velocity (right panels) for inflowing and outflowing \ion{Si}{II} absorption. Open regions again represent non-detections, while completely filtered regions are denoted with green stars, same as Figure \ref{fig:SiIII_Characteristics}.}
    \label{fig:SiII_Characteristics}
\end{figure*} 
\begin{figure*}
    \centering
    \includegraphics[width=0.95\textwidth]{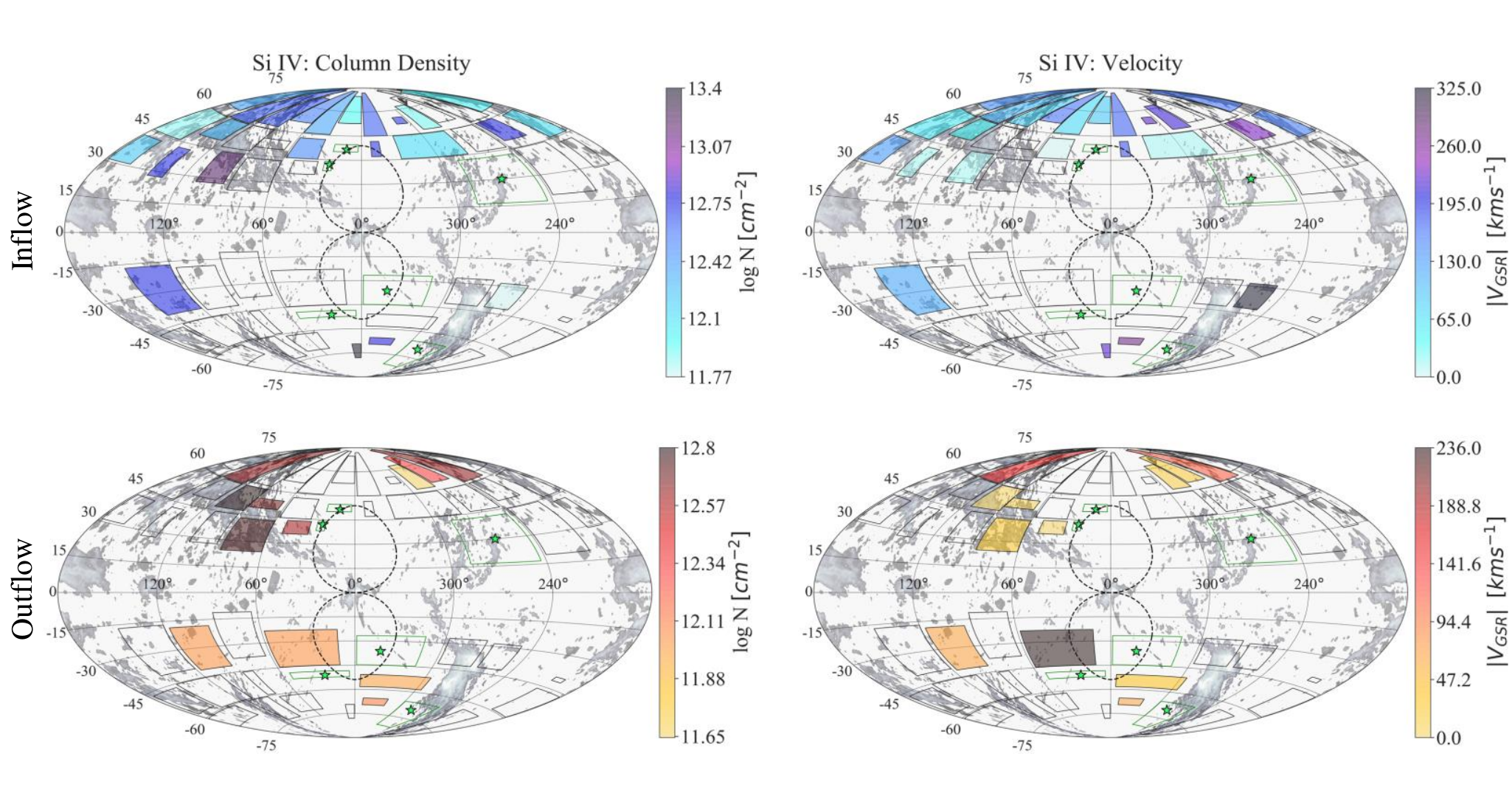}
    \caption{All-sky maps showing the column density (left panels) and absolute velocity (right panels) for inflowing (top) and outflowing (bottom) \ion{Si}{IV} absorption. Open regions represent non-detections, while completely filtered regions are denoted with green stars, same as Figure \ref{fig:SiIII_Characteristics}.}
    \label{fig:SiIV_Characteristics}
\end{figure*} 
\begin{figure*}
    \centering
    \includegraphics[width=0.95\textwidth]{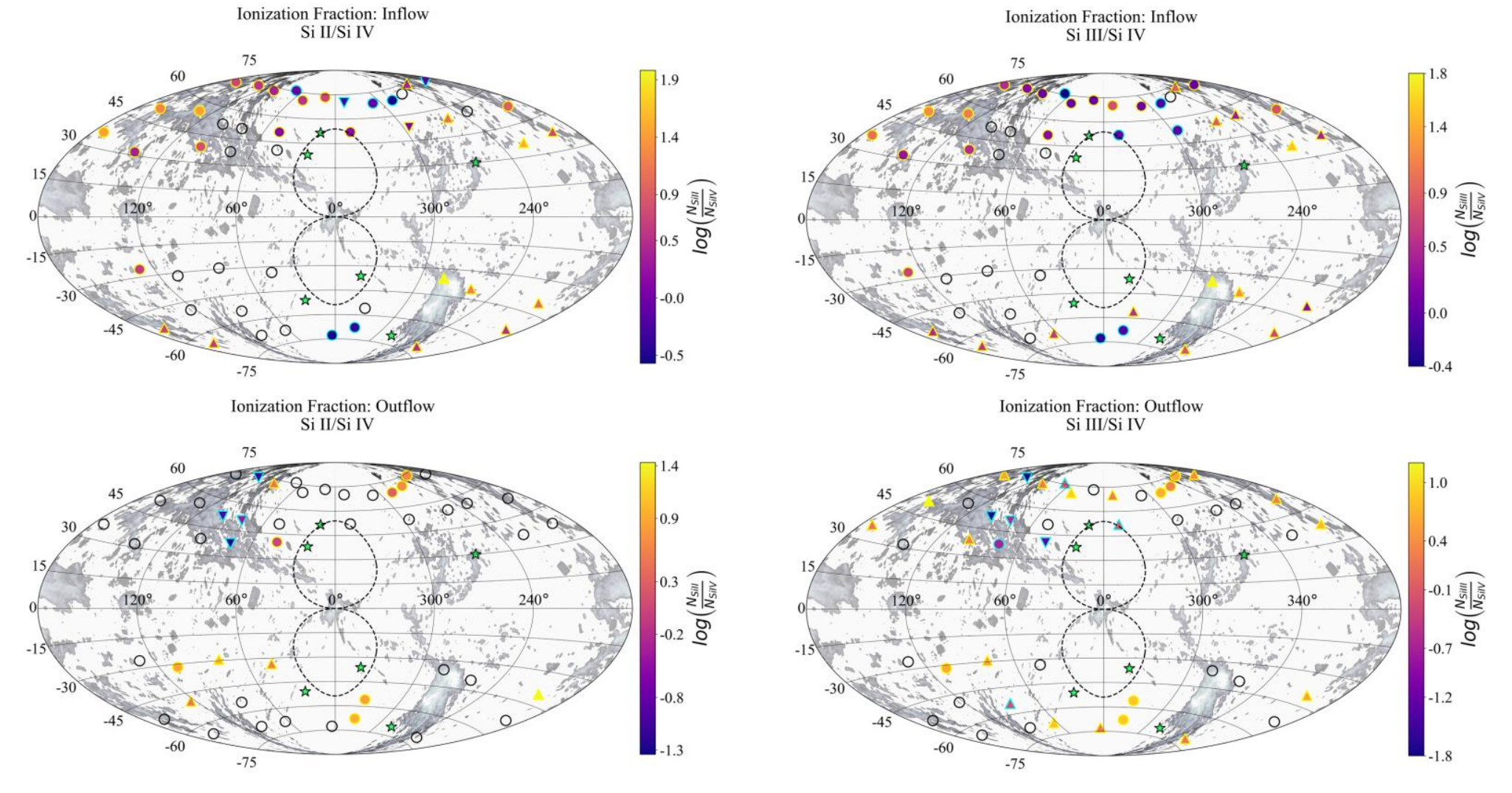}
    \caption{Ionization fractions of primary ions analyzed in this analysis. Upper triangles denote upper limits, while lower triangles denote lower limits. Any object showing higher ionization (negative values) are given a cyan border, while lower ionization (positive values) were given a yellow border. Open-faced circles represent when both ionization states yielded non-detections or if either of the ionization states velocities changed signs during the GSR transformation.}
    \label{fig:column_density_ratios}
\end{figure*}

\begin{figure*}
    \centering
    \includegraphics[width=0.95\textwidth]{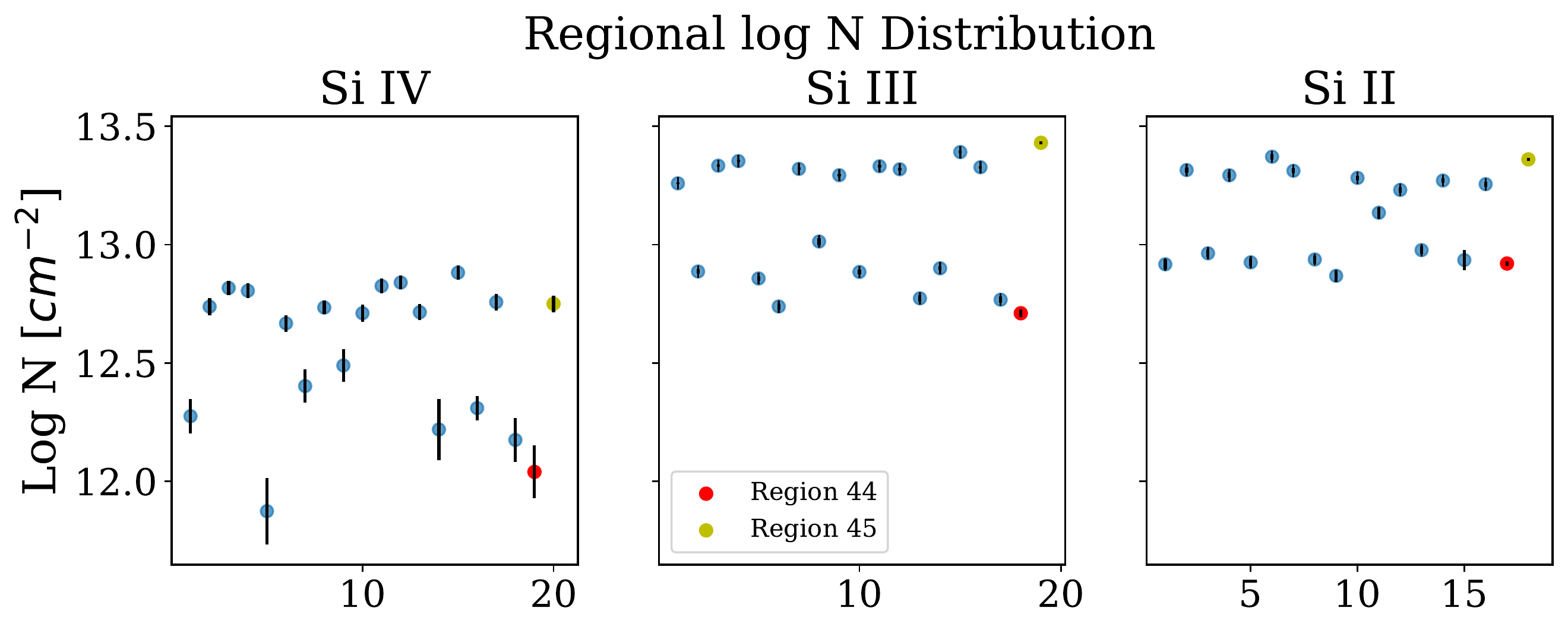}
    \caption{Test to check for sensitivity of column density measurements on the selection of the local regions. The quasar lines of sight  within regions 44 and 45 are randomly  sub-divided into two separate regions and co-added. The blue data points in each panel correspond to the Si II/III/IV column densities measured in each realization. The red circle corresponds to the column densities of region 44 as selected in this analysis, while the yellow circle corresponds to region 45.}
    \label{fig:N Distribution}
\end{figure*}

\begin{figure*}
    \centering
    \includegraphics[width=0.95\textwidth]{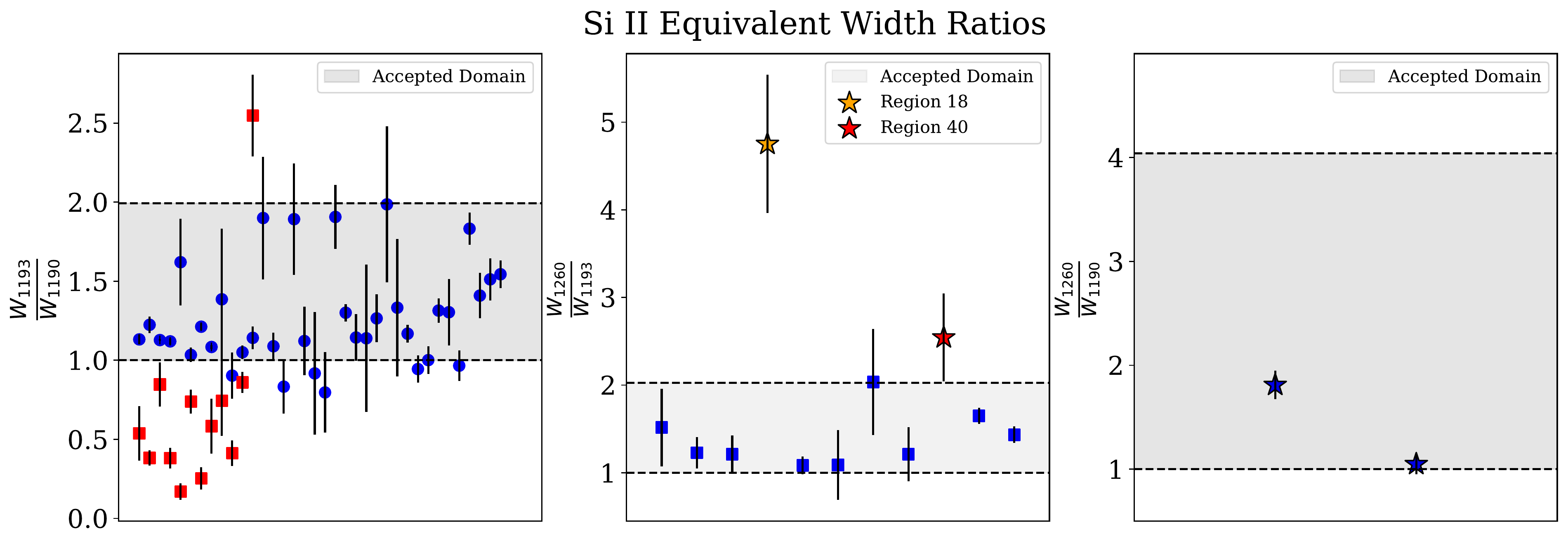}
    \caption{Equivalent widths ratios of Si II. For all images, the gray region denotes the accepted domain for the ratios of each transition. The lower limit is a 1:1 ratio and the upper limit is defined by the ratio of the transition's oscillator strengths. Red points denote ratios that fell beyond the accepted domain and were investigated amongst another transition.
    The left plot is the ratio between 1193 and 1190. The middle shows the ratio of the 1260/1193 transitions, for all points that fell outside of the domain for the doublet comparison. The two data points denoted by orange and red markers are the transitions for which 1193 was replaced with the 1260 absorption. The right plot verifies that the 1260/1190 ratios are within the accepted region and are valid for replacement.}
    \label{fig:EW_Ratio}
\end{figure*}

\section{Co-added Spectra}\label{Coadds}

In this section we present all the stacked \emph{HST}/COS  QSO spectra showing the \ion{Sii} {II}, \ion{Si} {III}, \ion{Si} {IV}, and \ion{C} {II} transitions, respectively. In each case, the blue shaded regions represent blueshifted HVCs detected in the $v_{\rm LSR}$ frame, whereas red shaded regions depict redshifted HVCs. The blue and red crosses in each plot represent the integration limits used to compute absorption strength in each co-added spectra. The measured rest frame equivalent widths are in m\AA. For non-detections, equivalent widths are 2$\sigma$ upper limits. The measured AOD column densities and velocities in GSR frame are presented in Table \ref{tab:regional_characteristics}. The light blue lines in each plot denote the co-added spectrum for that region and the light orange line in the bottom of the plot denote the associated error spectrum of that co-add.

\begin{figure*}
    \includegraphics[width=0.65\columnwidth,height = 0.45\textheight]{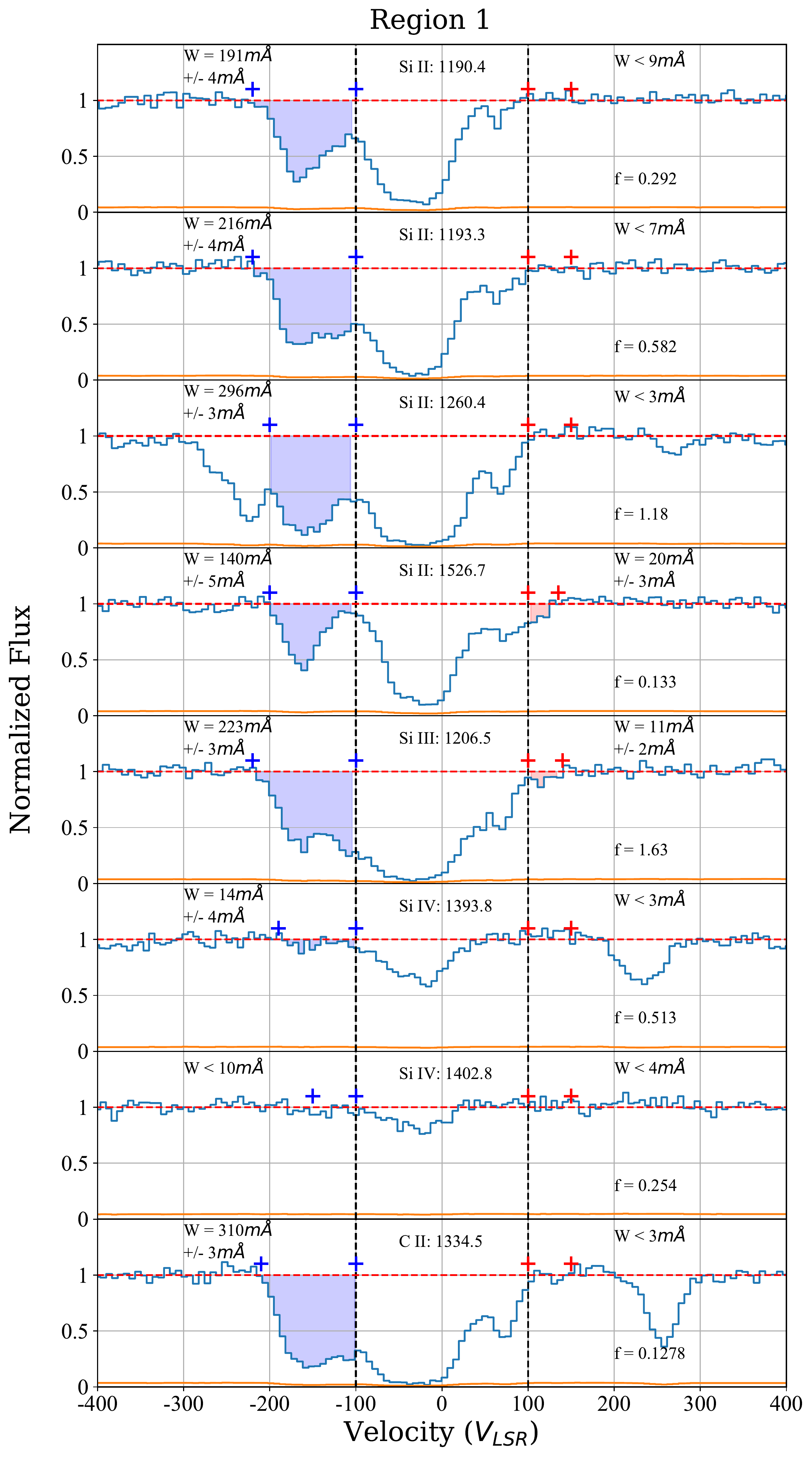}
    \includegraphics[width=0.65\columnwidth,height = 0.45\textheight]{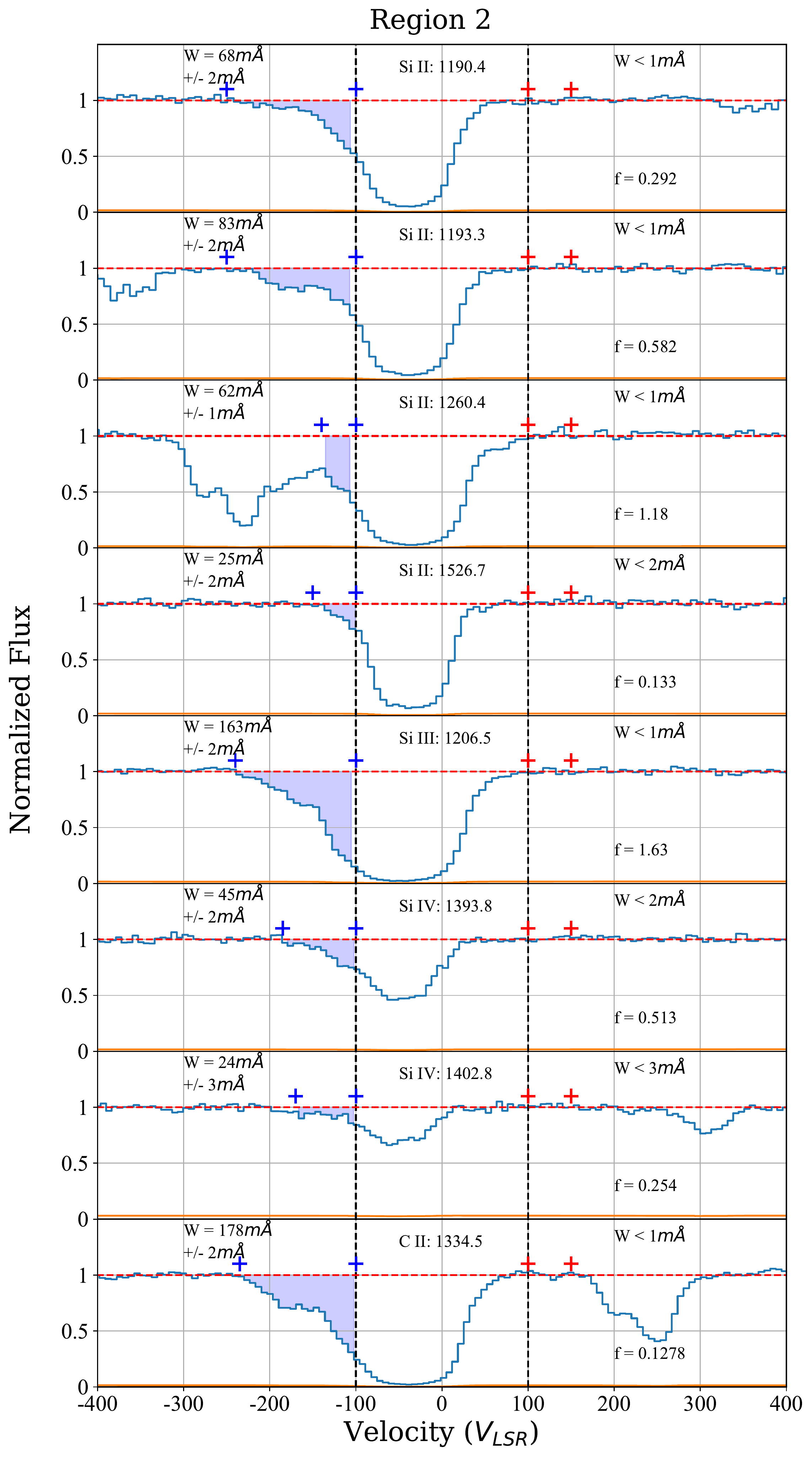}
    \includegraphics[width=0.65\columnwidth,height = 0.45\textheight]{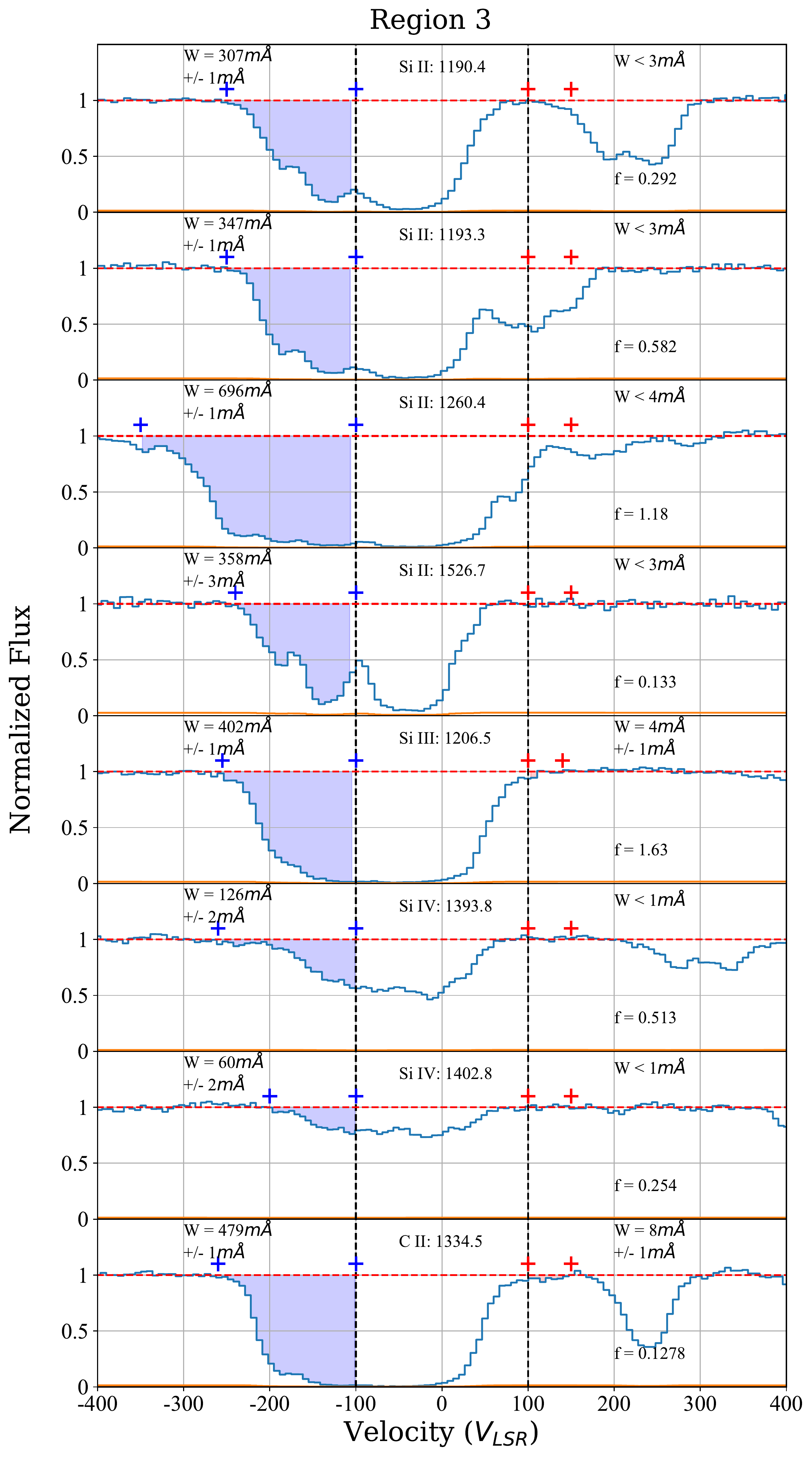}
    \includegraphics[width=0.65\columnwidth,height = 0.45\textheight]{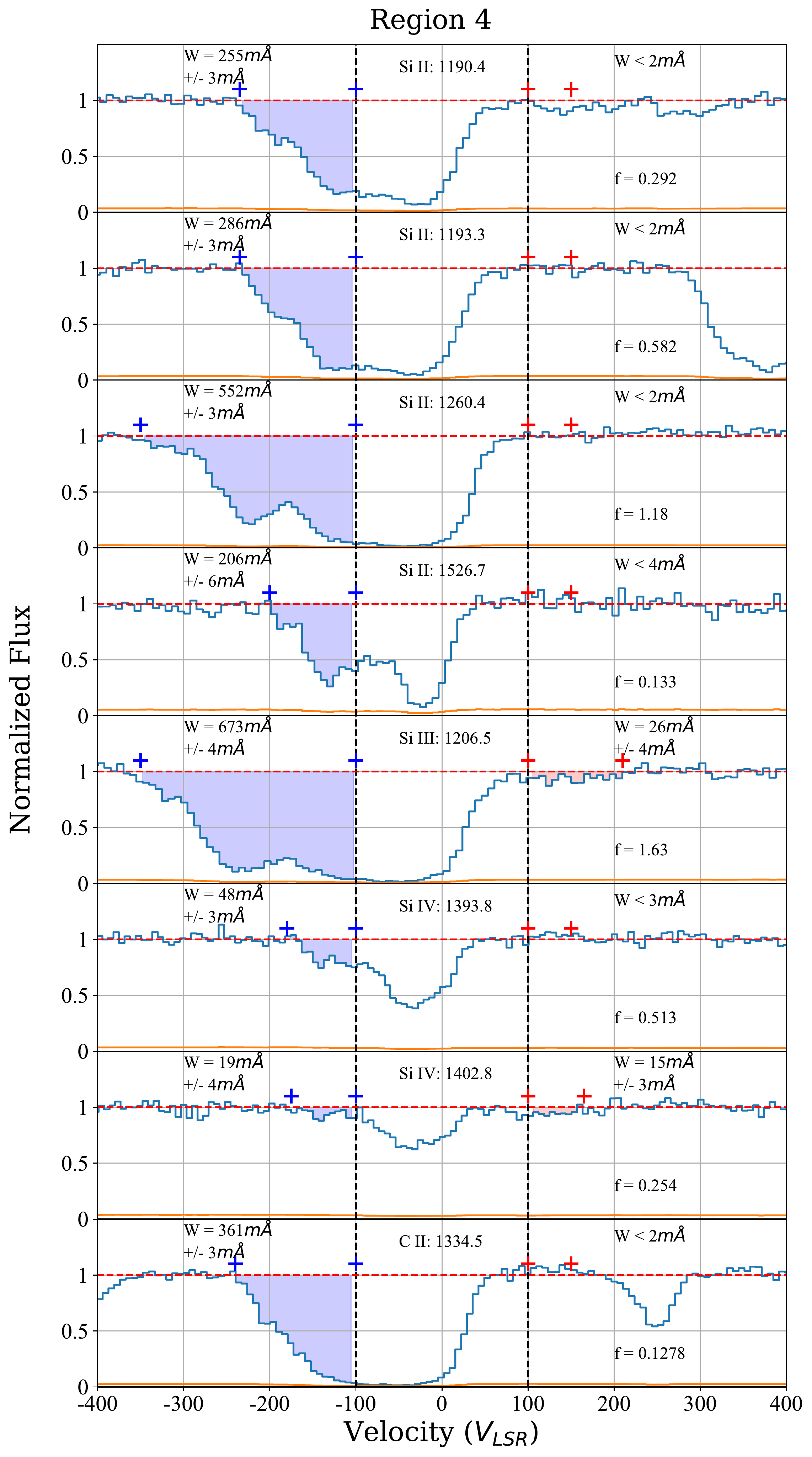}
    \includegraphics[width=0.65\columnwidth,height = 0.45\textheight]{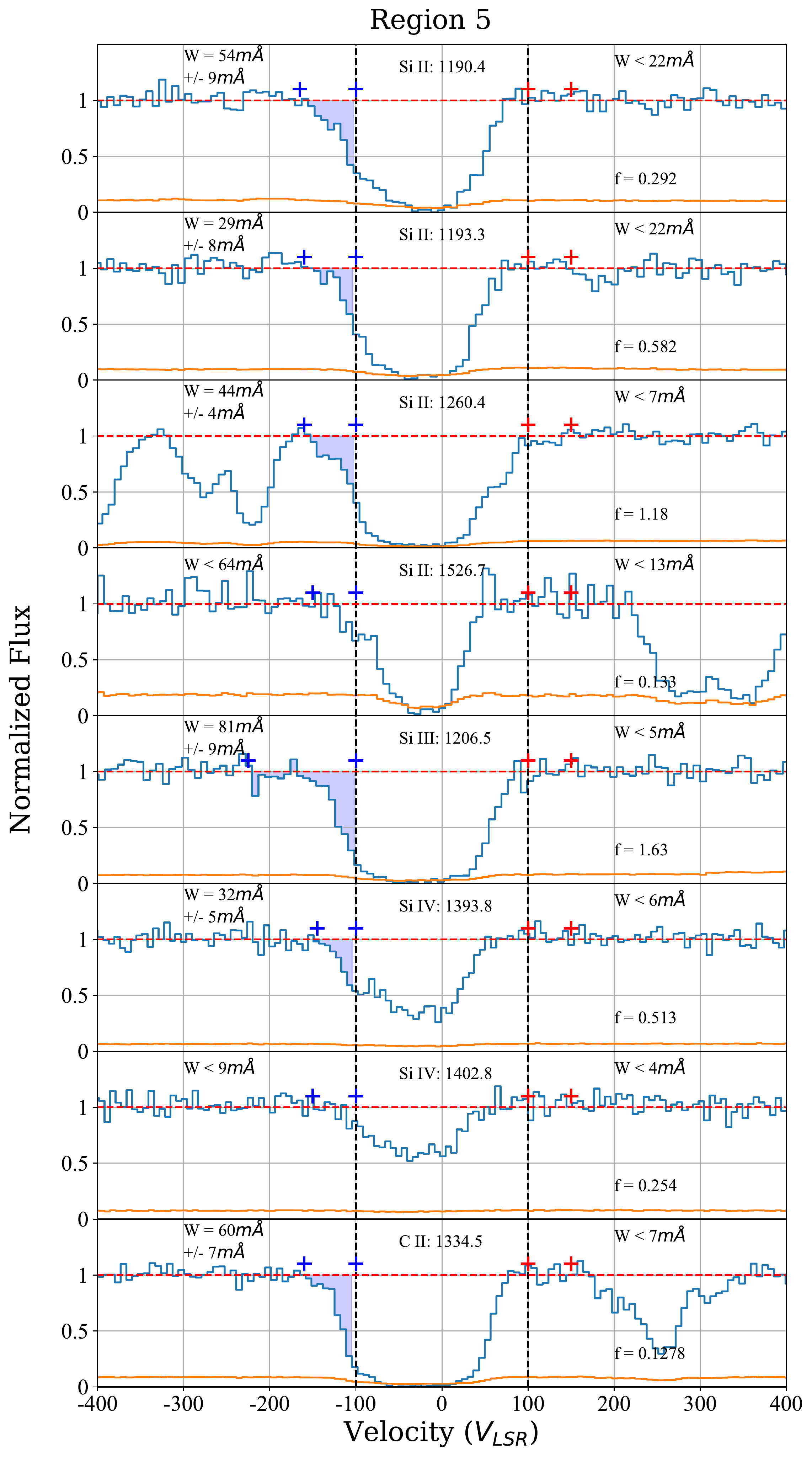}
    \includegraphics[width=0.65\columnwidth,height = 0.45\textheight]{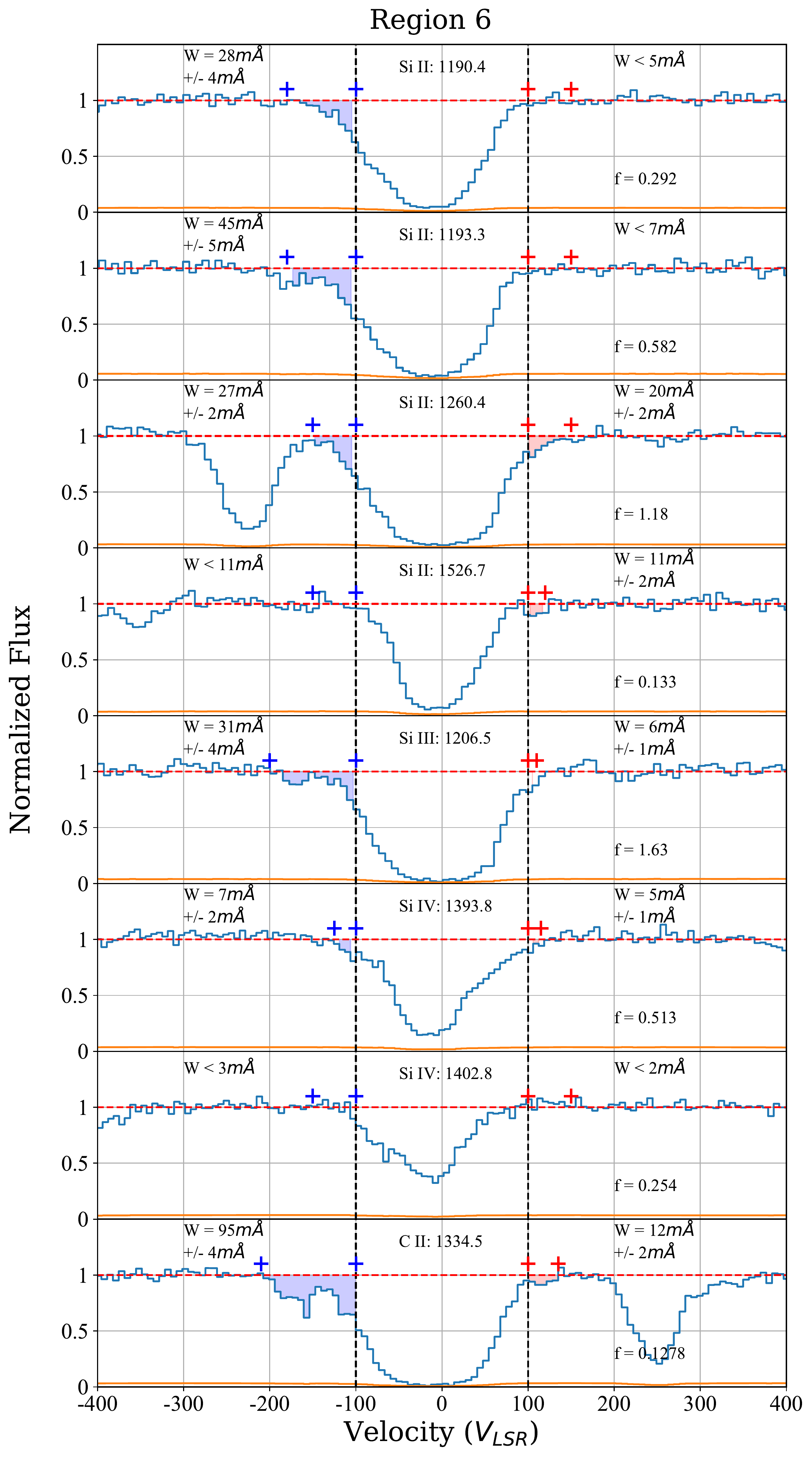}
    
    \caption{Co-added spectrum of different regions on the sky. Same as Figure \ref{fig:Patch31}.}
    \label{fig:AppendixB}
\end{figure*}

\begin{figure*}
    \includegraphics[width=0.65\columnwidth,height = 0.45\textheight]{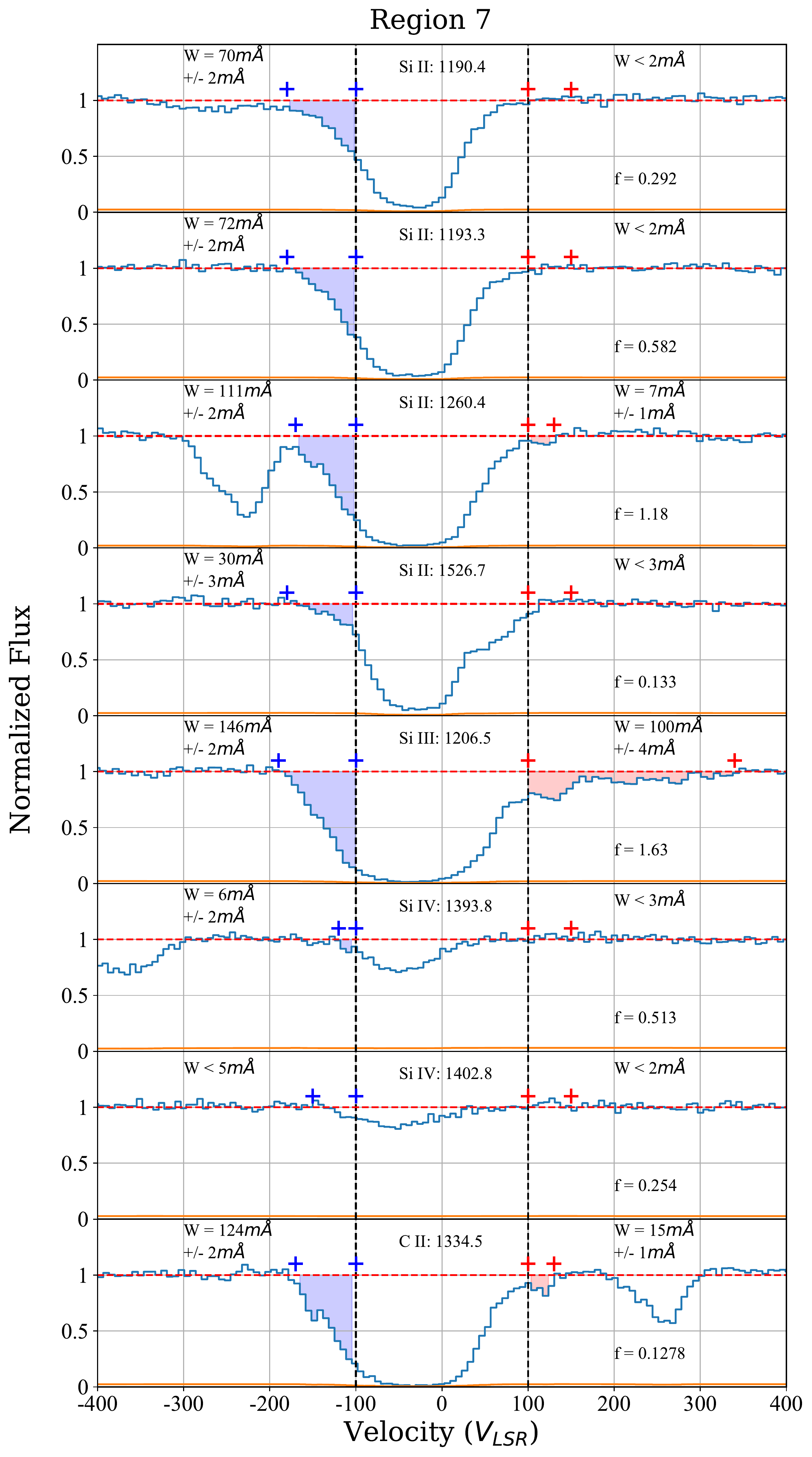}
    \includegraphics[width=0.65\columnwidth,height = 0.45\textheight]{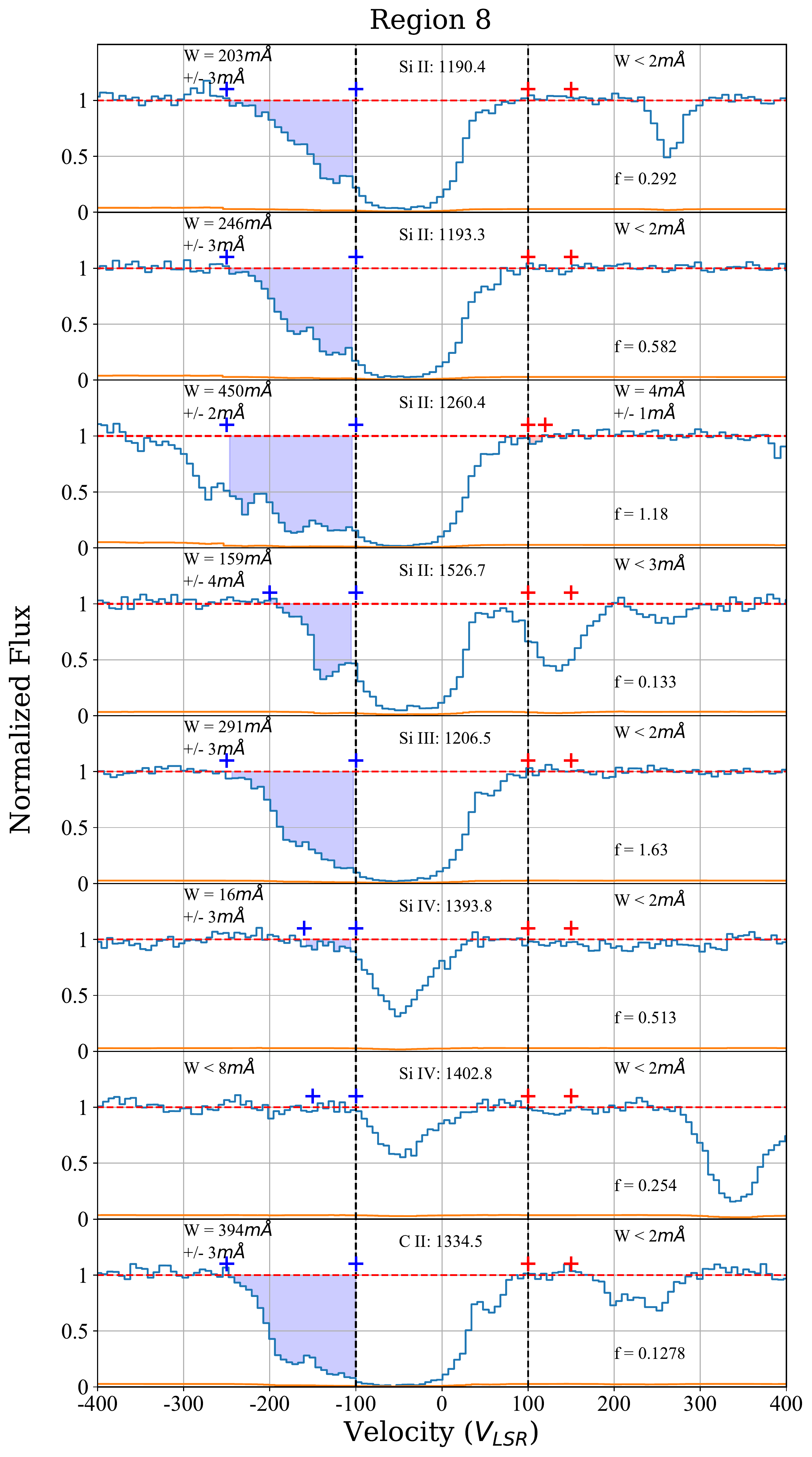}
    \includegraphics[width=0.65\columnwidth,height = 0.45\textheight]{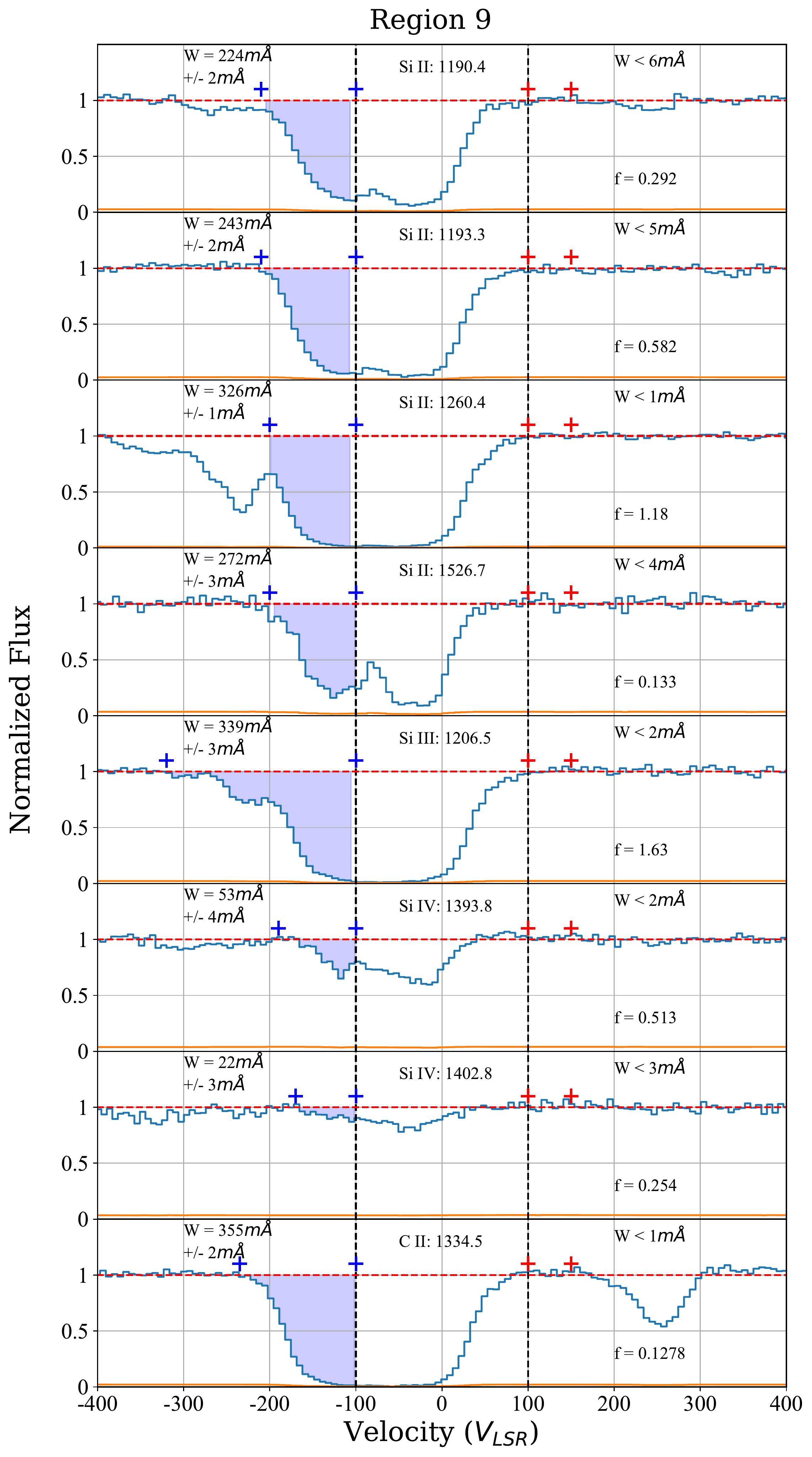}
    \includegraphics[width=0.65\columnwidth,height = 0.45\textheight]{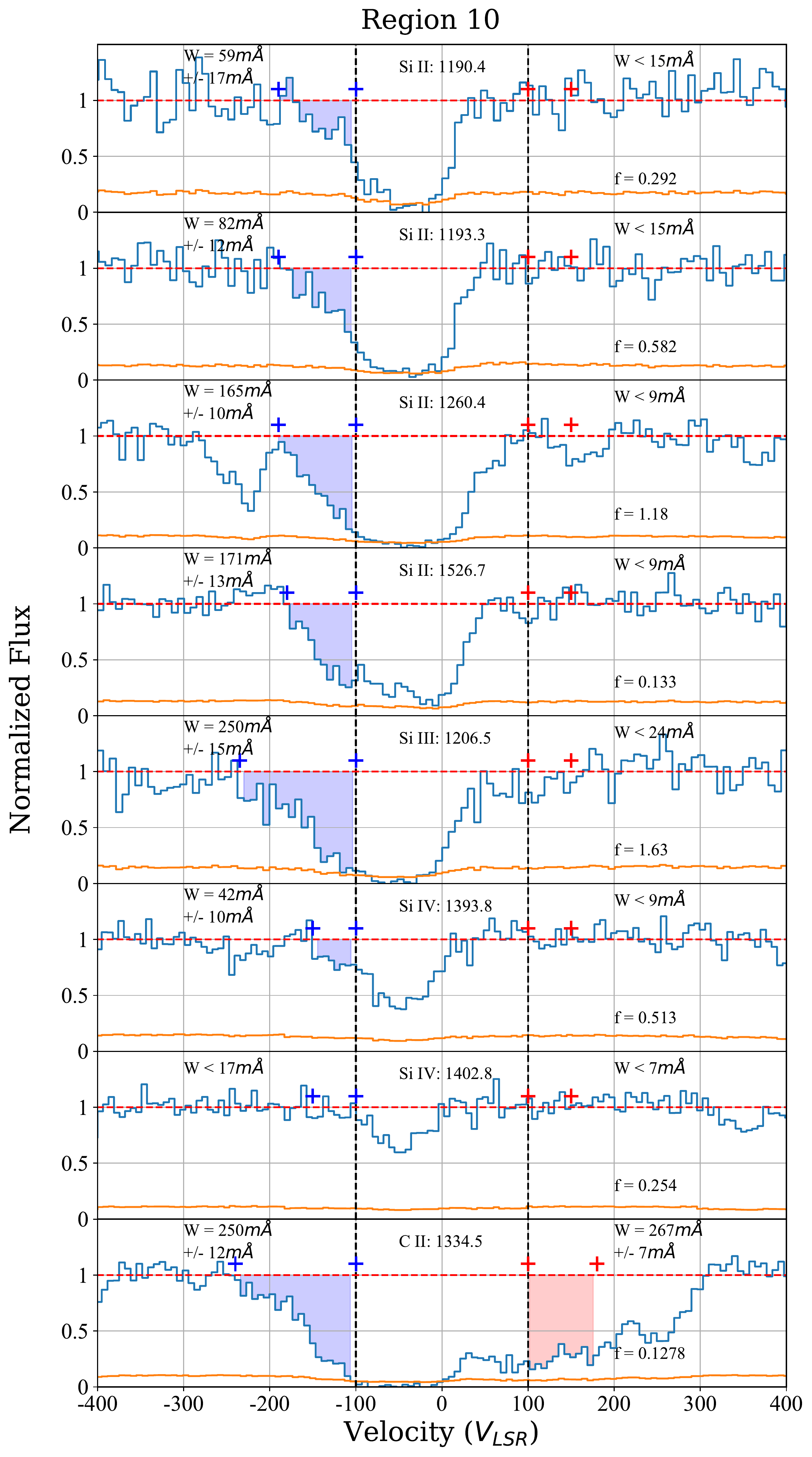}
    \includegraphics[width=0.65\columnwidth,height = 0.45\textheight]{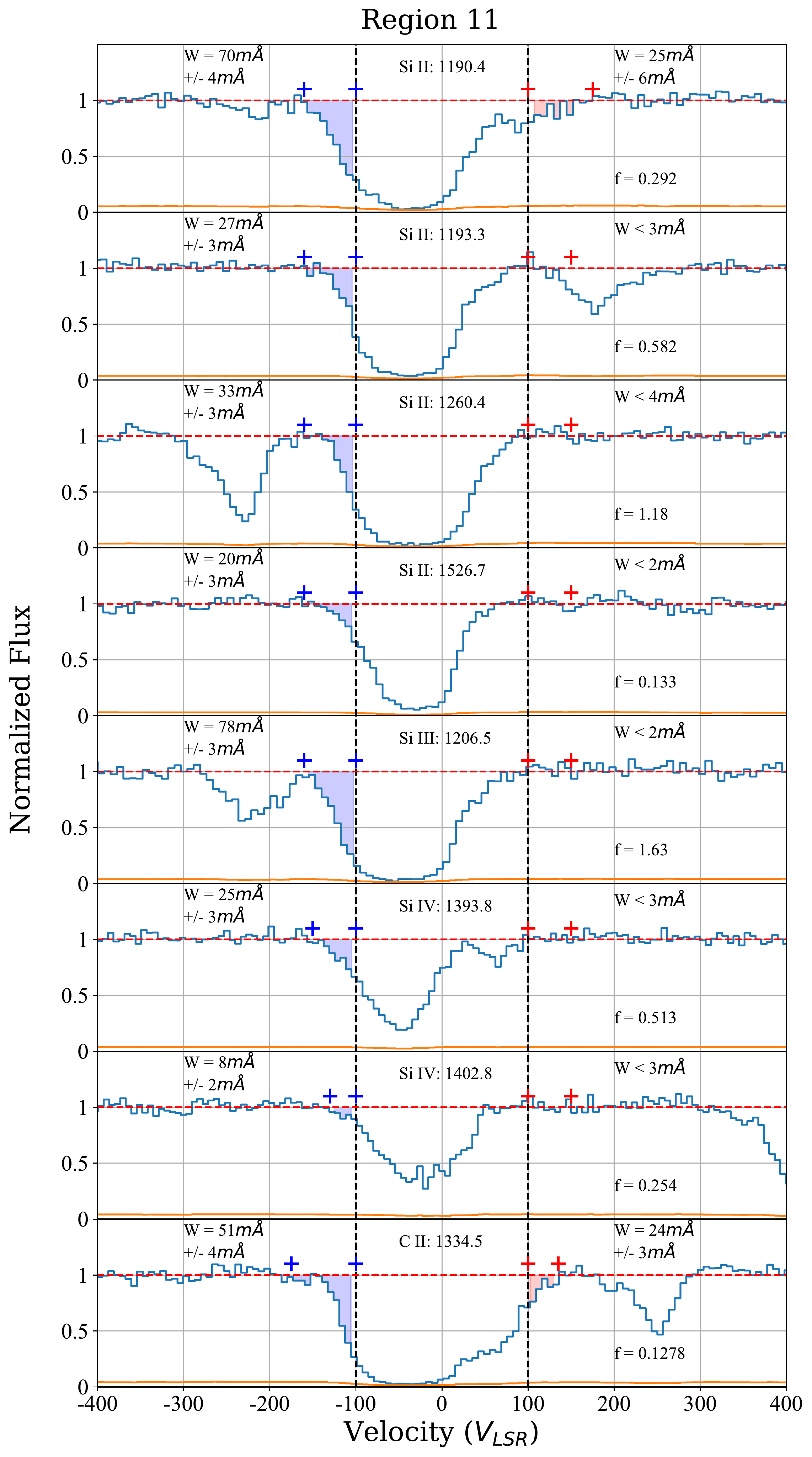}
    \includegraphics[width=0.65\columnwidth,height = 0.45\textheight]{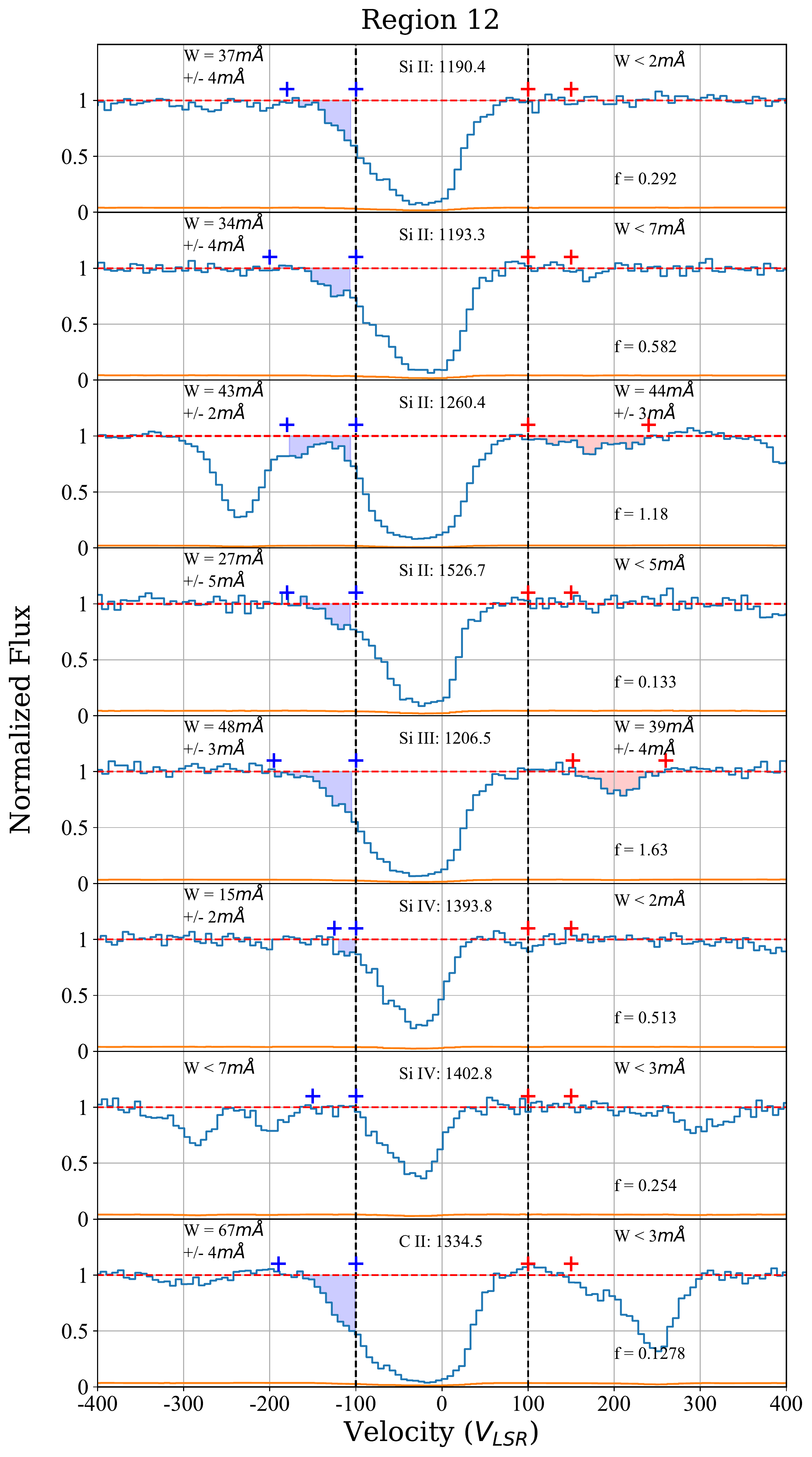}
    
    \contcaption{from Figure \ref{fig:AppendixB}.}
\end{figure*}

\begin{figure*}
    \includegraphics[width=0.65\columnwidth,height = 0.45\textheight]{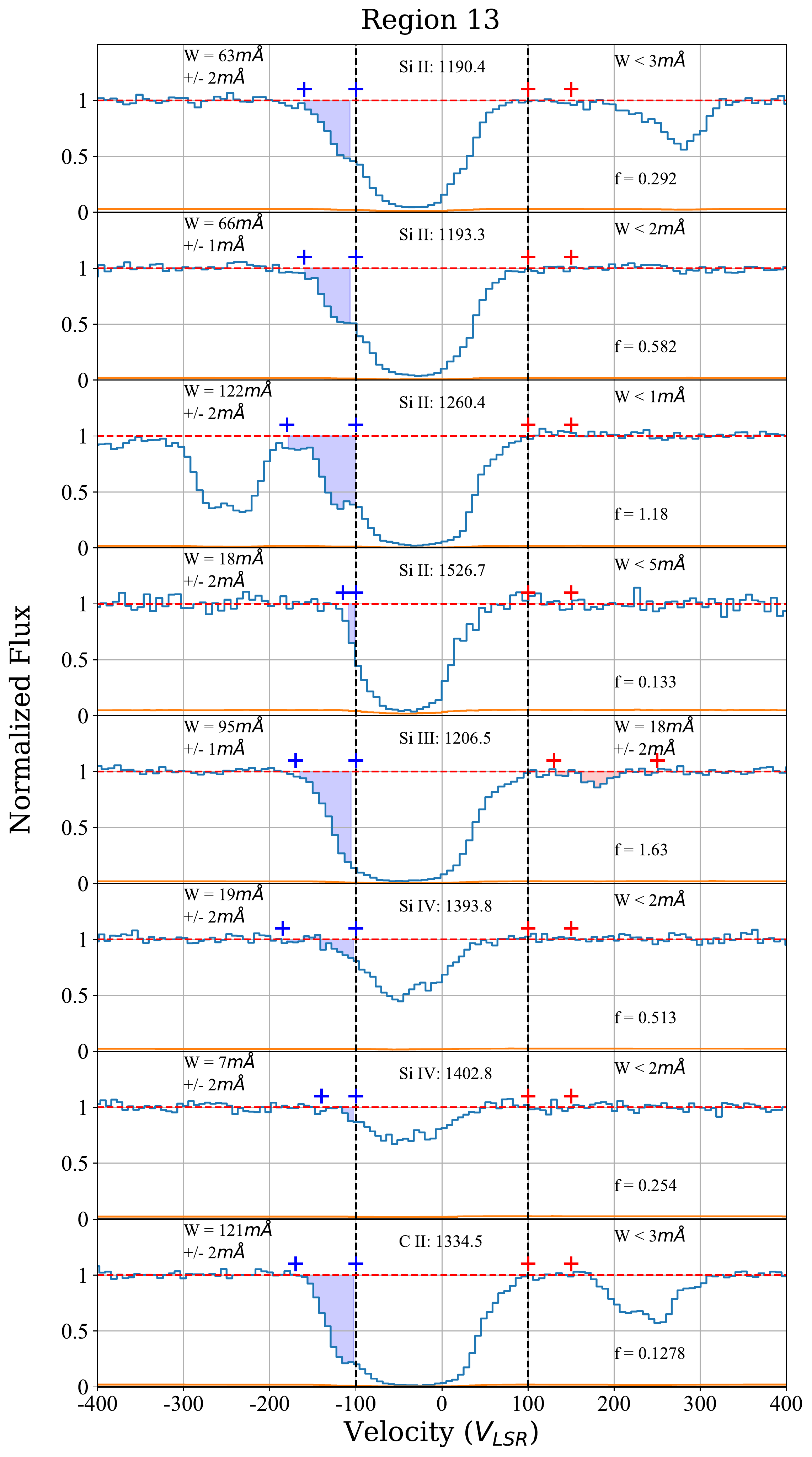}
    \includegraphics[width=0.65\columnwidth,height = 0.45\textheight]{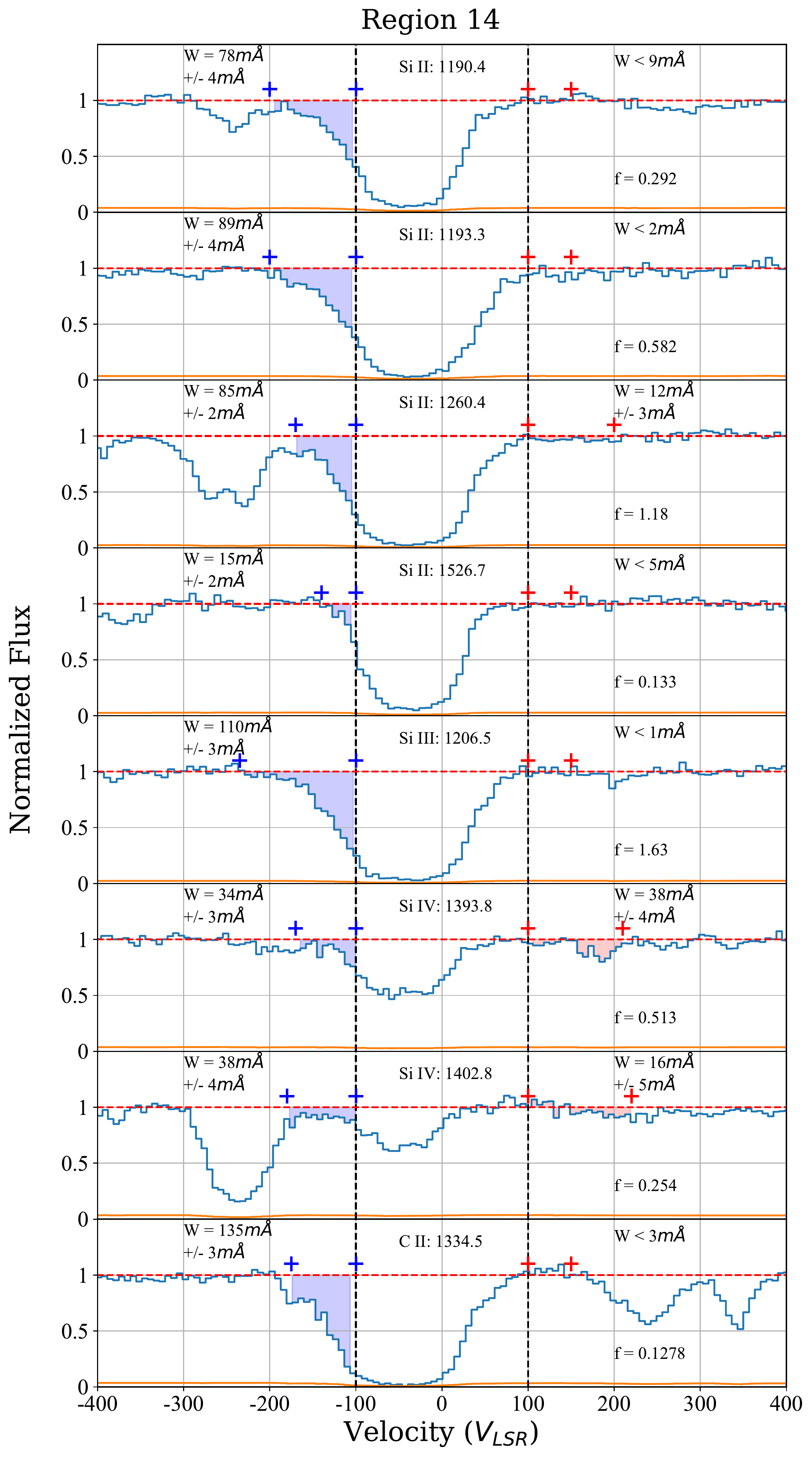}
    \includegraphics[width=0.65\columnwidth,height = 0.45\textheight]{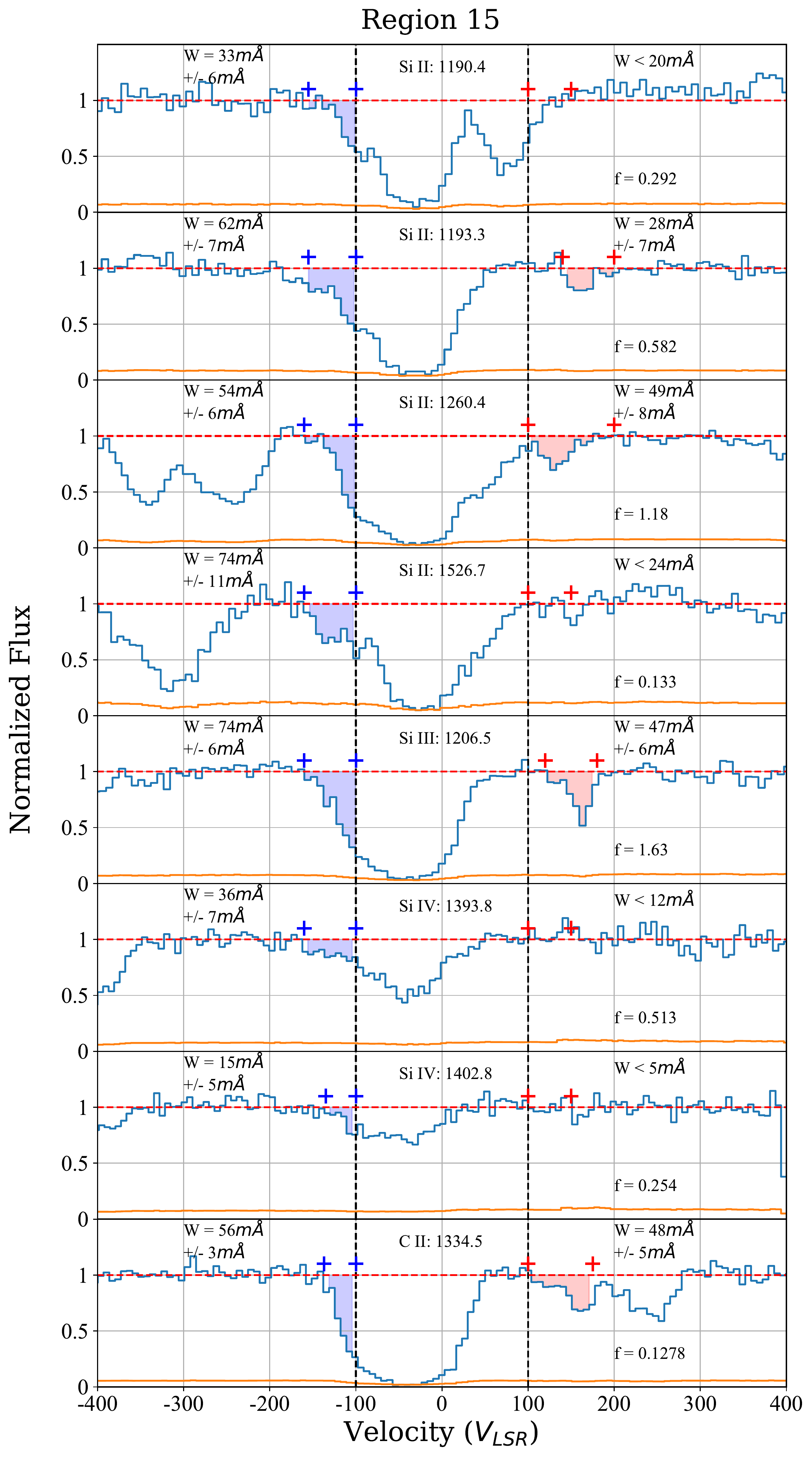}
    \includegraphics[width=0.65\columnwidth,height = 0.45\textheight]{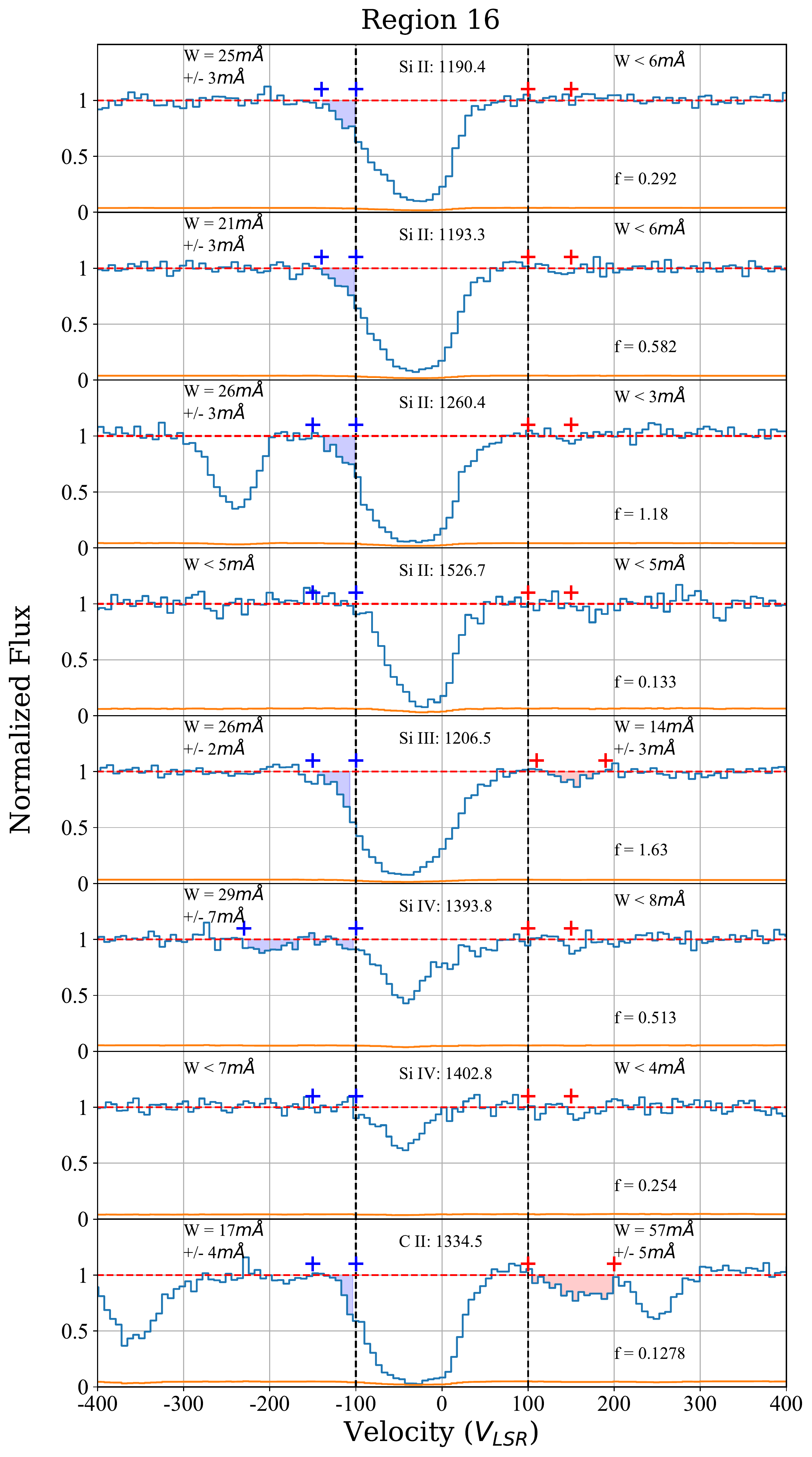}
    \includegraphics[width=0.65\columnwidth,height = 0.45\textheight]{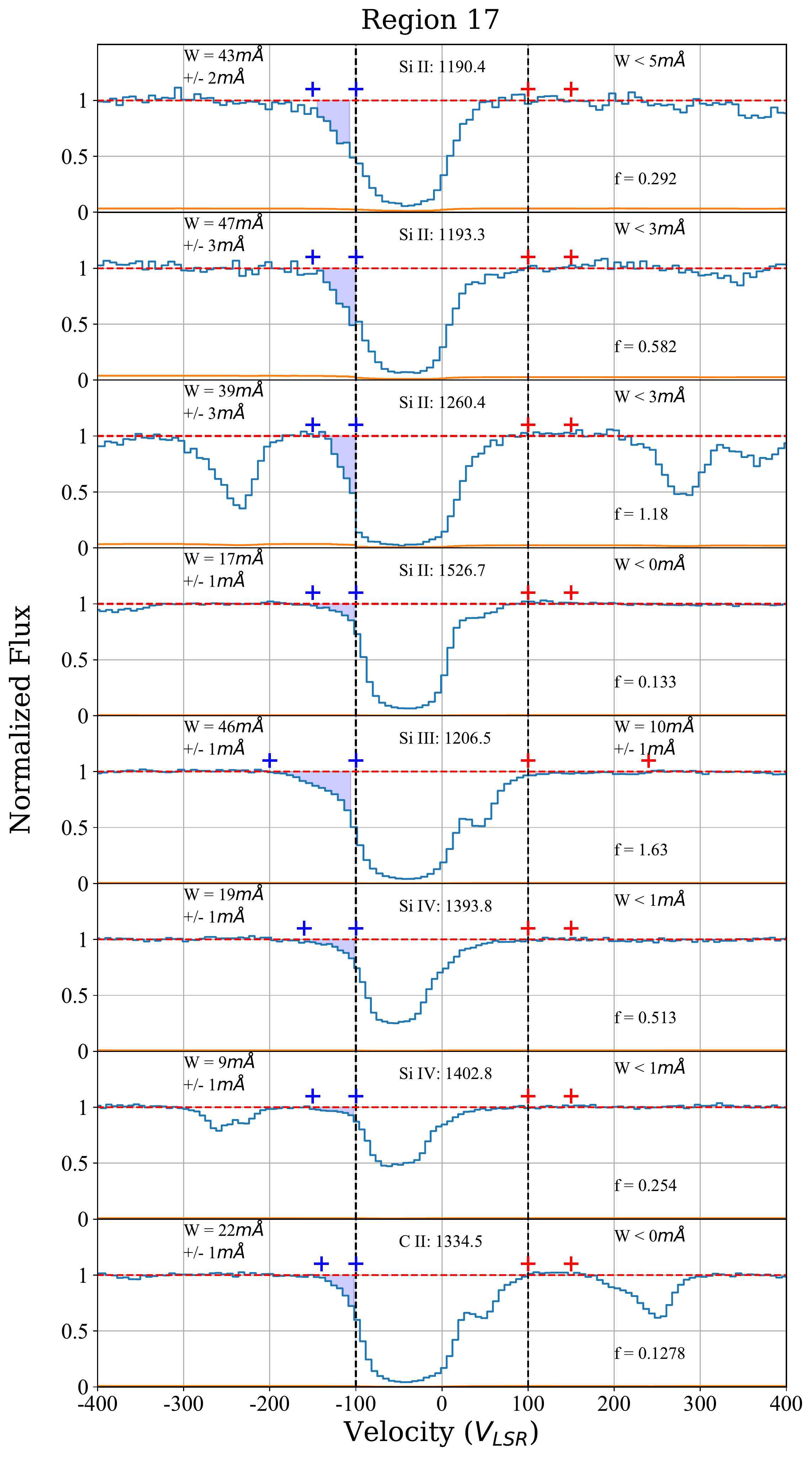}
    \includegraphics[width=0.65\columnwidth,height = 0.45\textheight]{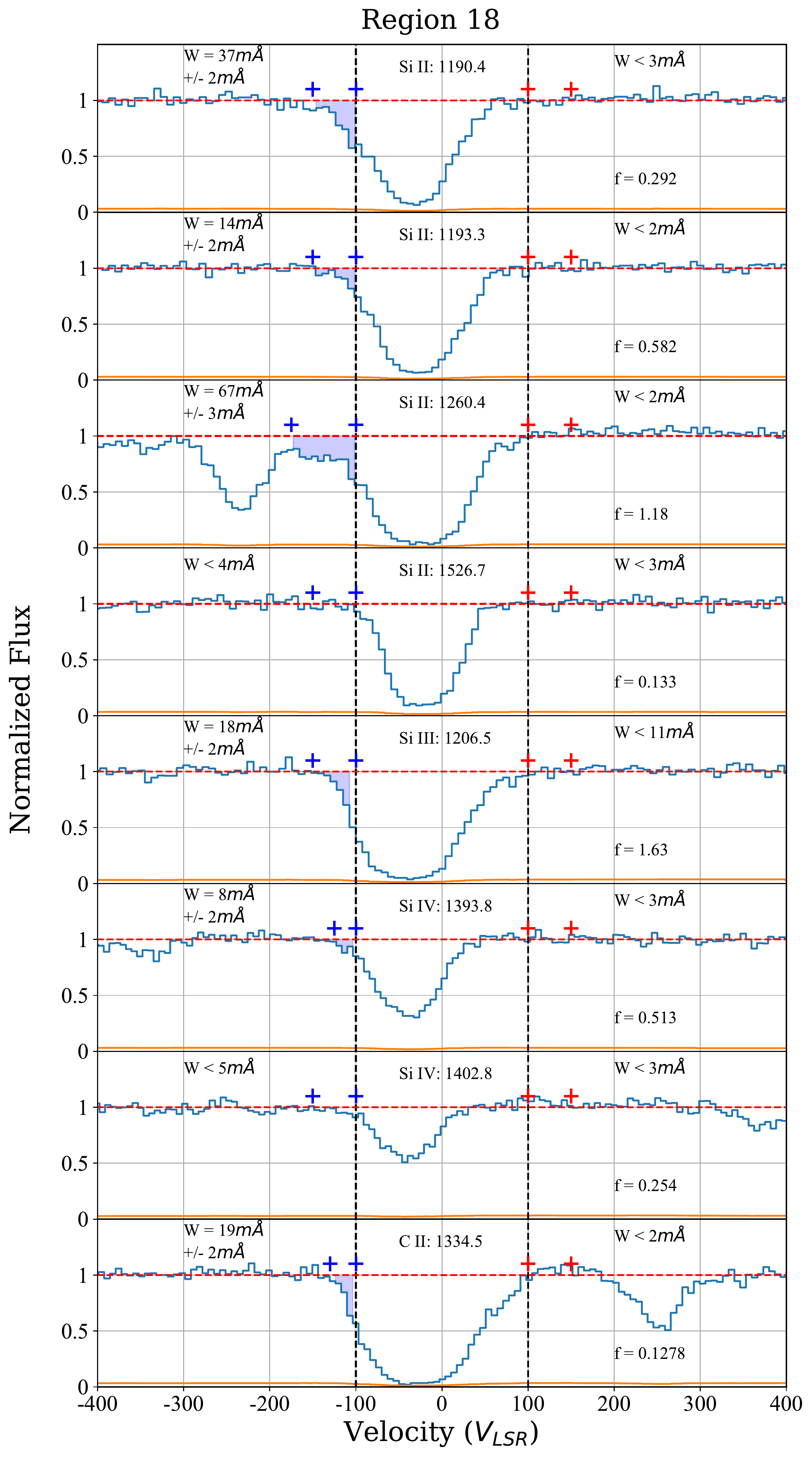}
    
    \contcaption{from Figure \ref{fig:AppendixB}.}
\end{figure*}

\begin{figure*}
    \includegraphics[width=0.65\columnwidth,height = 0.45\textheight]{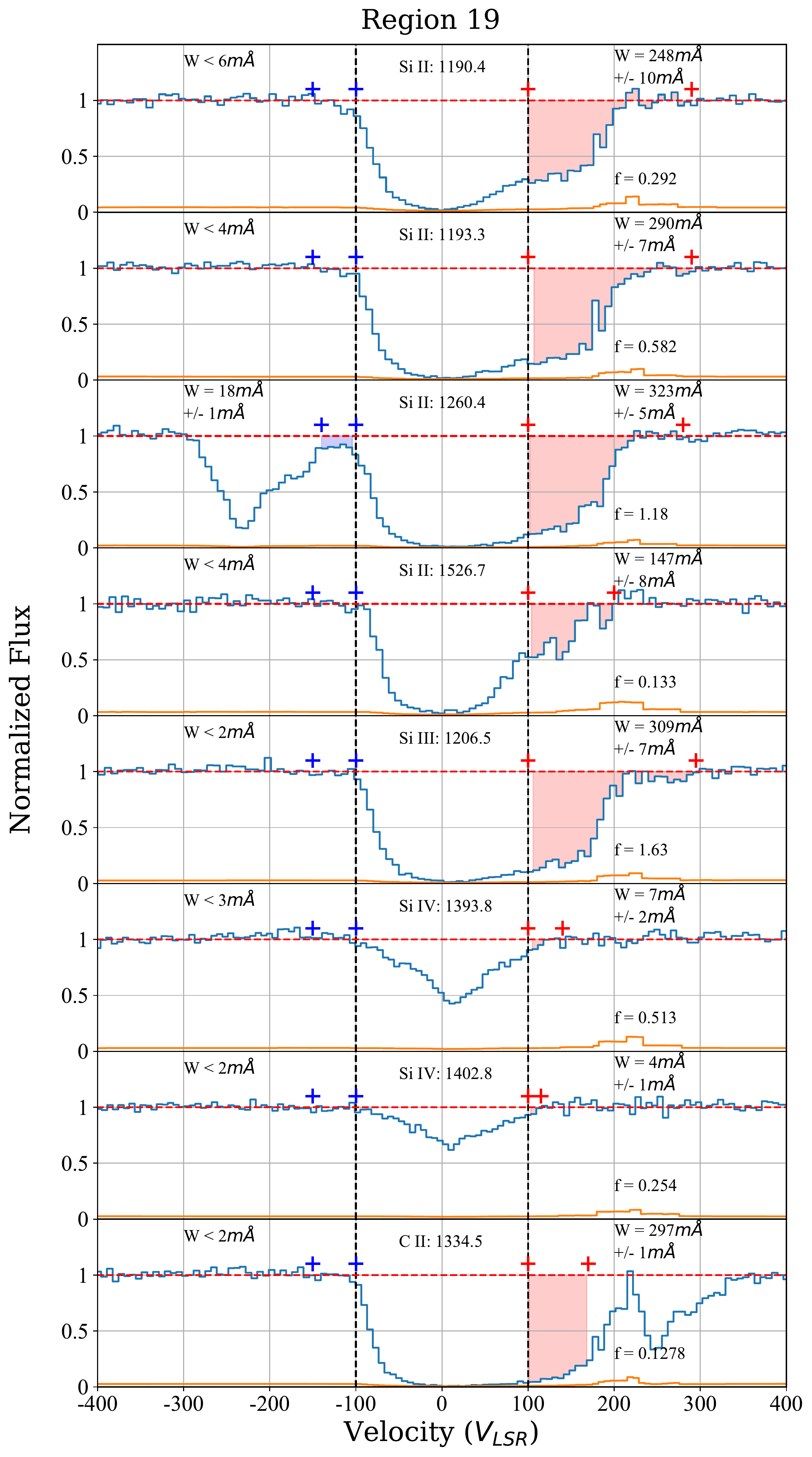}
    \includegraphics[width=0.65\columnwidth,height = 0.45\textheight]{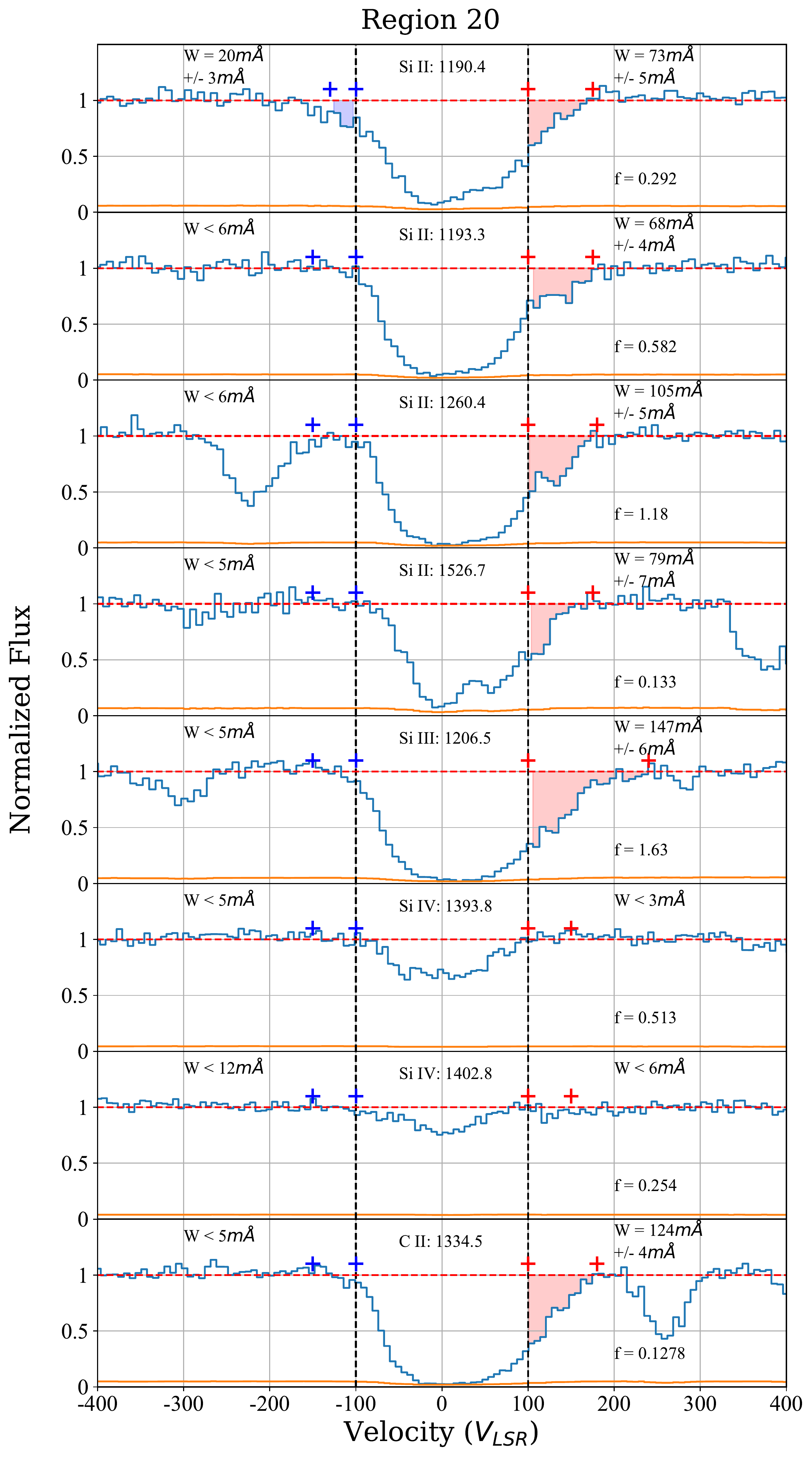}
    \includegraphics[width=0.65\columnwidth,height = 0.45\textheight]{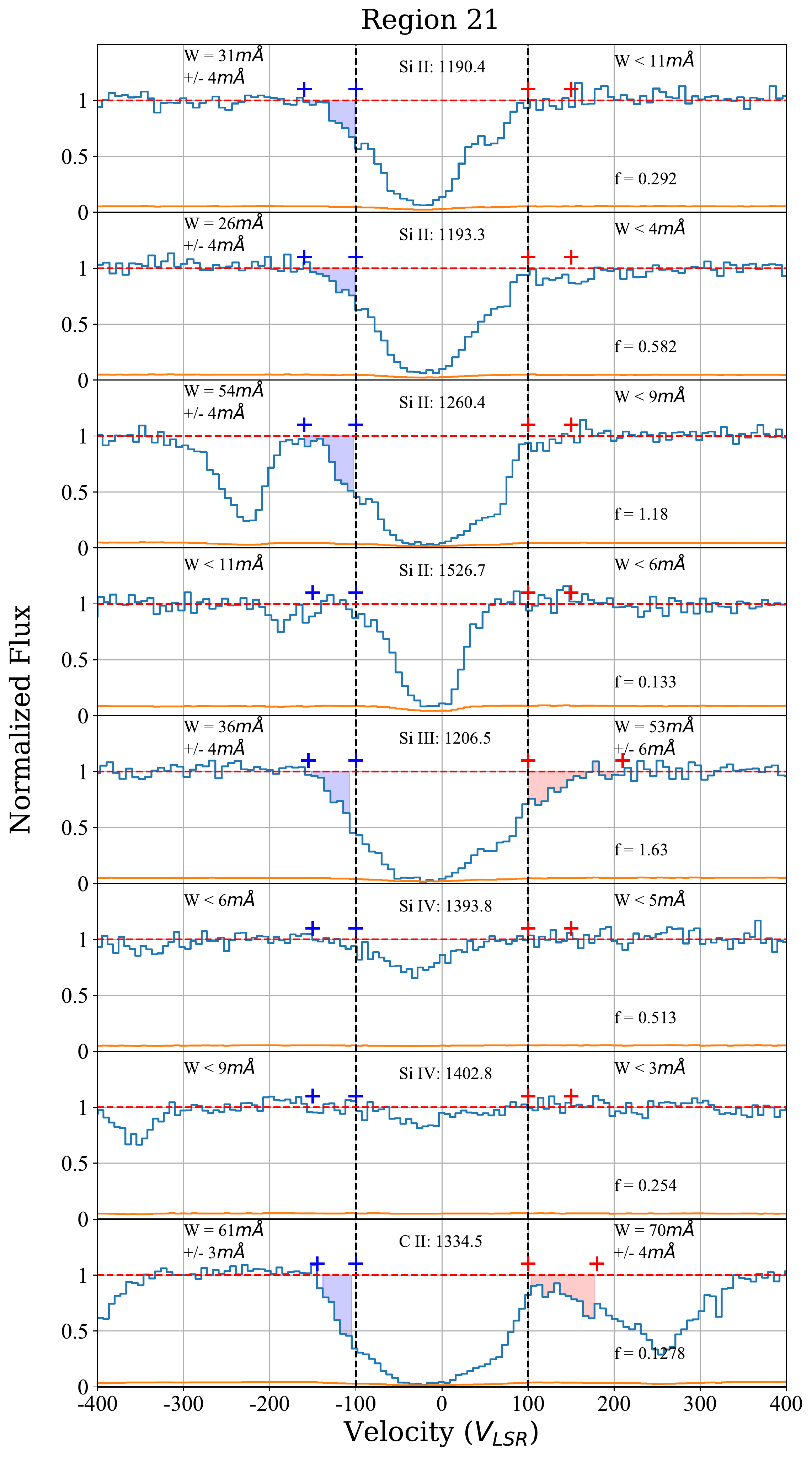}
    \includegraphics[width=0.65\columnwidth,height = 0.45\textheight]{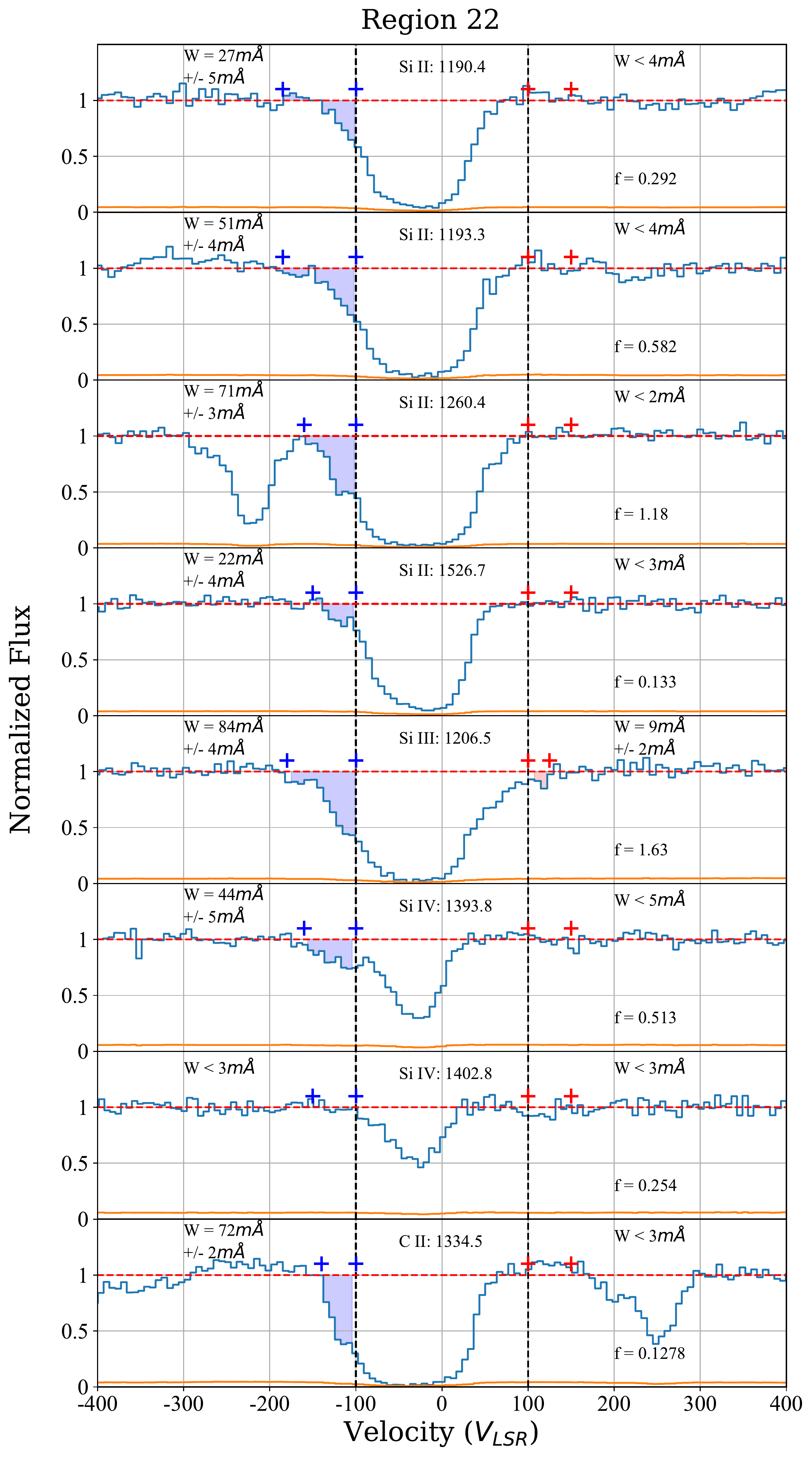}
    \includegraphics[width=0.65\columnwidth,height = 0.45\textheight]{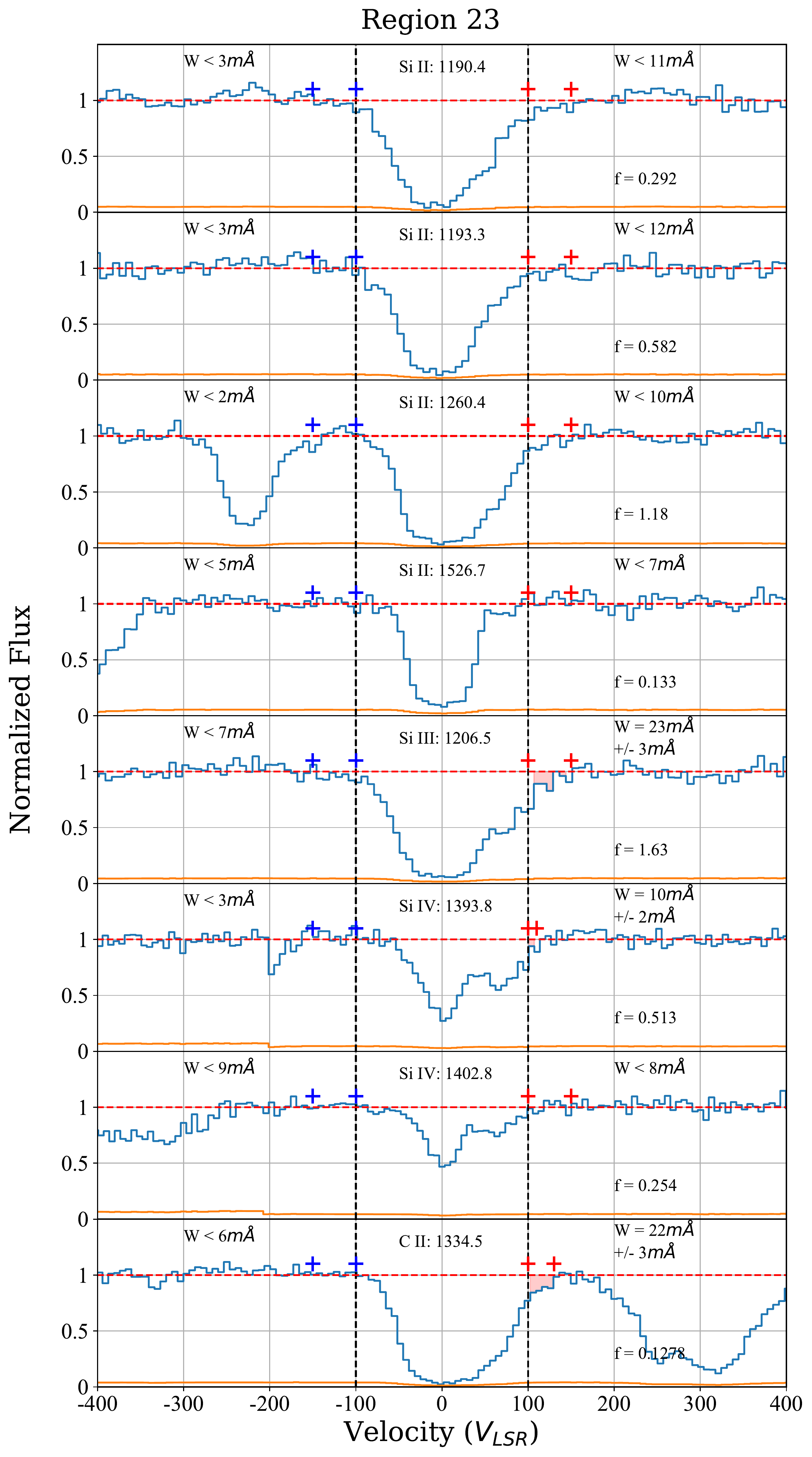}
    \includegraphics[width=0.65\columnwidth,height = 0.45\textheight]{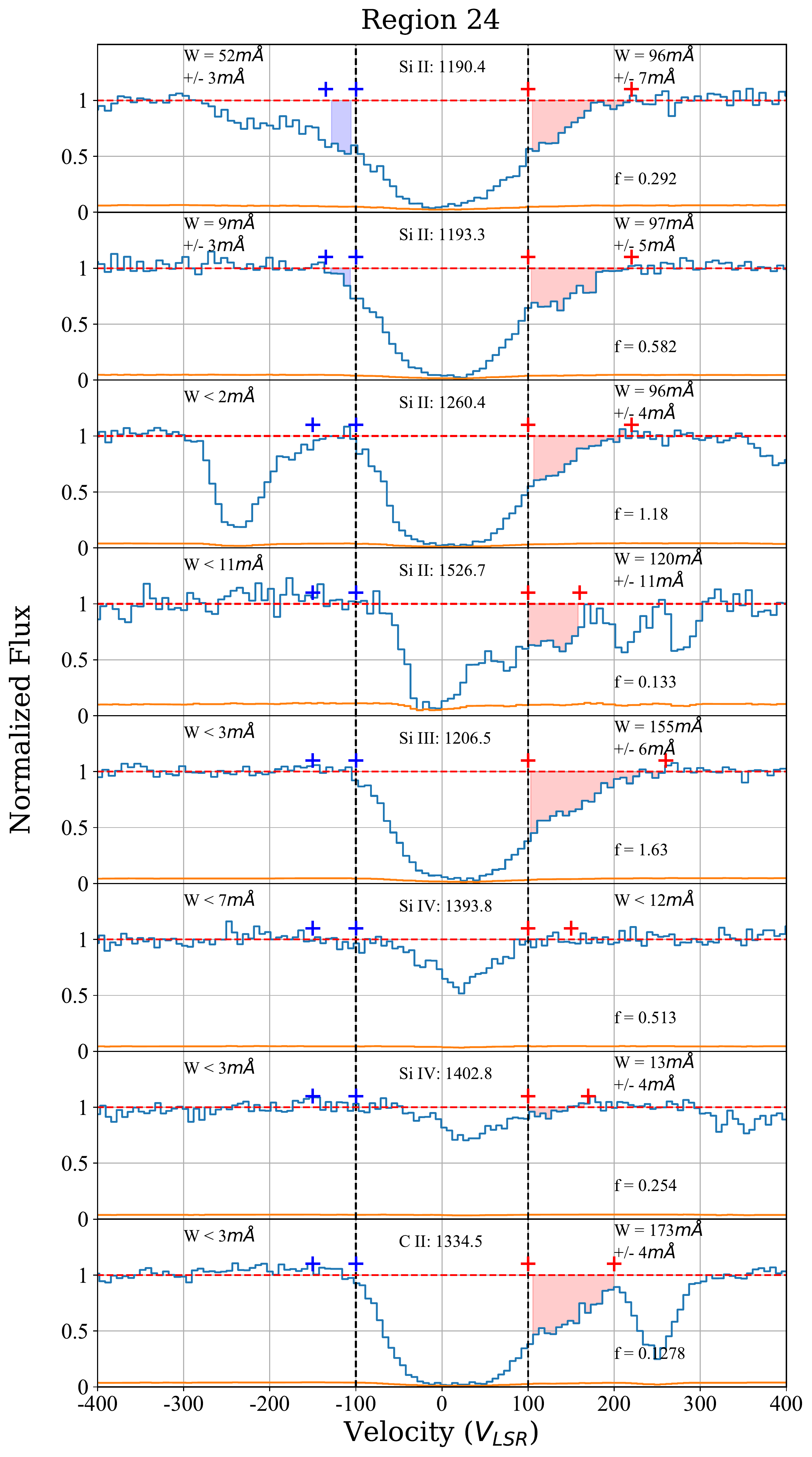}
    
    \contcaption{from Figure \ref{fig:AppendixB}.}
\end{figure*}

\begin{figure*}
    \includegraphics[width=0.65\columnwidth,height = 0.45\textheight]{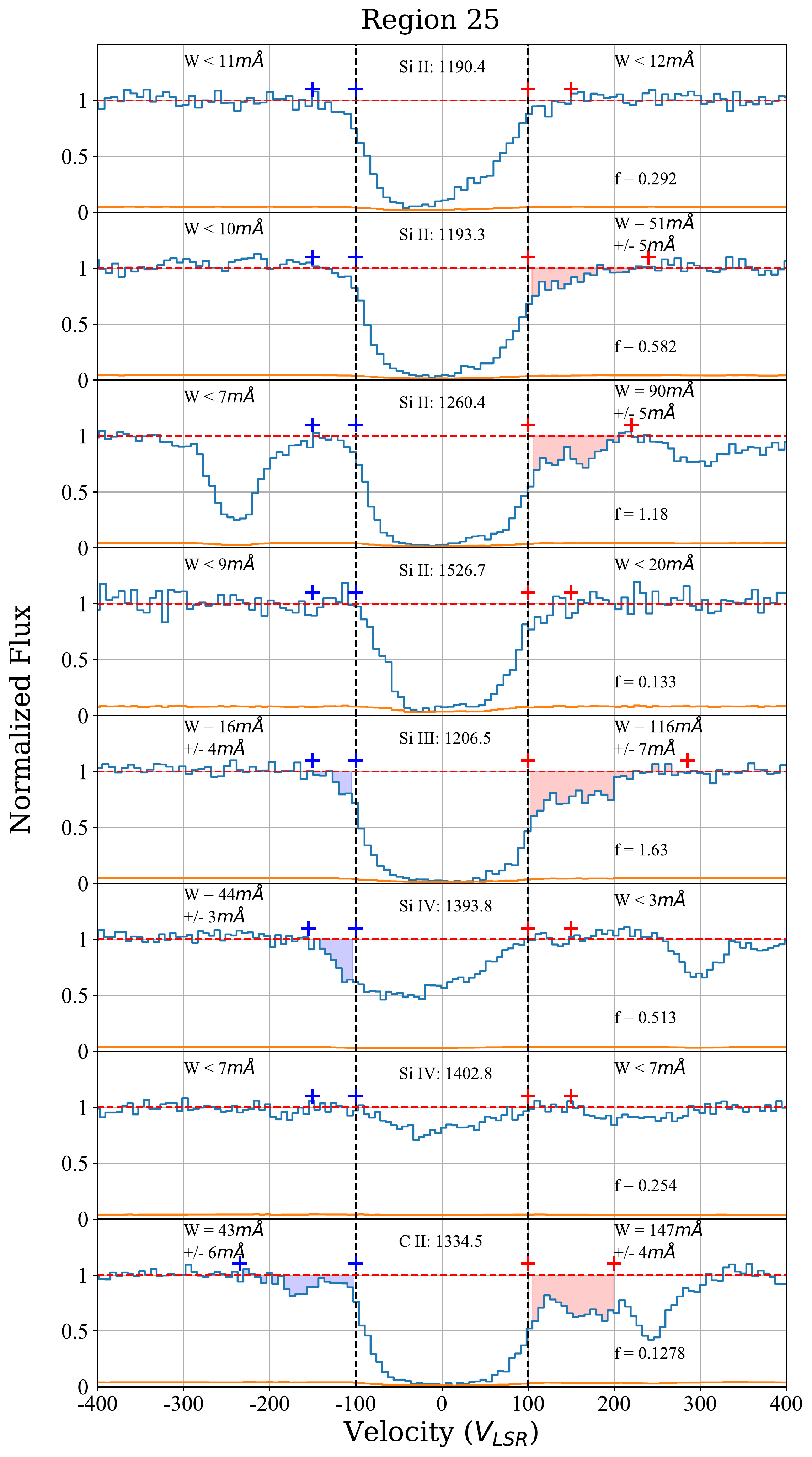}
    \includegraphics[width=0.65\columnwidth,height = 0.45\textheight]{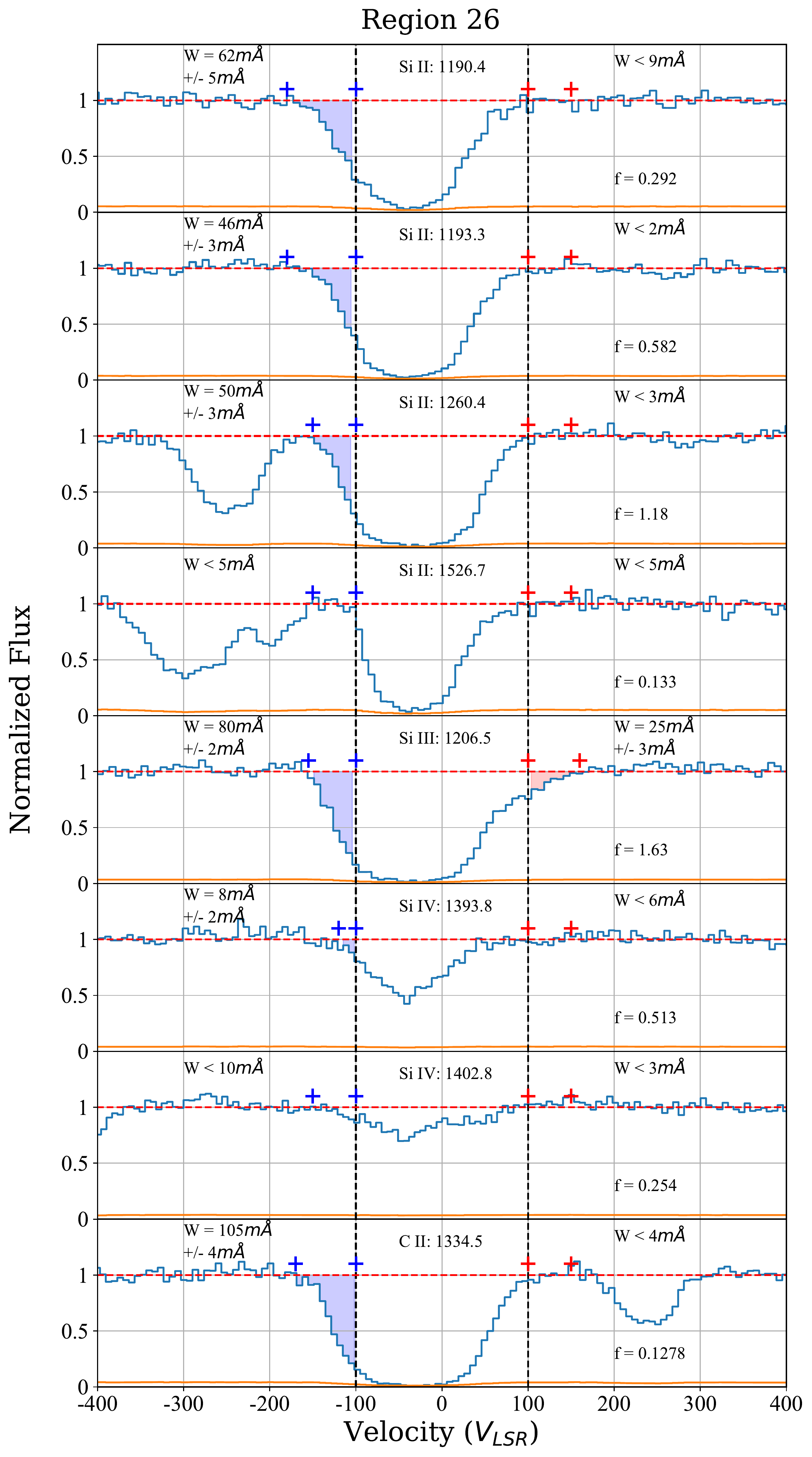}
    \includegraphics[width=0.65\columnwidth,height = 0.45\textheight]{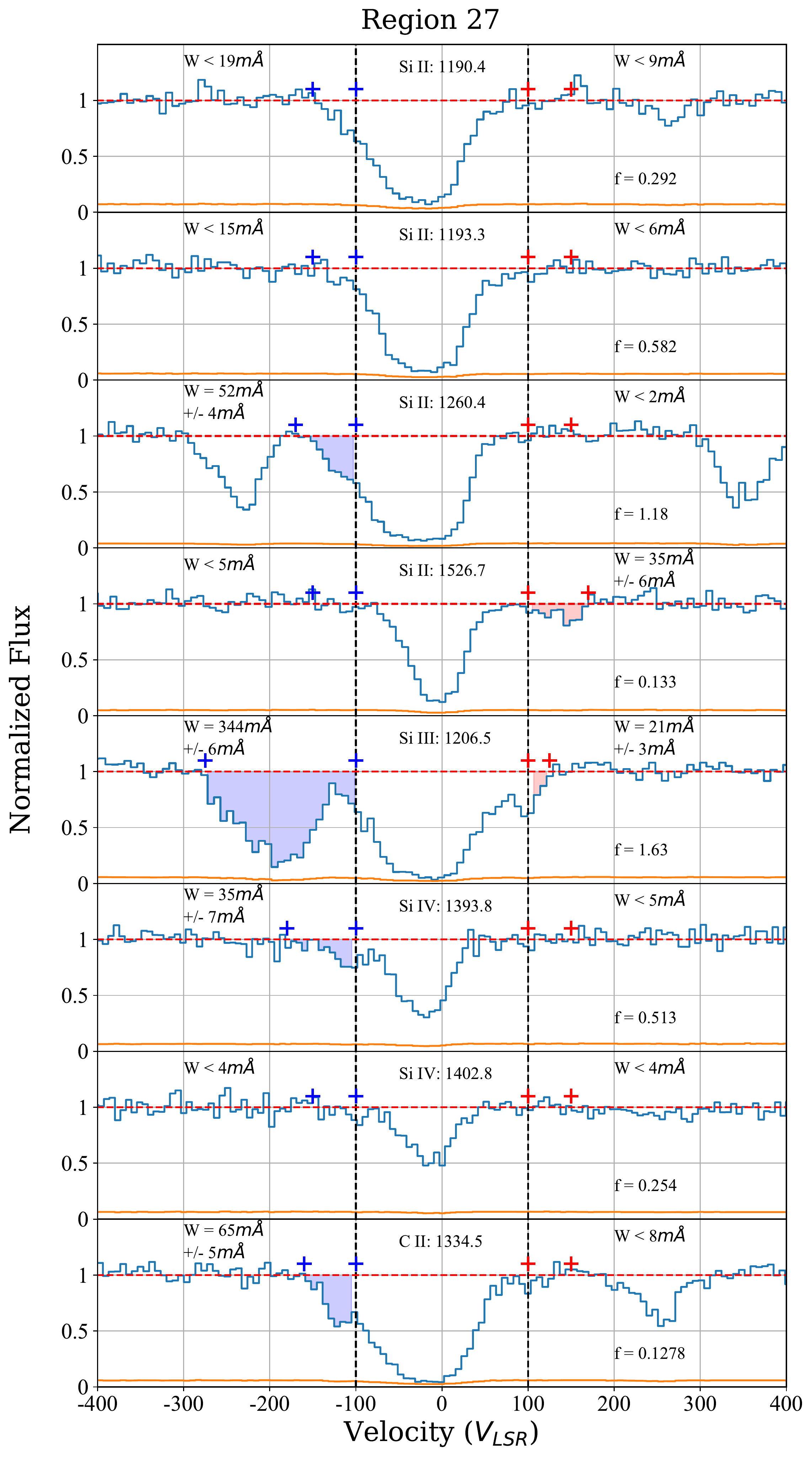}
    \includegraphics[width=0.65\columnwidth,height = 0.45\textheight]{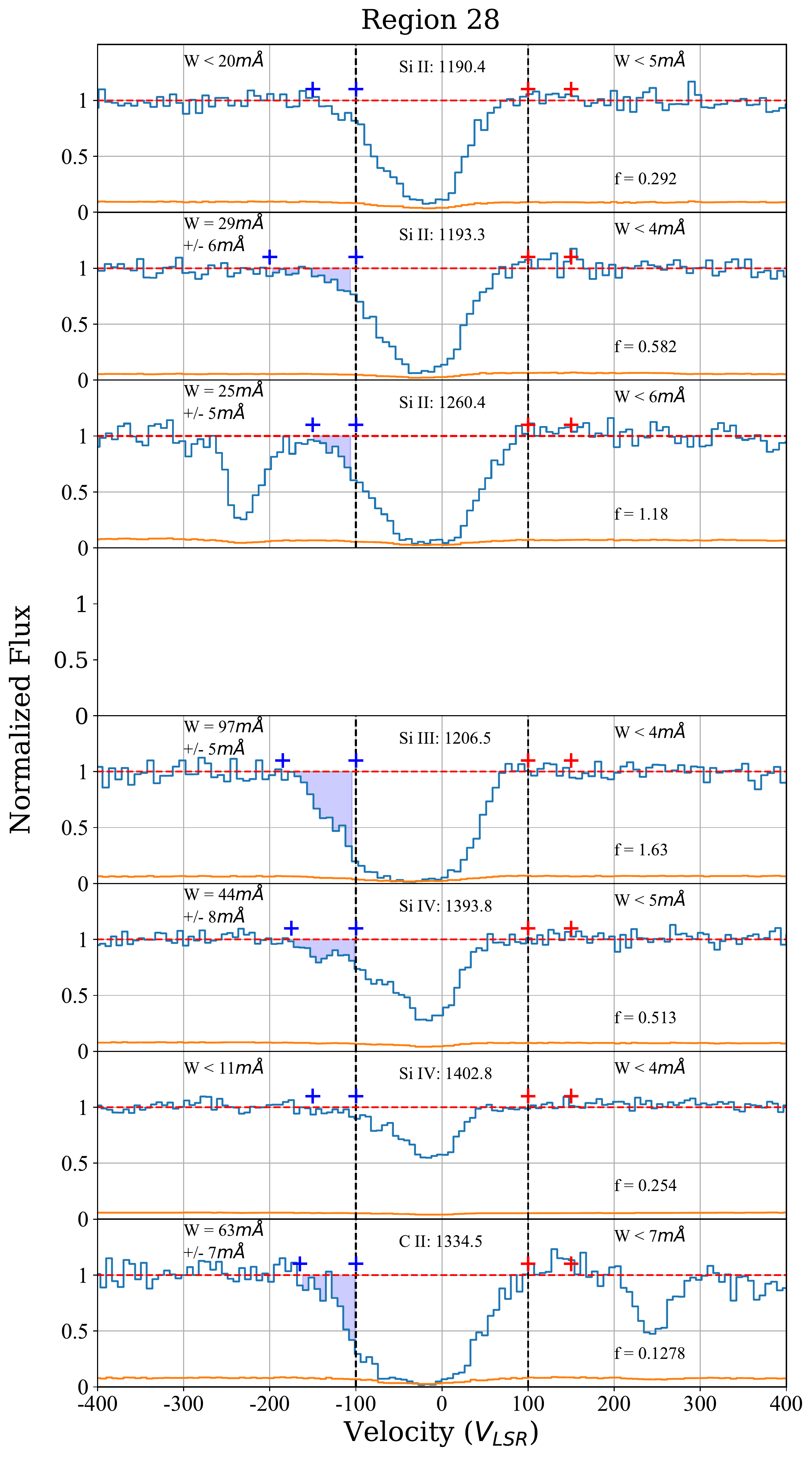}
    \includegraphics[width=0.65\columnwidth,height = 0.45\textheight]{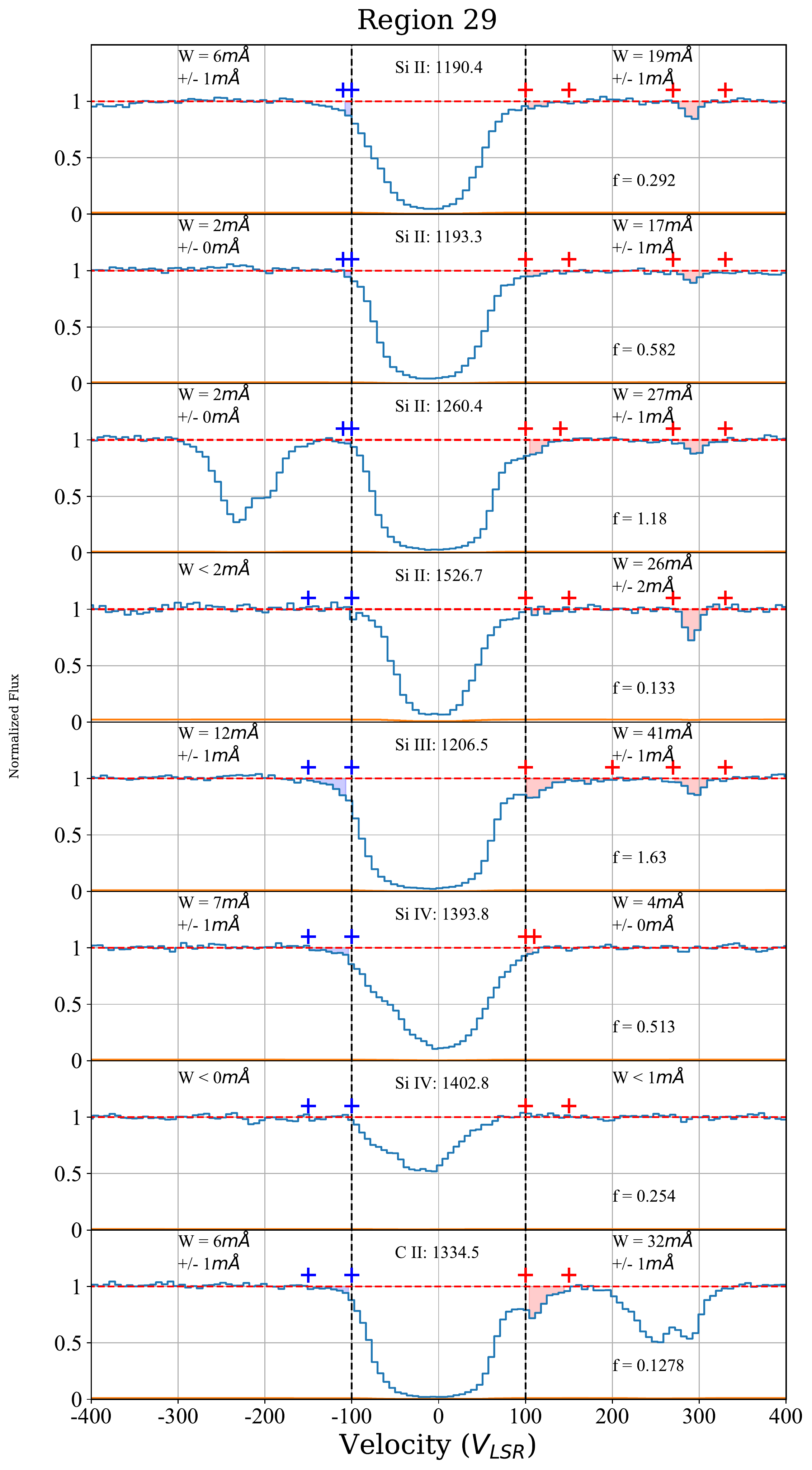}
    \includegraphics[width=0.65\columnwidth,height = 0.45\textheight]{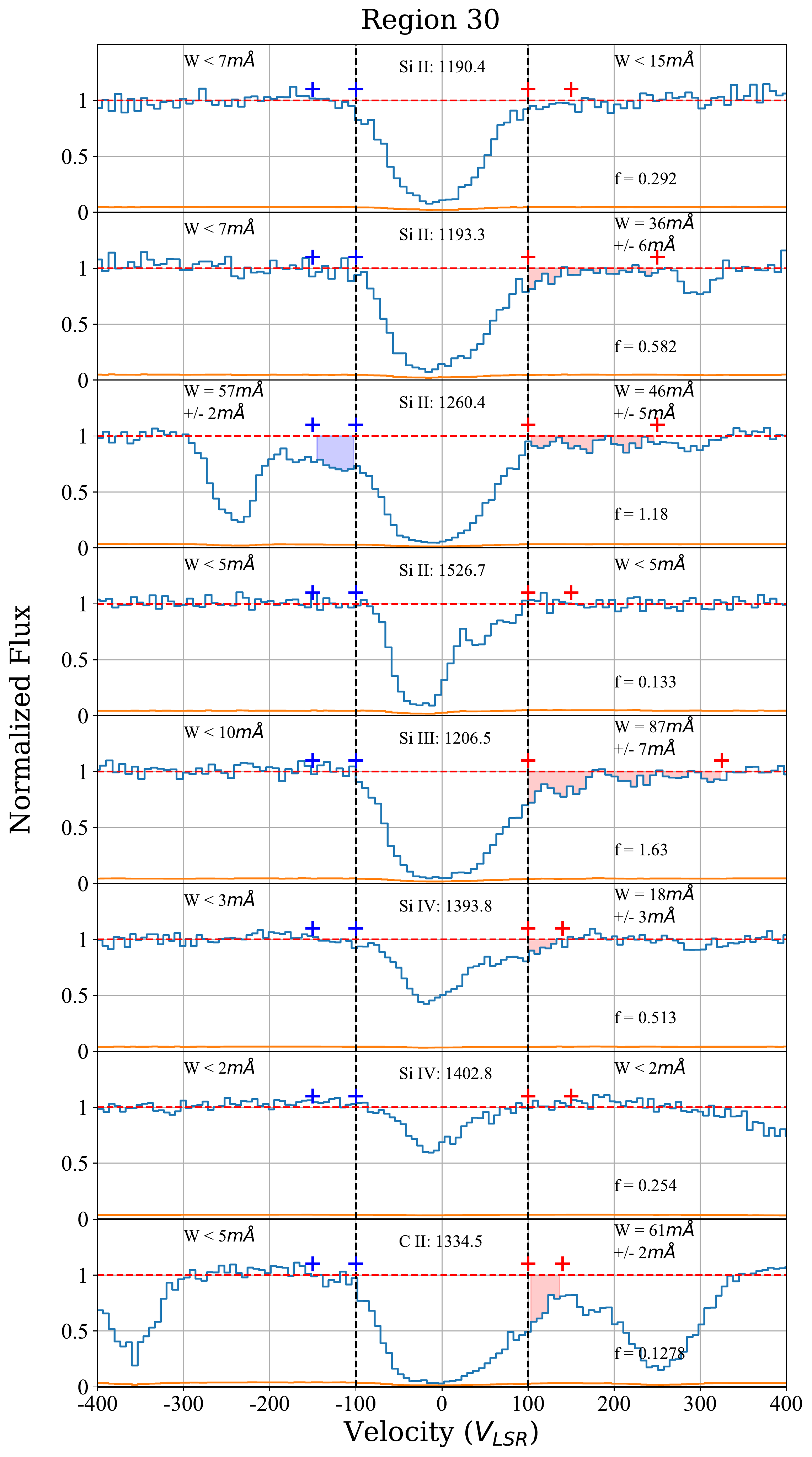}
    
    \contcaption{from Figure \ref{fig:AppendixB}.}
\end{figure*}

\begin{figure*}
    \includegraphics[width=0.65\columnwidth,height = 0.45\textheight]{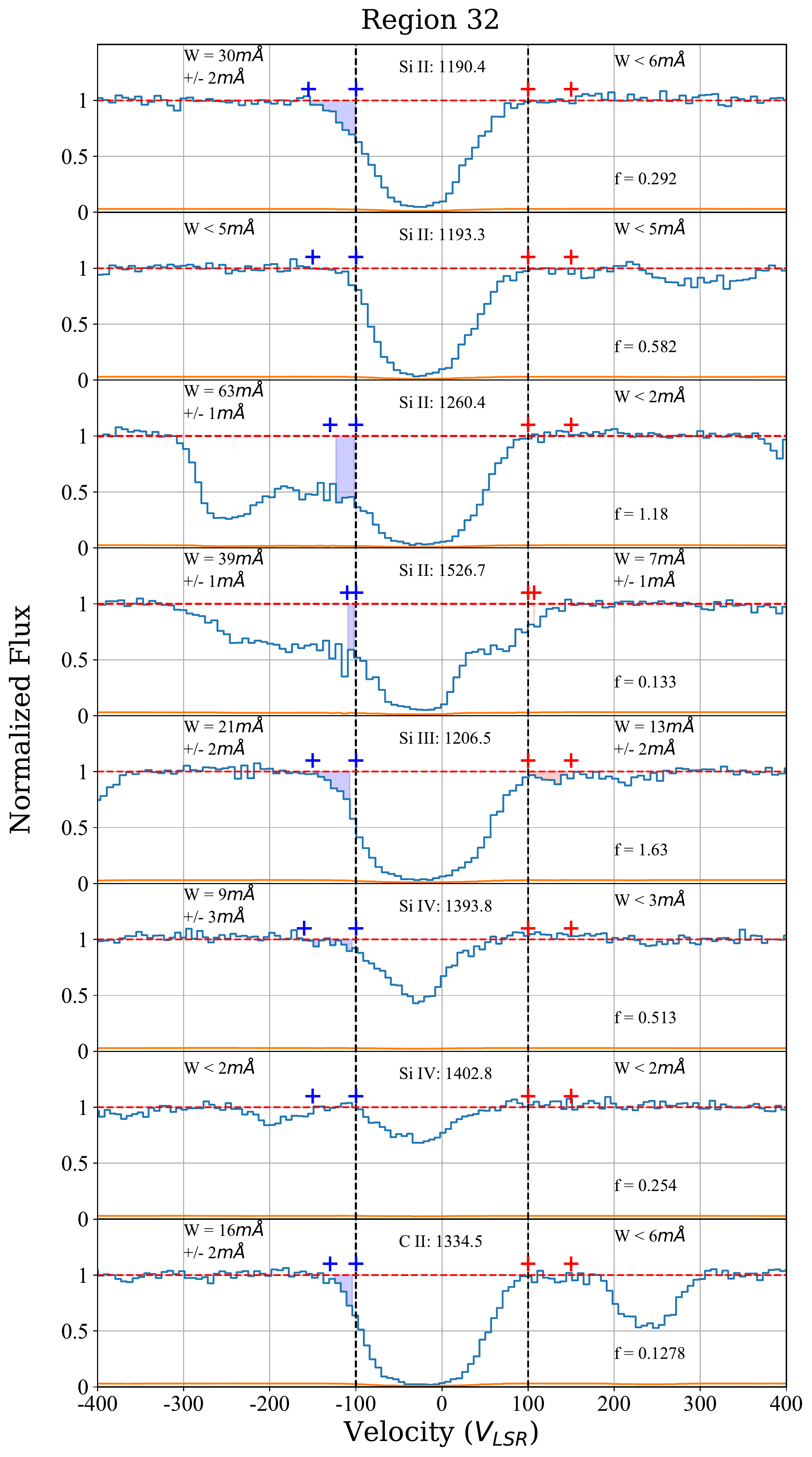}
    \includegraphics[width=0.65\columnwidth,height = 0.45\textheight]{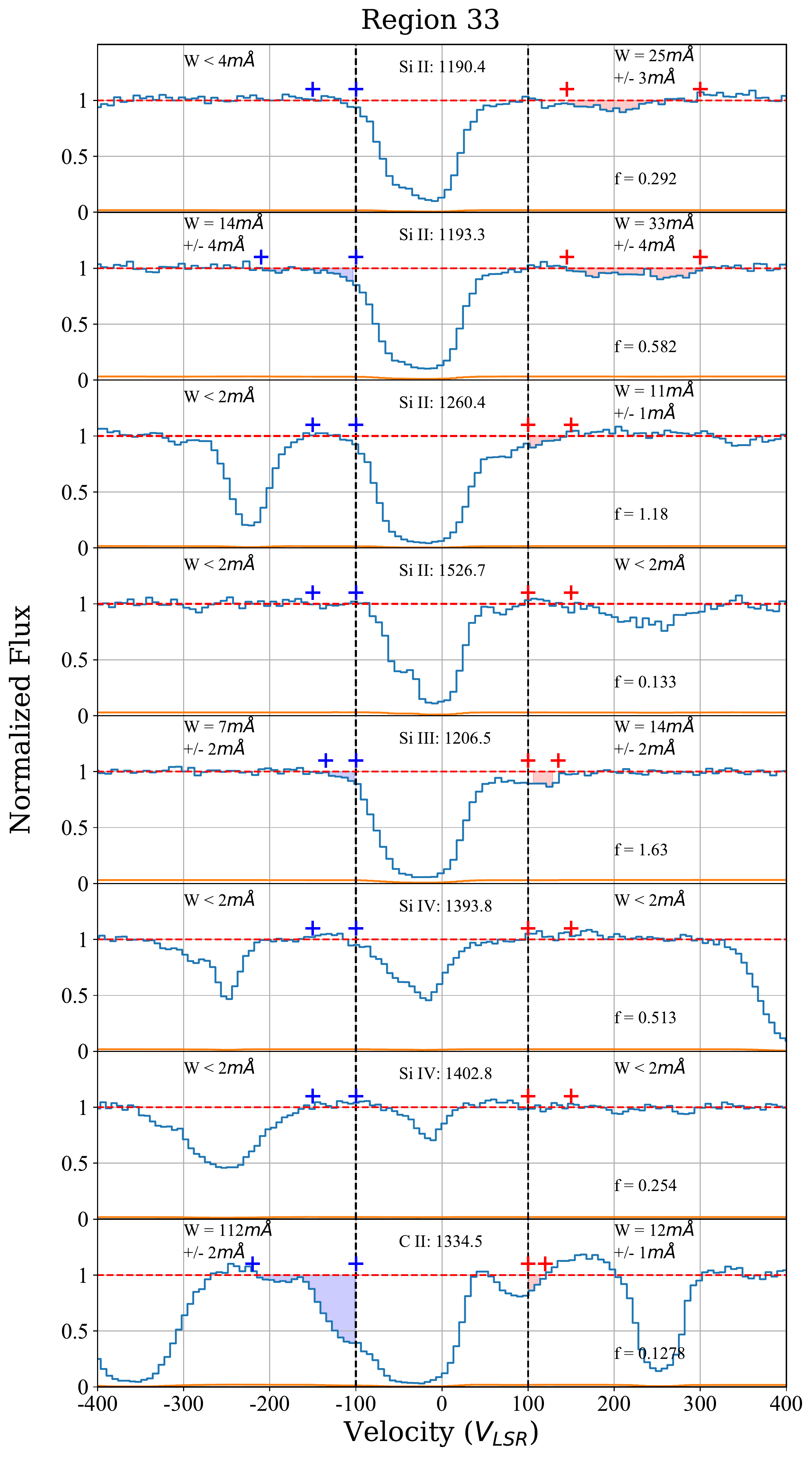}
    \includegraphics[width=0.65\columnwidth,height = 0.45\textheight]{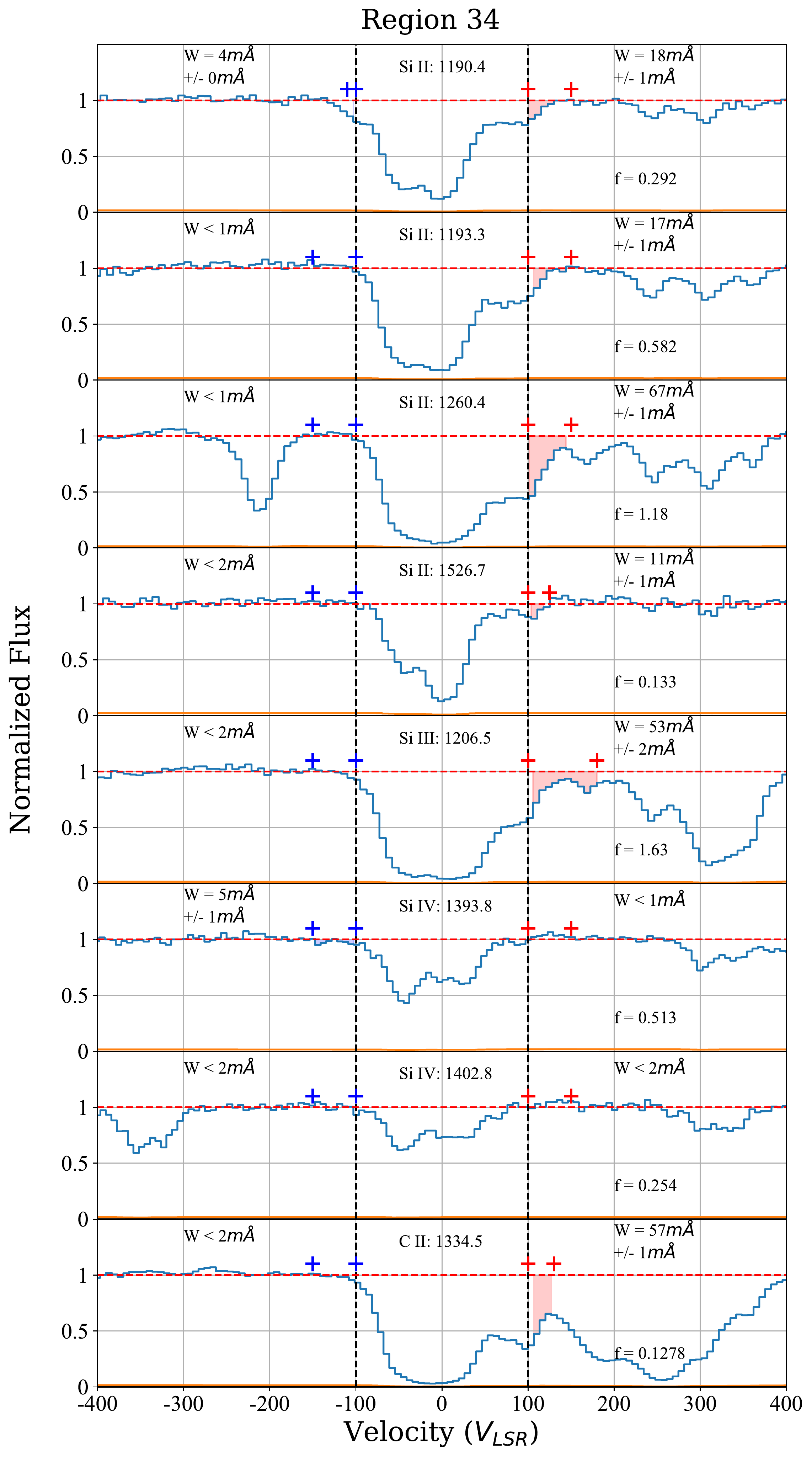}
    \includegraphics[width=0.65\columnwidth,height = 0.45\textheight]{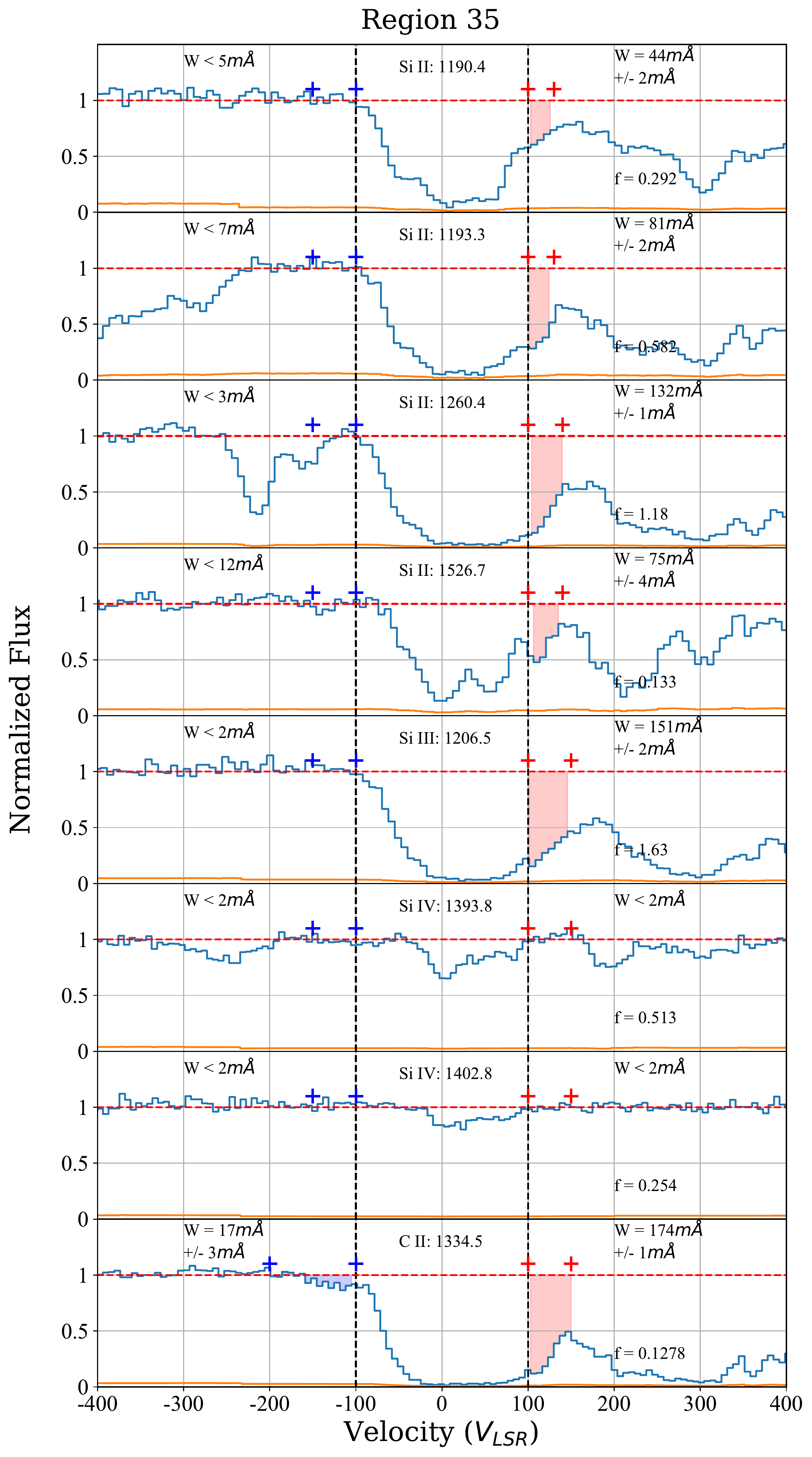}
    \includegraphics[width=0.65\columnwidth,height = 0.45\textheight]{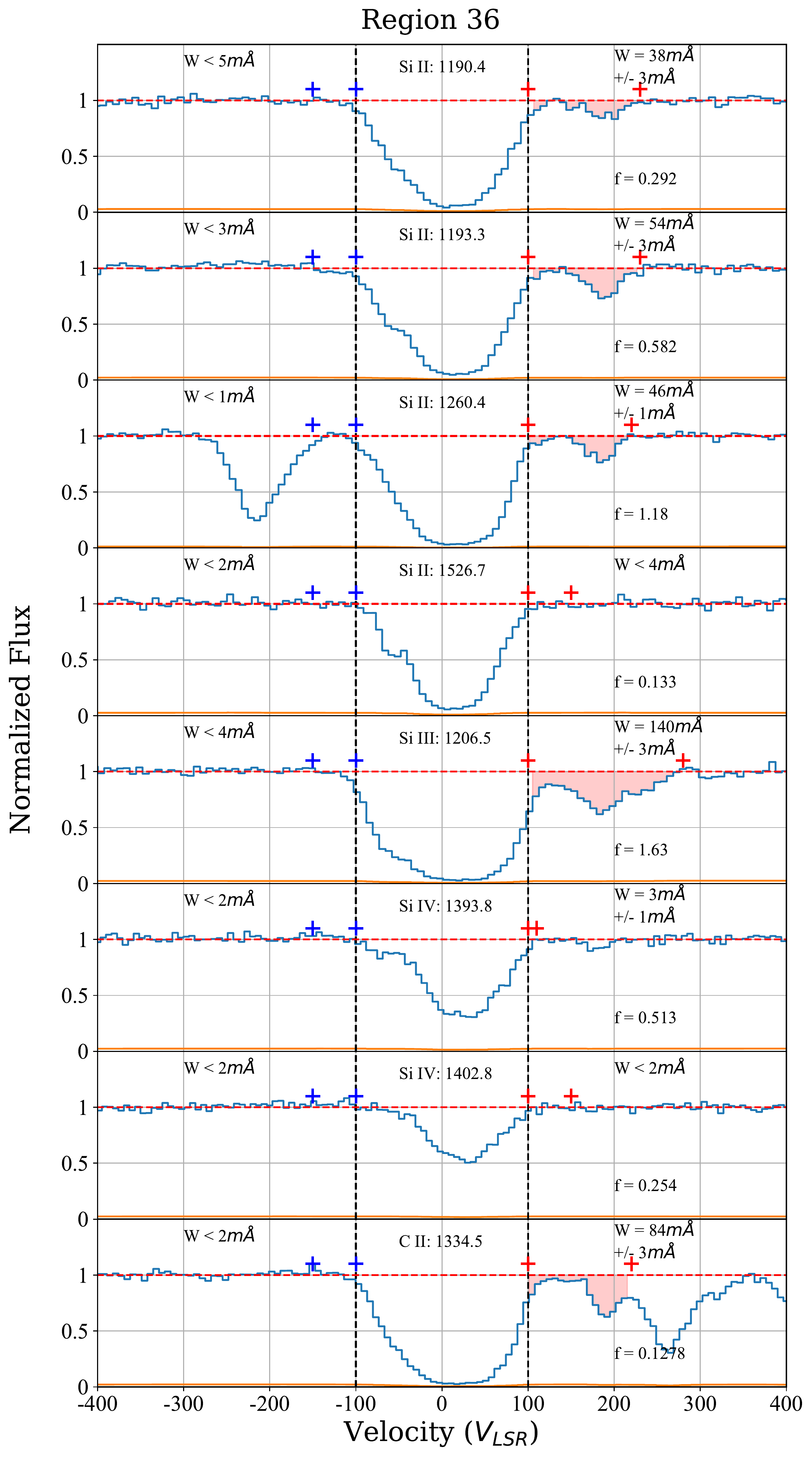}
    \includegraphics[width=0.65\columnwidth,height = 0.45\textheight]{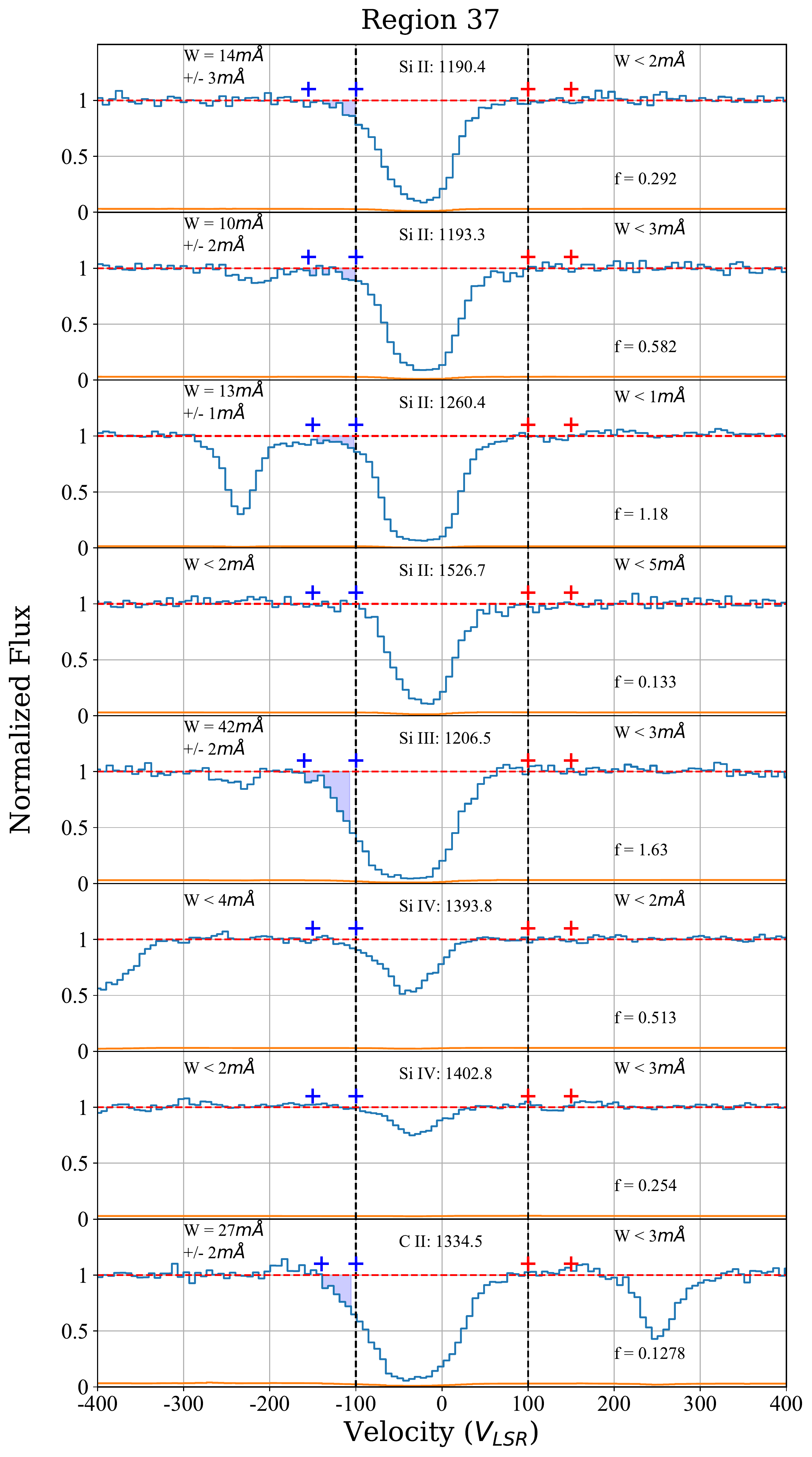}
    
    \contcaption{from Figure \ref{fig:AppendixB}.}
\end{figure*}

\begin{figure*}
    \includegraphics[width=0.65\columnwidth,height = 0.45\textheight]{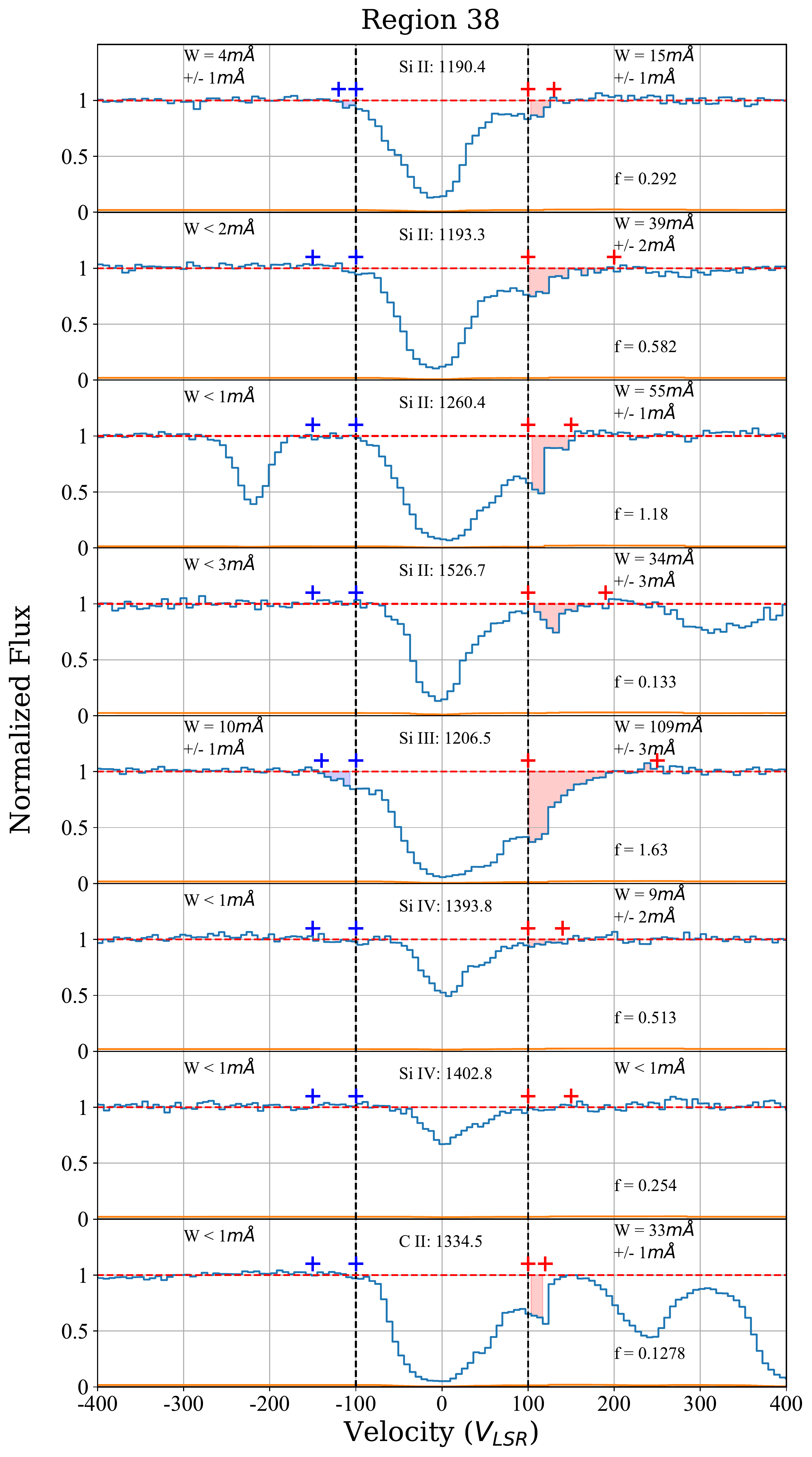}
    \includegraphics[width=0.65\columnwidth,height = 0.45\textheight]{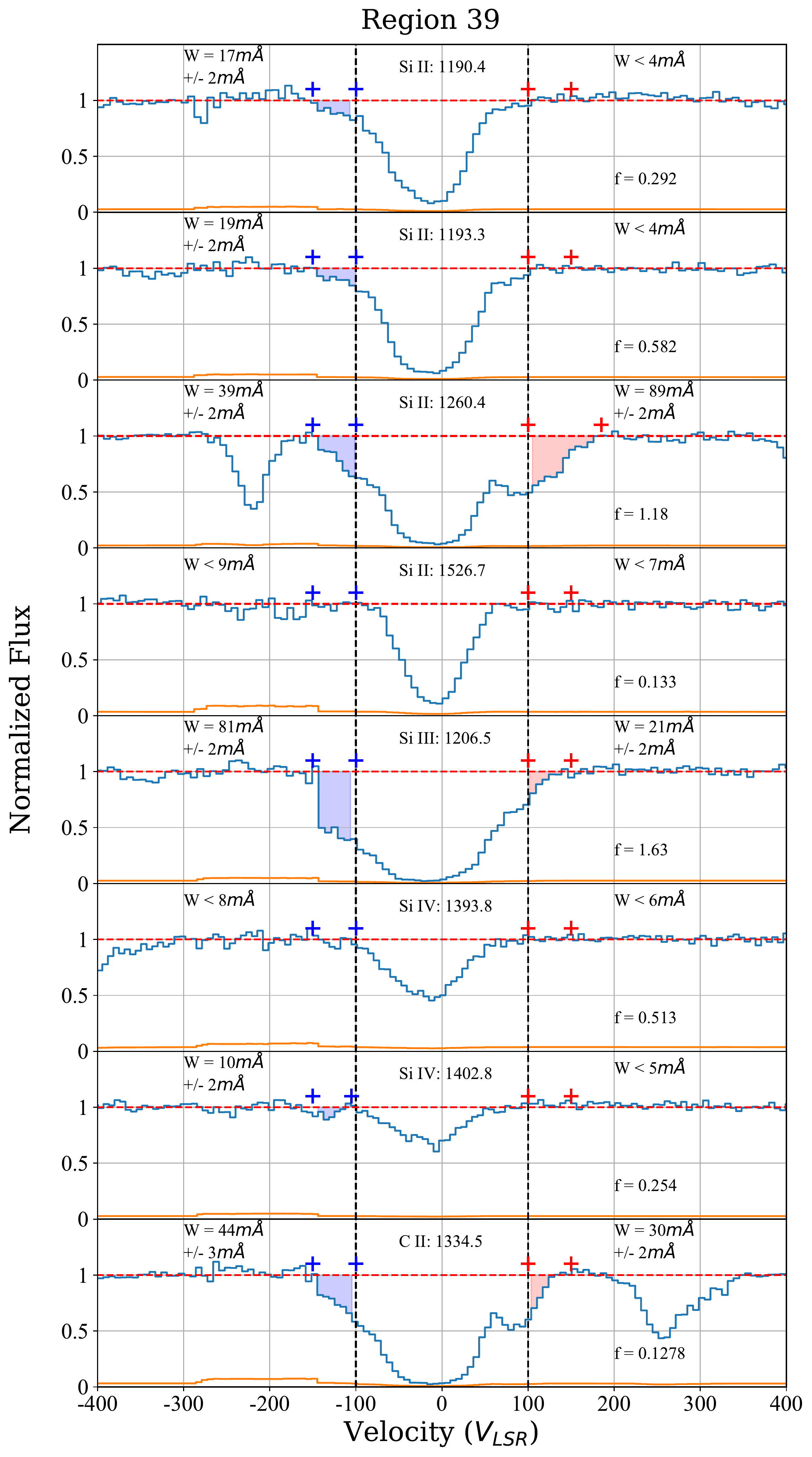}
    \includegraphics[width=0.65\columnwidth,height = 0.45\textheight]{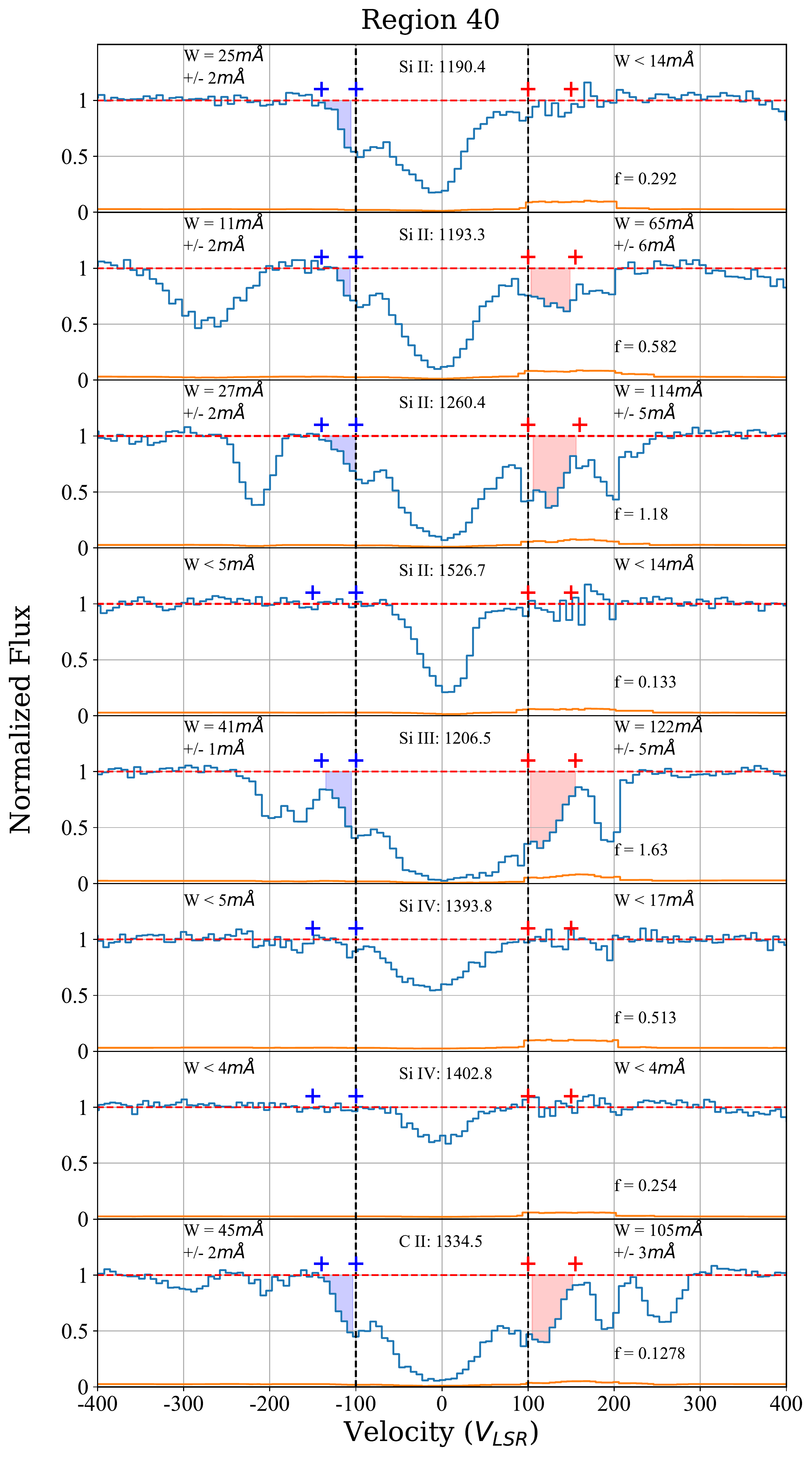}
    \includegraphics[width=0.65\columnwidth,height = 0.45\textheight]{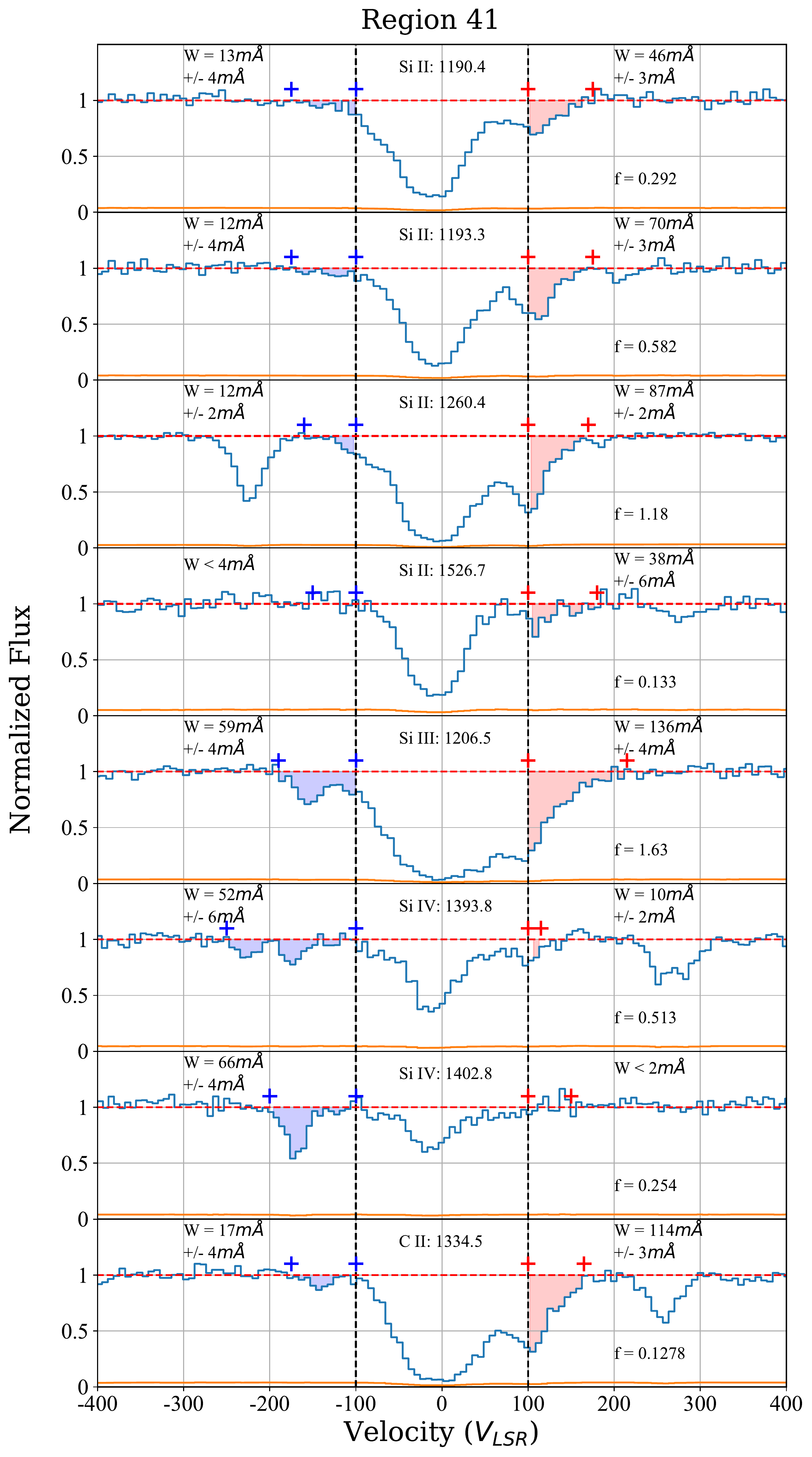}
    \includegraphics[width=0.65\columnwidth,height = 0.45\textheight]{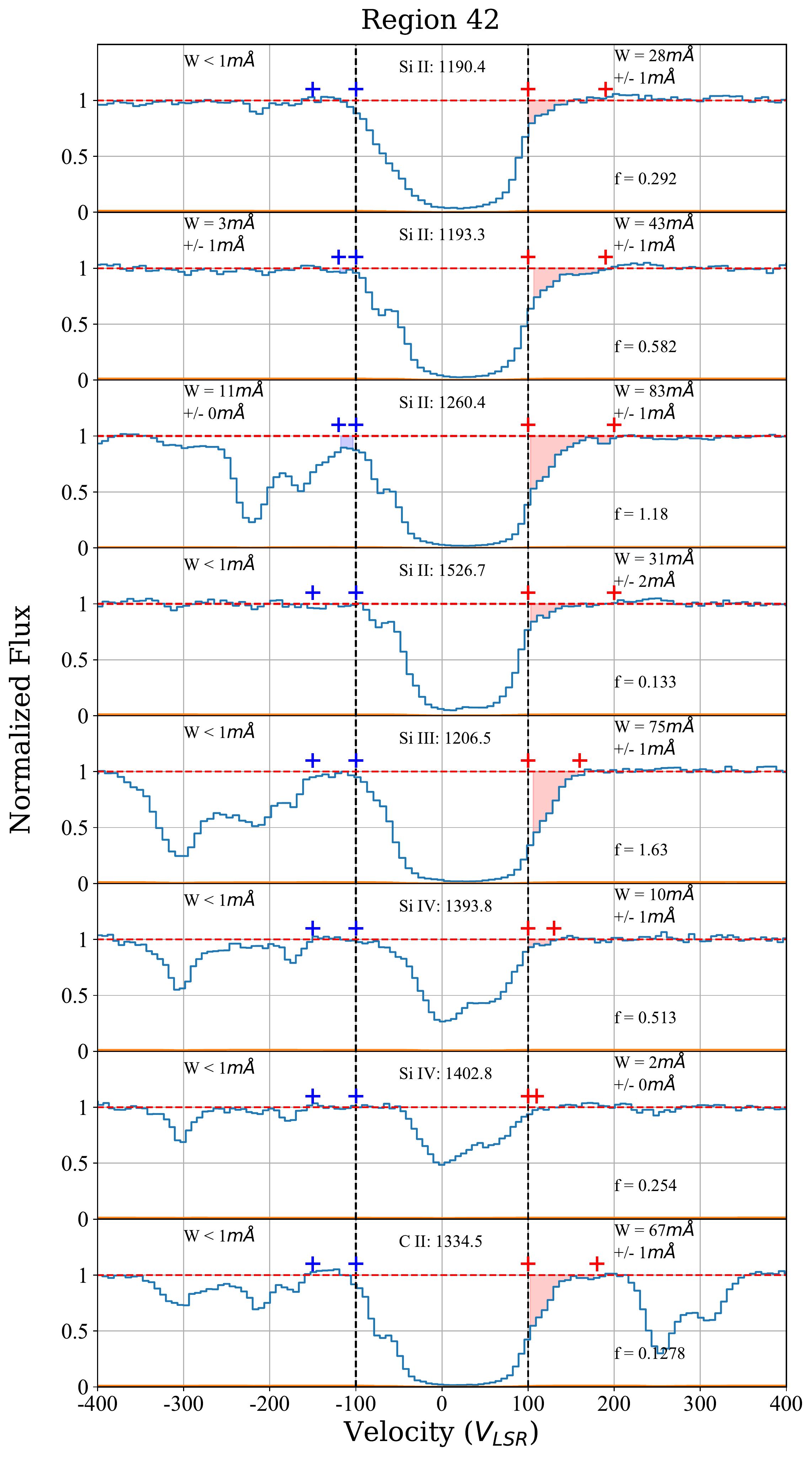}
    \includegraphics[width=0.65\columnwidth,height = 0.45\textheight]{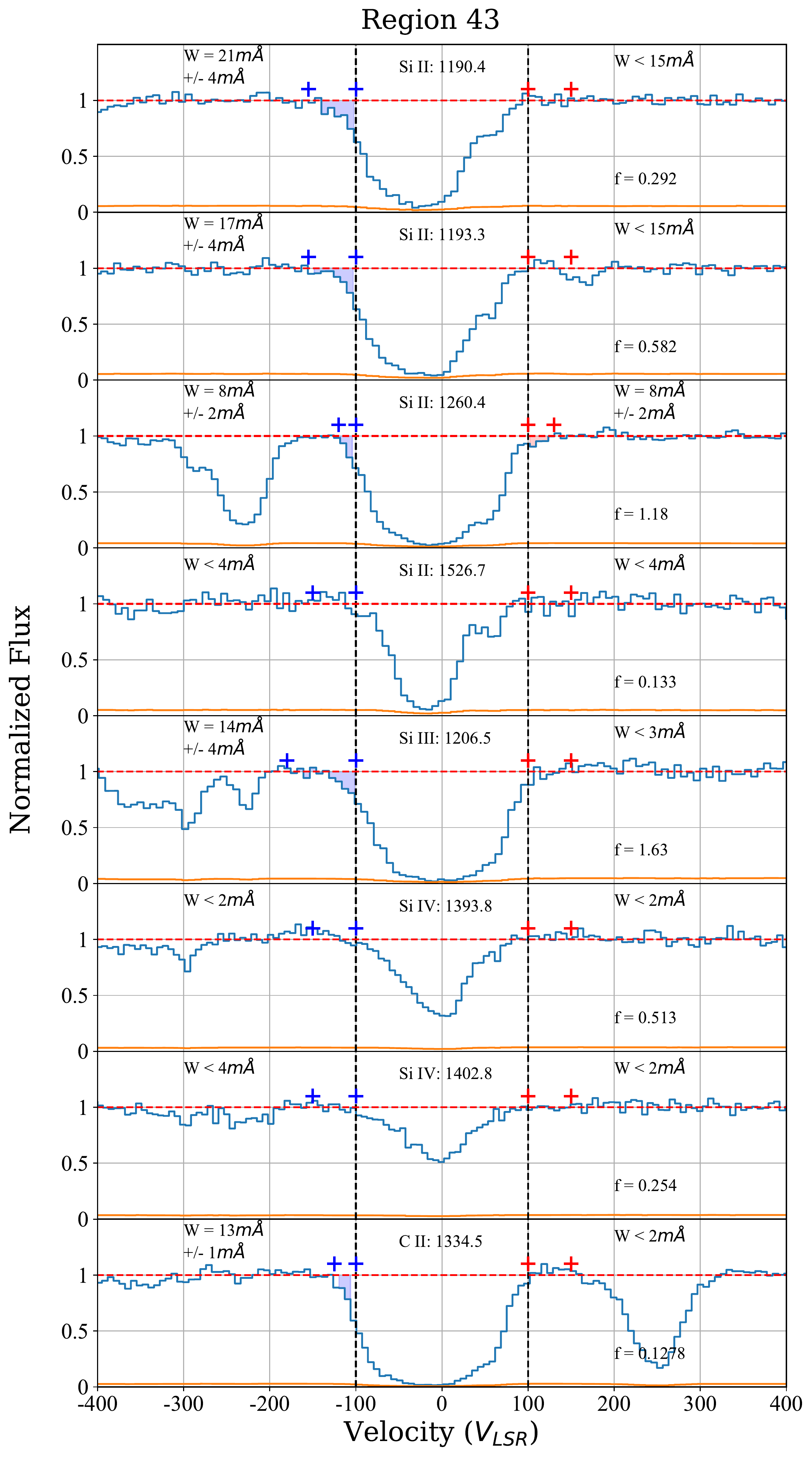}
    
    \contcaption{from Figure \ref{fig:AppendixB}.}
\end{figure*}

\begin{figure*}
    \includegraphics[width=0.65\columnwidth,height = 0.45\textheight]{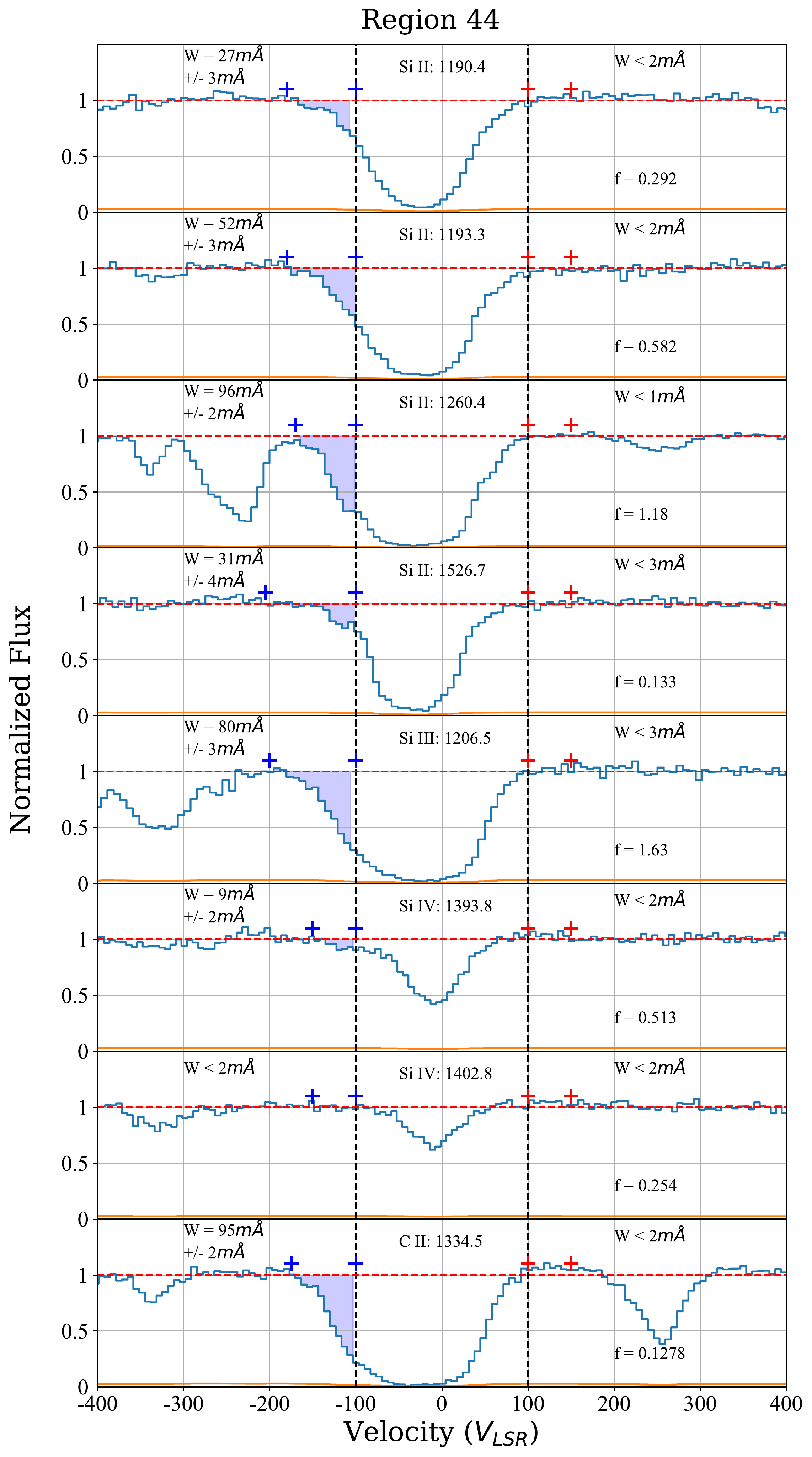}
    \includegraphics[width=0.65\columnwidth,height = 0.45\textheight]{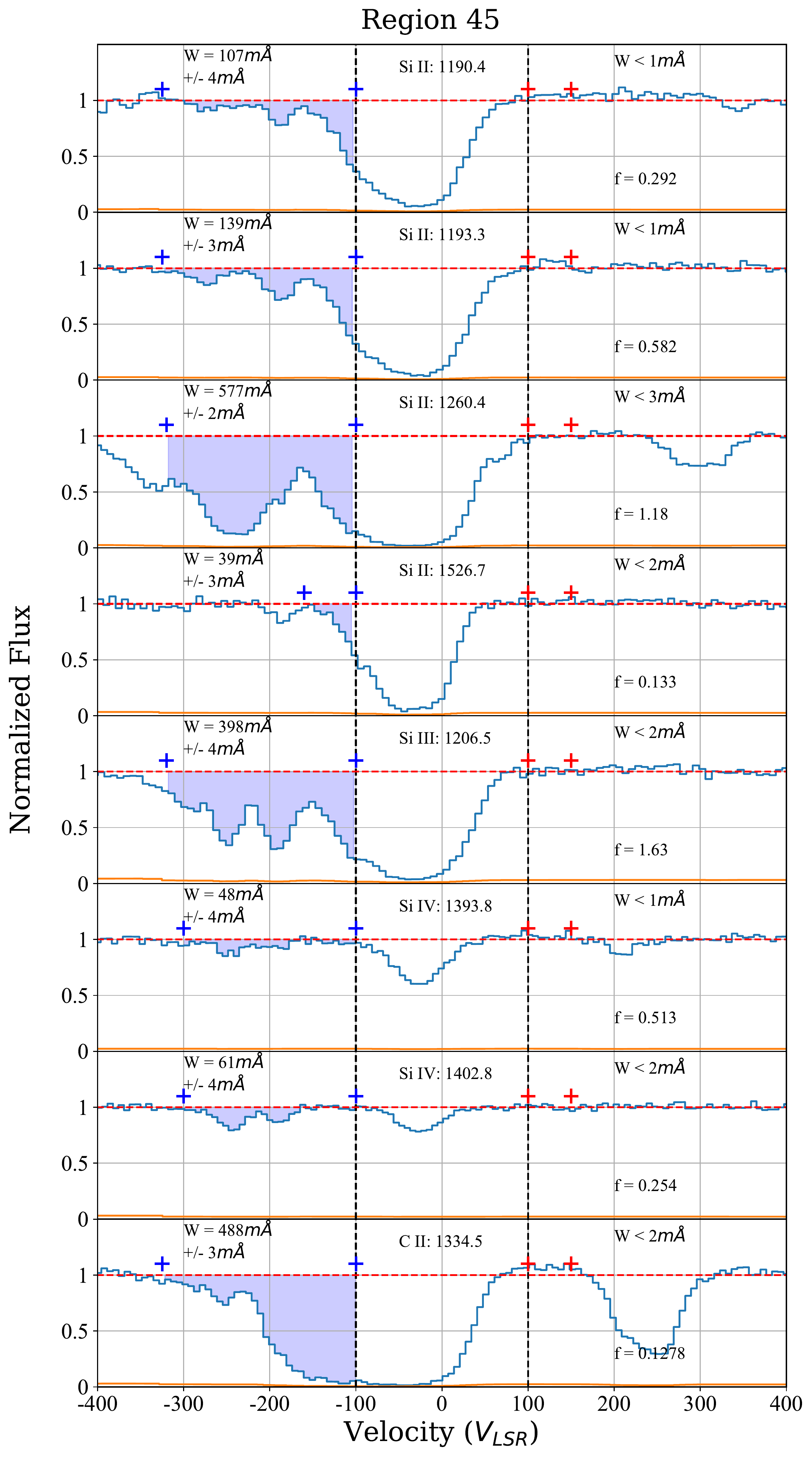}
    \includegraphics[width=0.65\columnwidth,height = 0.45\textheight]{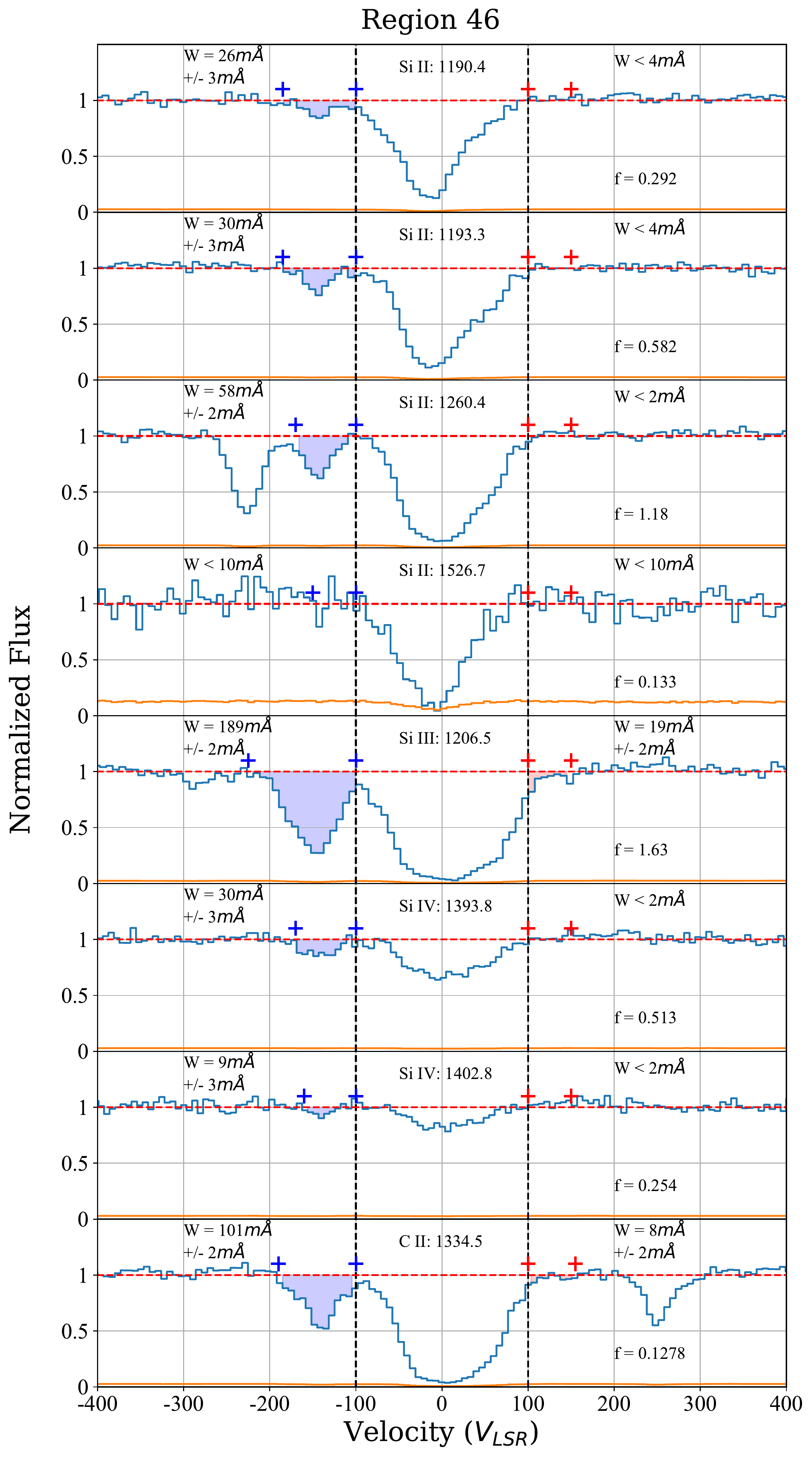}
    \includegraphics[width=0.65\columnwidth,height = 0.45\textheight]{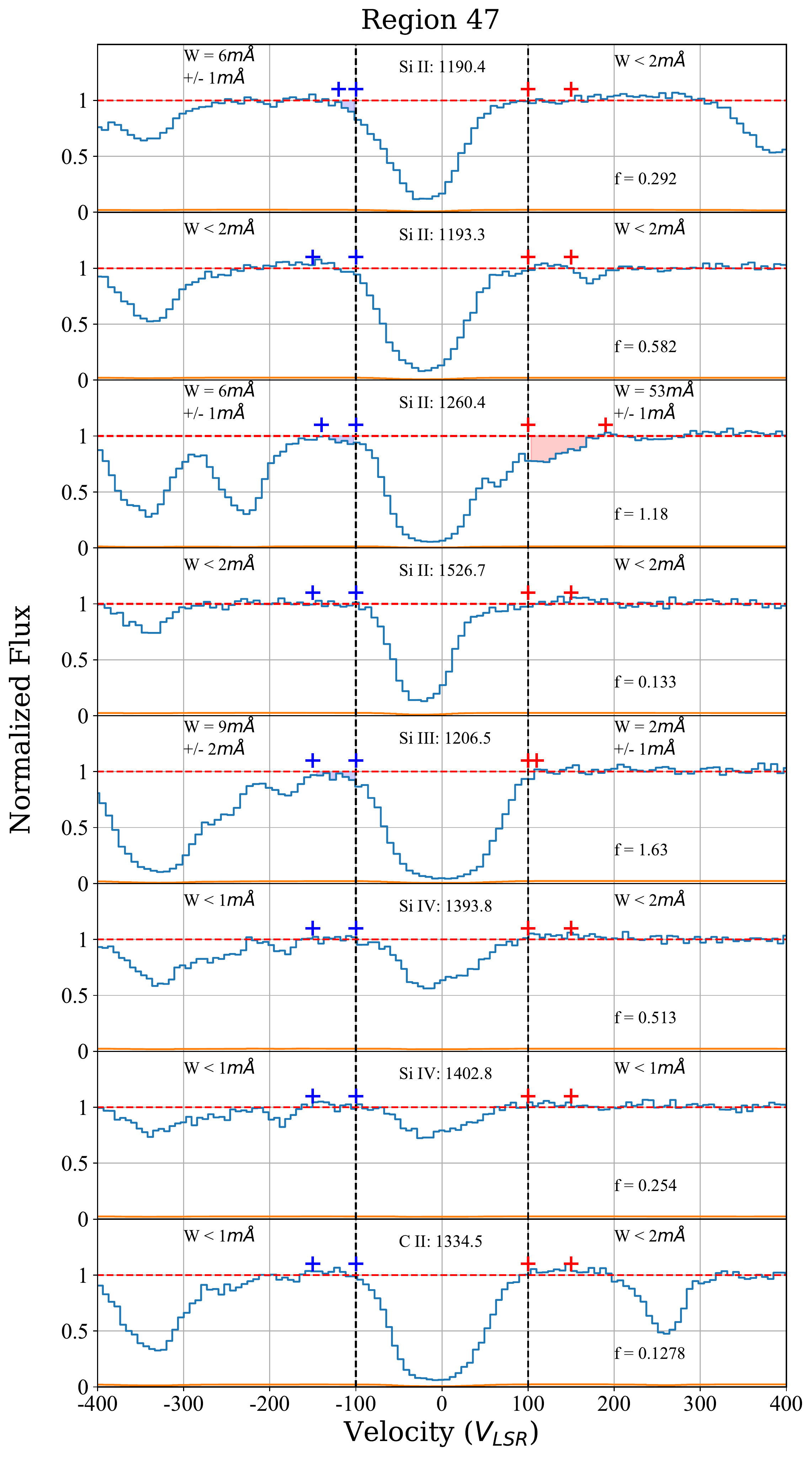}
    \includegraphics[width=0.65\columnwidth,height = 0.45\textheight]{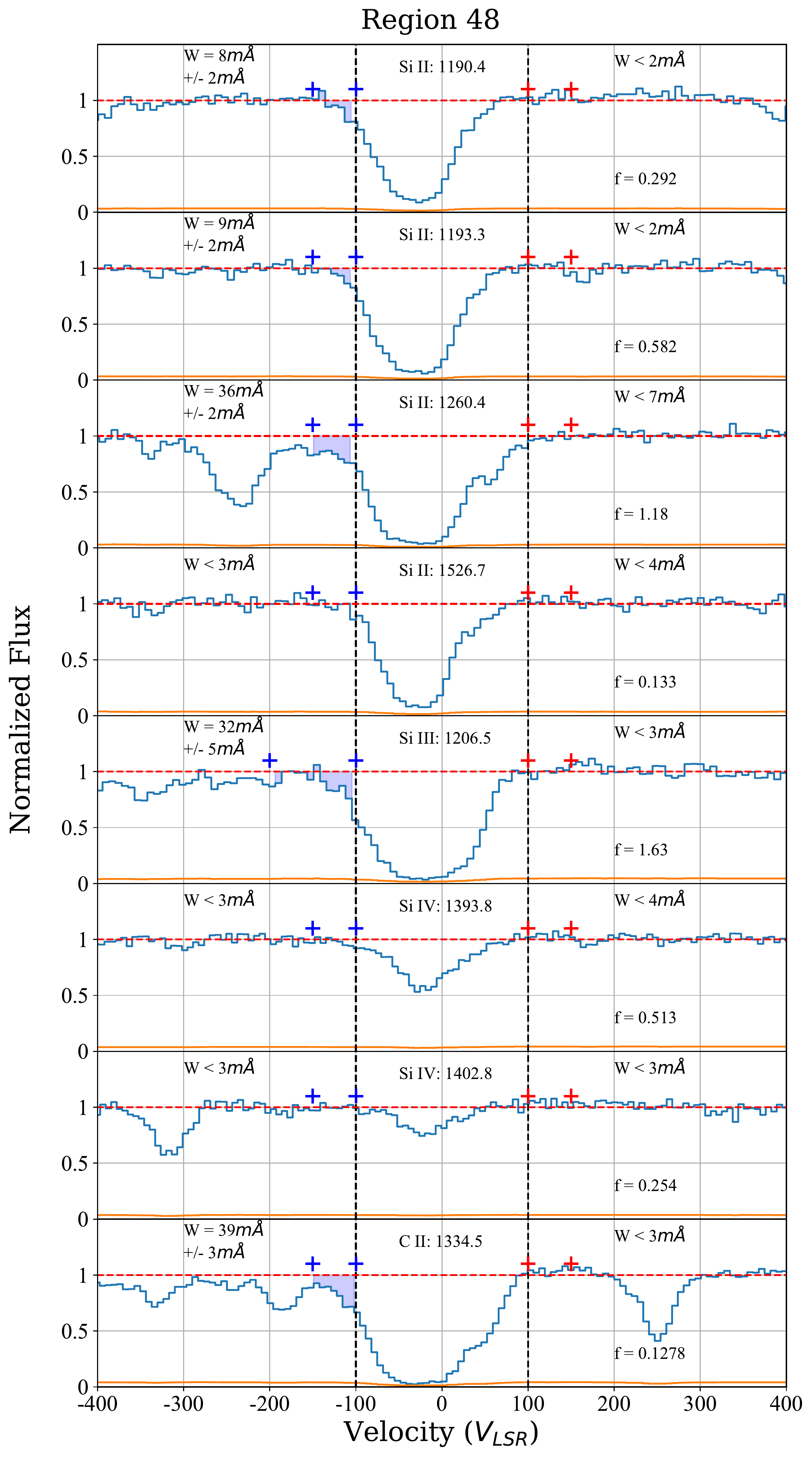}
    \includegraphics[width=0.65\columnwidth,height = 0.45\textheight]{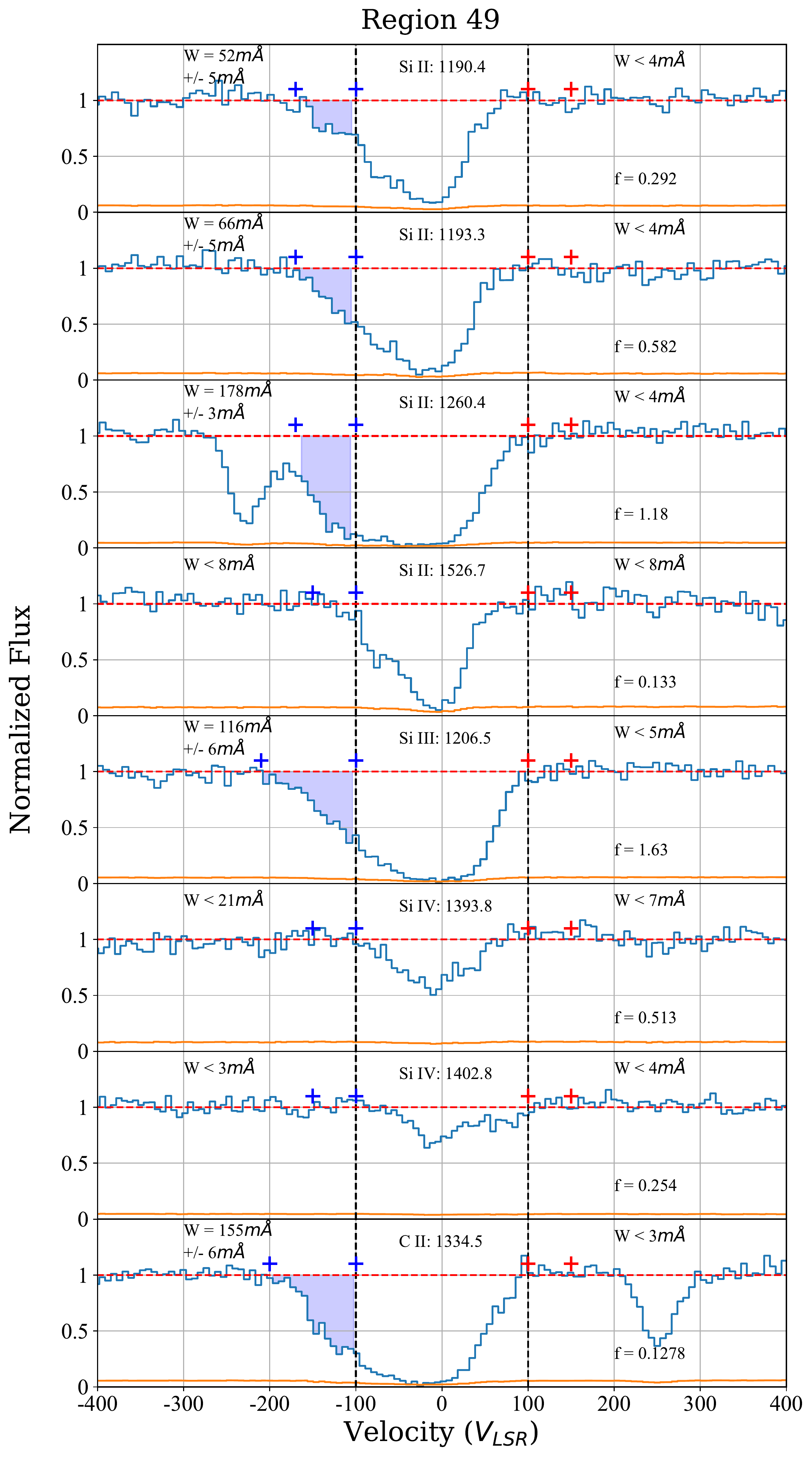}
    
    \contcaption{from Figure \ref{fig:AppendixB}.}
\end{figure*}

\begin{figure*}
    \includegraphics[width=0.65\columnwidth,height = 0.45\textheight]{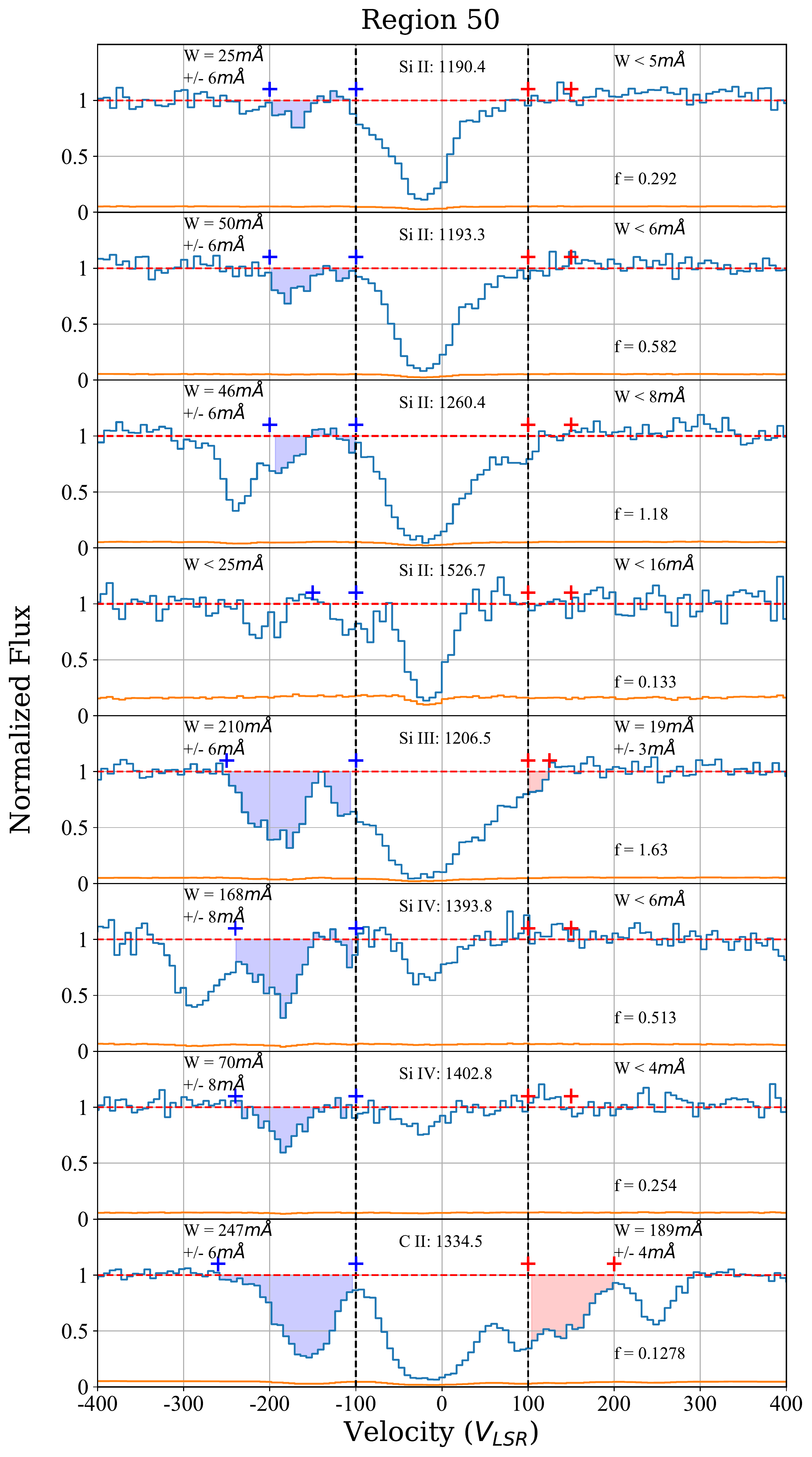}
    \includegraphics[width=0.65\columnwidth,height = 0.45\textheight]{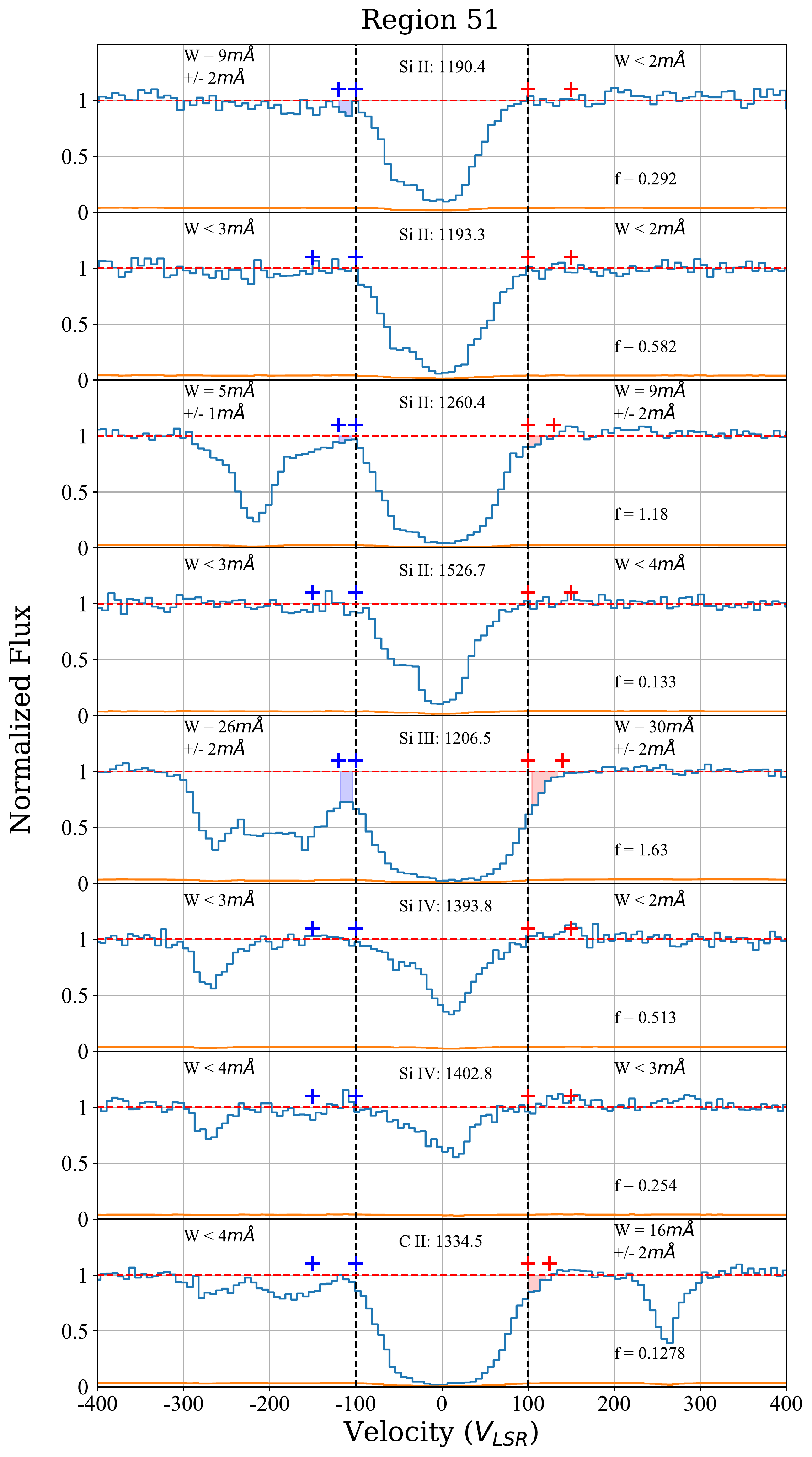}
    \includegraphics[width=0.65\columnwidth,height = 0.45\textheight]{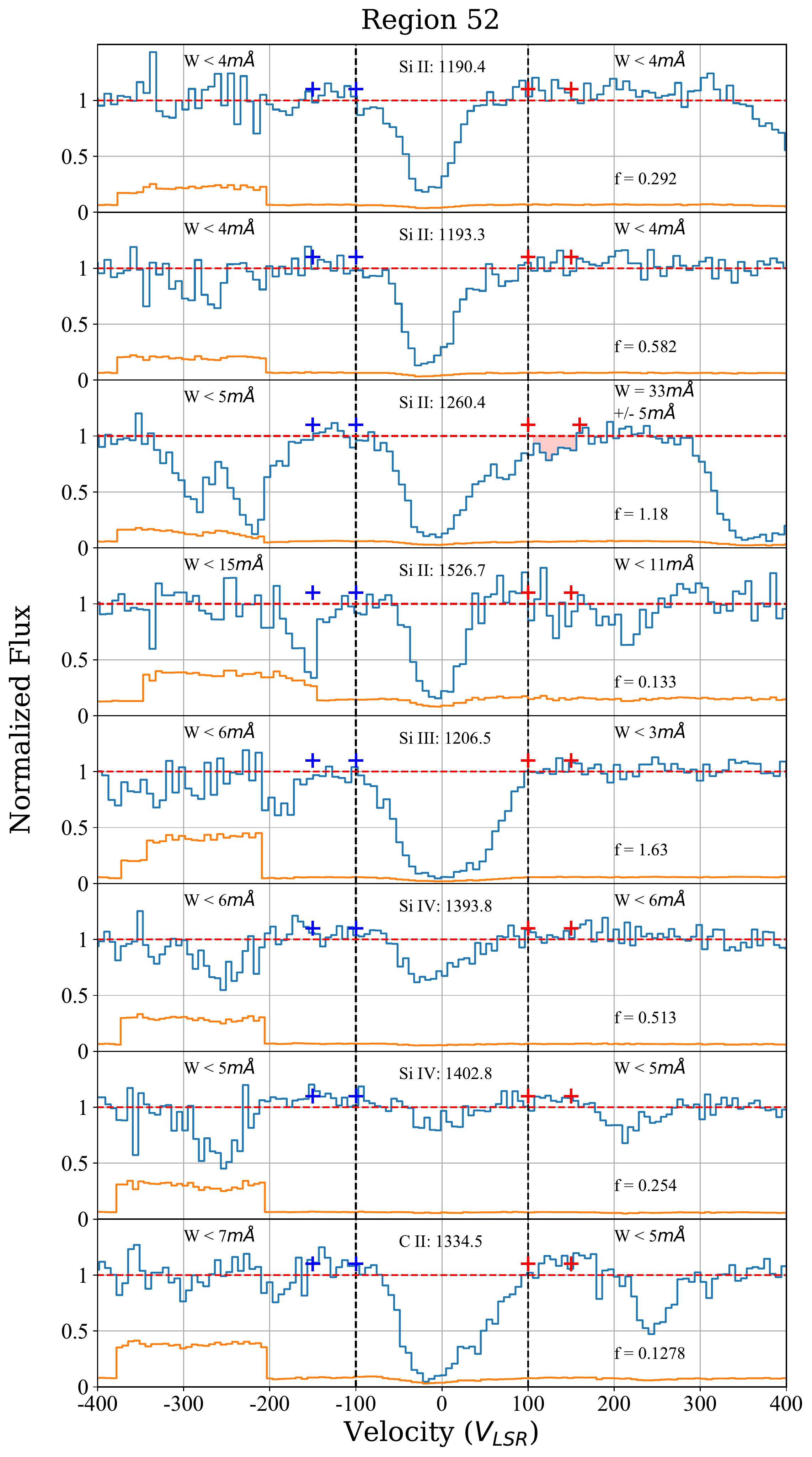}
    \includegraphics[width=0.65\columnwidth,height = 0.45\textheight]{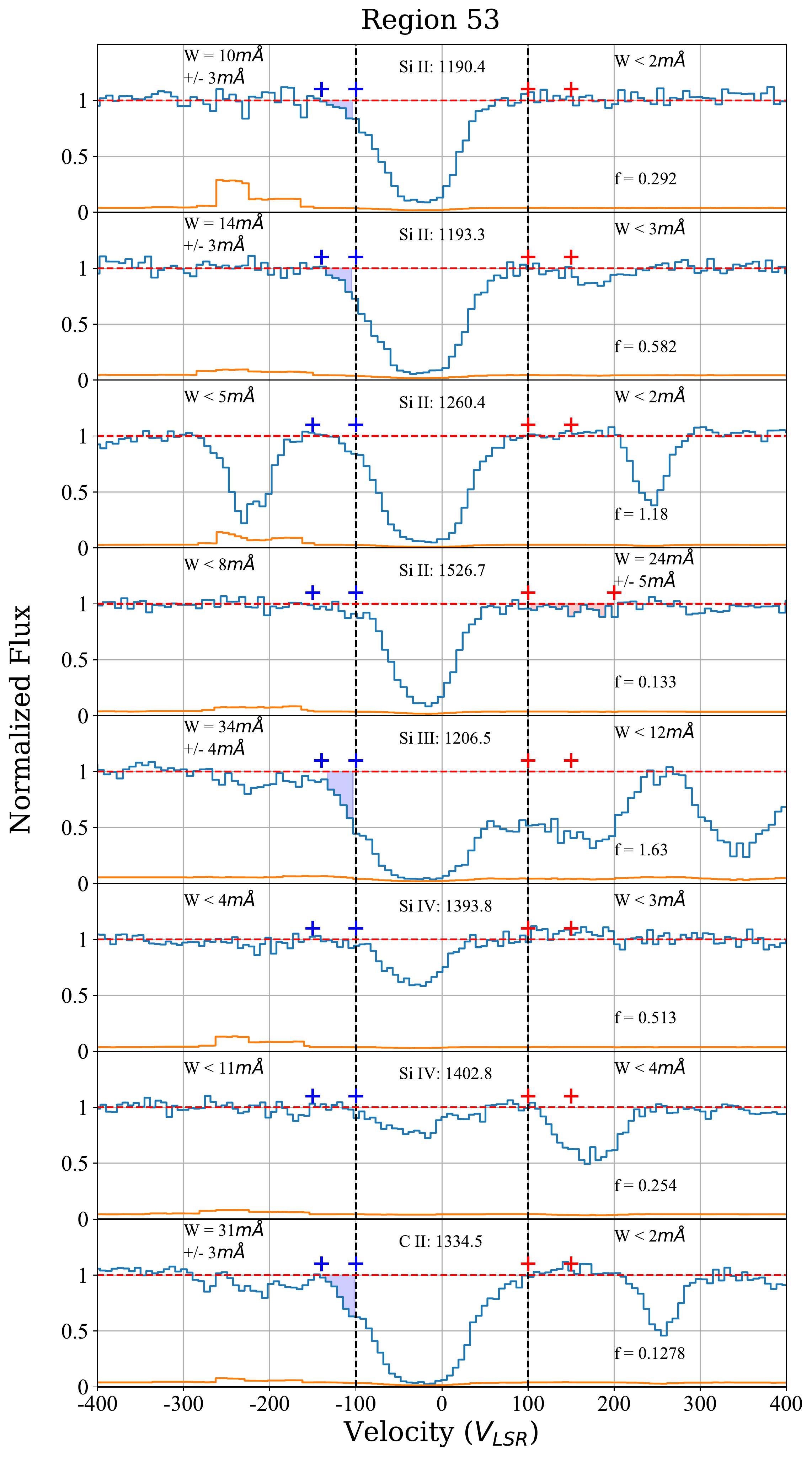}

    \contcaption{from Figure \ref{fig:AppendixB}.}
\end{figure*}


\bsp	
\label{lastpage}
\end{document}

%% file: Patch_Identity_table.tex
\begin{table*}
\centering 
\caption{Mean Galactic coordinates of each spatial region used in this work. The number of QSO spectra used in each co-add for each transition is also presented. The solid angle subtended on the sky by each spatial region is presented in Sr.}
\label{tab:SolidAngProperties}
\begin{tabular}{lccccccc}
Region ID & Galactic $l$ & Galactic $b$ & QSO$_{\rm Si\,IV}$ & QSO$_{\rm SiIII}$ & QSO$_{\rm SiII}$ & $\Omega$ (sr) \\
\hline
1 & 173.1 & 36.53 & 5 & 5 & 5 & 0.063 \\
2 & 138.9 & 31.61 & 5 & 6 & 6 & 0.048 \\
3 & 98.25 & 38.73 & 9 & 9 & 9 & 0.097 \\
4 & 75.16 & 37.77 & 8 & 5 & 6 & 0.154 \\
5 & 42.52 & 40.14 & 3 & 3 & 3 & 0.045 \\
6 & 24.04 & 42.67 & 8 & 5 & 6 & 0.011 \\
7 & 163.89 & 51.12 & 16 & 17 & 16 & 0.126 \\
8 & 133.13 & 55.54 & 5 & 5 & 5 & 0.07 \\
9 & 99.14 & 52.04 & 12 & 13 & 13 & 0.098 \\
10 & 80.17 & 51.68 & 5 & 3 & 3 & 0.041 \\
11 & 47.15 & 50.96 & 10 & 10 & 9 & 0.091 \\
12 & 11.33 & 51.99 & 9 & 7 & 9 & 0.028 \\
13 & 169.0 & 69.97 & 10 & 10 & 10 & 0.181 \\
14 & 137.92 & 72.9 & 13 & 11 & 11 & 0.223 \\
15 & 103.36 & 71.67 & 7 & 6 & 5 & 0.154 \\
16 & 72.41 & 74.75 & 8 & 10 & 8 & 0.183 \\
17 & 46.48 & 70.9 & 12 & 13 & 12 & 0.225 \\
18 & 16.53 & 73.72 & 10 & 9 & 9 & 0.15 \\
19 & -94.66 & 29.55 & 8 & 9 & 8 & 0.346 \\
20 & -137.45 & 37.35 & 12 & 9 & 10 & 0.217 \\
21 & -165.1 & 39.06 & 10 & 11 & 11 & 0.046 \\
22 & -11.82 & 51.72 & 3 & 3 & 3 & 0.022 \\
23 & -64.85 & 52.77 & 4 & 5 & 3 & 0.23 \\
24 & -105.91 & 54.78 & 10 & 9 & 9 & 0.107 \\
25 & -130.47 & 56.79 & 11 & 7 & 7 & 0.073 \\
26 & -165.8 & 52.24 & 9 & 9 & 9 & 0.076 \\
27 & -11.92 & 70.41 & 8 & 7 & 7 & 0.192 \\
28 & -49.52 & 68.69 & 3 & 3 & 3 & 0.017 \\
29 & -77.31 & 68.84 & 16 & 17 & 18 & 0.097 \\
30 & -101.06 & 71.18 & 10 & 8 & 8 & 0.125 \\
31 & -140.39 & 74.63 & 9 & 11 & 11 & 0.208 \\
32 & -163.6 & 72.01 & 17 & 15 & 14 & 0.154 \\
33 & -159.33 & -40.86 & 3 & 2 & 2 & 0.005 \\
34 & -100.29 & -39.62 & 7 & 7 & 7 & 0.079 \\
35 & -76.27 & -35.36 & 7 & 7 & 6 & 0.117 \\
36 & -17.97 & -36.22 & 7 & 7 & 7 & 0.251 \\
37 & -168.96 & -53.0 & 7 & 5 & 5 & 0.103 \\
38 & -27.33 & -55.85 & 6 & 6 & 6 & 0.137 \\
39 & -136.58 & -71.55 & 8 & 8 & 8 & 0.166 \\
40 & -80.27 & -70.12 & 4 & 4 & 4 & 0.21 \\
41 & -24.54 & -67.68 & 4 & 4 & 4 & 0.037 \\
42 & 43.74 & -33.1 & 8 & 7 & 7 & 0.312 \\
43 & 76.72 & -28.05 & 6 & 4 & 4 & 0.189 \\
44 & 107.76 & -31.6 & 8 & 8 & 8 & 0.175 \\
45 & 129.29 & -26.6 & 15 & 14 & 14 & 0.206 \\
46 & 25.16 & -50.96 & 3 & 3 & 3 & 0.056 \\
47 & 85.1 & -54.17 & 4 & 4 & 4 & 0.036 \\
48 & 122.79 & -49.26 & 7 & 6 & 7 & 0.172 \\
49 & 164.89 & -53.03 & 6 & 6 & 5 & 0.088 \\
50 & 4.9 & -72.67 & 4 & 4 & 4 & 0.032 \\
51 & 64.18 & -67.88 & 3 & 3 & 3 & 0.105 \\
52 & 99.7 & -68.2 & 3 & 3 & 3 & 0.093 \\
53 & 159.61 & -64.39 & 7 & 6 & 6 & 0.094 \\
\hline
\end{tabular}
\end{table*}

%% file: PatchProperty_Final.tex
\begin{table*}
\tiny
\caption{Mean column density and velocity measurements from the co-added QSO spectra in each spatial region. '*' denotes filtered regions not considered in the analysis, which belong to the Fermi Bubbles or the Magellanic Stream. '<' denotes non-detection and the corresponding log\,$N$ values are 2$\sigma$ upper limits of non-detection. '-' denotes regions that were shifted from blueshifted to redshifted absorption or vise-versa during transformations to the GSR frame.}
\label{tab:regional_characteristics}
\begin{tabular}{ccccccc | cccccc}
\toprule
      & \multicolumn{6}{c|}{Inflow} & \multicolumn{6}{c}{Outflow} \\
\midrule
     ID & \multicolumn{2}{c|}{Si IV} & \multicolumn{2}{c|}{Si III} & \multicolumn{2}{c|}{Si II} & \multicolumn{2}{c|}{Si IV} & \multicolumn{2}{c|}{Si III} & \multicolumn{2}{c|}{Si II}\\
\hline
- & log\,$N$ & $v_{\rm GSR}$ & log\,$N$ & $v_{\rm GSR}$ & log\,$N$ & $v_{\rm GSR}$ & log\,$N$ & $v_{\rm GSR}$ & log\,$N$ & $v_{\rm GSR}$ & log\,$N$ & $v_{\rm GSR}$ \\
- & $\left(\mathrm{cm}^{-2}\right)$ & $\left(\mathrm{km\,s}^{-1}\right)$ & $\left(\mathrm{cm}^{-2}\right)$ & $\left(\mathrm{km\,s}^{-1}\right)$ & $\left(\mathrm{cm}^{-2}\right)$ & $\left(\mathrm{km\,s}^{-1}\right)$ & $\left(\mathrm{cm}^{-2}\right)$ & $\left(\mathrm{km\,s}^{-1}\right)$ & $\left(\mathrm{cm}^{-2}\right)$ & $\left(\mathrm{km\,s}^{-1}\right)$ & $\left(\mathrm{cm}^{-2}\right)$ & $\left(\mathrm{km\,s}^{-1}\right)$ \\
\bottomrule
	1 & 12.21 $\pm$0.144 & -142 $\pm$3 & 13.22 $\pm$0.009 & -164 $\pm$7 & 13.66 $\pm$0.003 & -164 $\pm$6 & \textless11.58 & - & 11.73 $\pm$0.105 & 129 $\pm$5 & \textless12.0 & - \\
2 & 12.75 $\pm$0.018 & -29 $\pm$1 & 13.06 $\pm$0.006 & -46 $\pm$2 & 13.11 $\pm$0.005 & -71 $\pm$3 & \textless11.43 & - & \textless10.86 & - & \textless11.16 & - \\
3 & 13.23 $\pm$0.006 & -29 $\pm$1 & 13.73 $\pm$0.004 & -18 $\pm$1 & 14.02 $\pm$0.002 & -17 $\pm$1 & \textless10.94 & - & 11.24 $\pm$0.143 & 288 $\pm$1 & \textless11.8 & - \\
4 & - & - & 13.86 $\pm$0.006 & -113 $\pm$4 & 13.89 $\pm$0.005 & -10 $\pm$1 & 12.79 $\pm$0.031 & 29 $\pm$1 & 12.12 $\pm$0.066 & 290 $\pm$23 & \textless11.45 & - \\
5 & - & - & 12.73 $\pm$0.047 & -16 $\pm$1 & 12.5 $\pm$0.081 & -11 $\pm$1 & 12.63 $\pm$0.067 & 2 $\pm$1 & \textless11.37 & - & \textless11.67 & - \\
6 & *11.94 $\pm$0.116 & *-62 $\pm$2 & *12.2 $\pm$0.06 & *-100 $\pm$6 & *12.84 $\pm$0.014 & *-108 $\pm$8 & *11.75 $\pm$0.146 & *161 $\pm$1 & *11.46 $\pm$0.098 & *177 $\pm$1 & *\textless11.97 & *- \\
7 & 11.87 $\pm$0.11 & -72 $\pm$2 & 13.06 $\pm$0.007 & -107 $\pm$5 & 13.1 $\pm$0.007 & -102 $\pm$5 & \textless11.47 & - & 12.42 $\pm$0.018 & 140 $\pm$10 & \textless11.45 & - \\
8 & 12.29 $\pm$0.068 & -59 $\pm$3 & 13.4 $\pm$0.005 & -98 $\pm$4 & 13.73 $\pm$0.002 & -97 $\pm$4 & \textless11.3 & - & \textless10.86 & - & \textless11.34 & - \\
9 & - & - & 13.57 $\pm$0.006 & -68 $\pm$3 & 13.88 $\pm$0.003 & -24 $\pm$1 & 12.84 $\pm$0.036 & 5 $\pm$1 & \textless10.97 & - & \textless11.88 & - \\
10 & - & - & 13.31 $\pm$0.042 & -59 $\pm$4 & 13.16 $\pm$0.06 & -14 $\pm$1 & 12.72 $\pm$0.113 & 9 $\pm$1 & \textless12.12 & - & \textless12.34 & - \\
11 & 12.5 $\pm$0.054 & -18 $\pm$1 & 12.74 $\pm$0.019 & -18 $\pm$1 & 12.63 $\pm$0.025 & -12 $\pm$1 & \textless11.57 & - & \textless11.05 & - & \textless11.5 & - \\
12 & *12.26 $\pm$0.072 & *-85 $\pm$2 & *12.42 $\pm$0.031 & *-119 $\pm$5 & *12.71 $\pm$0.016 & *-113 $\pm$4 & *\textless11.47 & *- & *12.31 $\pm$0.043 & *207 $\pm$6 & *\textless11.99 & *- \\
13 & 12.36 $\pm$0.055 & -148 $\pm$5 & 12.86 $\pm$0.008 & -119 $\pm$5 & 13.06 $\pm$0.005 & -126 $\pm$5 & \textless11.36 & - & 11.94 $\pm$0.058 & 179 $\pm$4 & \textless11.36 & - \\
14 & 12.63 $\pm$0.046 & -116 $\pm$8 & 12.85 $\pm$0.011 & -110 $\pm$5 & 13.18 $\pm$0.005 & -126 $\pm$7 & 12.66 $\pm$0.051 & 160 $\pm$17 & \textless10.85 & - & \textless11.5 & - \\
15 & 12.65 $\pm$0.085 & -92 $\pm$4 & 12.69 $\pm$0.039 & -58 $\pm$3 & 13.02 $\pm$0.019 & -83 $\pm$5 & \textless12.13 & - & 12.43 $\pm$0.06 & 220 $\pm$9 & 12.62 $\pm$0.039 & 213 $\pm$6 \\
16 & 12.54 $\pm$0.107 & -172 $\pm$4 & 12.14 $\pm$0.041 & -74 $\pm$4 & 12.51 $\pm$0.018 & -69 $\pm$3 & \textless11.96 & - & 11.83 $\pm$0.1 & 200 $\pm$6 & \textless11.9 & - \\
17 & 12.37 $\pm$0.018 & -87 $\pm$5 & 12.39 $\pm$0.006 & -101 $\pm$5 & 12.9 $\pm$0.002 & -67 $\pm$3 & \textless10.78 & - & 11.68 $\pm$0.037 & 159 $\pm$17 & \textless11.68 & - \\
18 & 11.96 $\pm$0.107 & -100 $\pm$3 & 11.99 $\pm$0.055 & -102 $\pm$4 & 12.66 $\pm$0.012 & -148 $\pm$7 & \textless11.48 & - & \textless11.72 & - & \textless11.06 & - \\
19 & *11.89 $\pm$0.146 & *-102 $\pm$-1 & *13.46 $\pm$0.009 & *-102 $\pm$-6 & *13.87 $\pm$0.003 & *-100 $\pm$-6 & *- & *- & *- & *- & *- & *- \\
20 & \textless11.48 & - & 12.98 $\pm$0.02 & -23 $\pm$-1 & 13.03 $\pm$0.018 & -25 $\pm$-2 & - & - & - & - & - & - \\
21 & \textless11.87 & - & 12.3 $\pm$0.048 & -181 $\pm$7 & 12.6 $\pm$0.024 & -179 $\pm$6 & \textless11.7 & - & 12.46 $\pm$0.047 & 53 $\pm$2 & \textless11.77 & - \\
22 & 12.74 $\pm$0.051 & -188 $\pm$8 & 12.72 $\pm$0.021 & -180 $\pm$9 & 12.91 $\pm$0.014 & -186 $\pm$9 & \textless11.74 & - & 11.66 $\pm$0.113 & 78 $\pm$3 & \textless11.72 & - \\
23 & 12.11 $\pm$0.081 & -31 $\pm$1 & 12.09 $\pm$0.064 & -25 $\pm$1 & \textless12.22 & - & - & - & - & - & - & - \\
24 & \textless12.13 & - & 12.97 $\pm$0.018 & -30 $\pm$-2 & 13.19 $\pm$0.011 & -29 $\pm$-2 & - & - & - & - & - & - \\
25 & \textless11.58 & - & 11.93 $\pm$0.098 & -223 $\pm$6 & \textless12.11 & - & - & - & 12.82 $\pm$0.026 & 3 $\pm$1 & 12.89 $\pm$0.022 & 1 $\pm$1 \\
26 & 11.97 $\pm$0.124 & -150 $\pm$3 & 12.75 $\pm$0.017 & -168 $\pm$8 & 12.9 $\pm$0.012 & -164 $\pm$6 & \textless11.81 & - & 12.11 $\pm$0.051 & 68 $\pm$2 & \textless11.49 & - \\
27 & 12.64 $\pm$0.086 & -180 $\pm$12 & 13.42 $\pm$0.012 & -270 $\pm$10 & \textless12.32 & - & \textless11.83 & - & 12.08 $\pm$0.063 & 90 $\pm$1 & \textless11.92 & - \\
28 & 12.73 $\pm$0.079 & -230 $\pm$7 & 12.8 $\pm$0.028 & -214 $\pm$9 & 12.63 $\pm$0.041 & -255 $\pm$18 & \textless11.7 & - & \textless11.29 & - & \textless11.69 & - \\
29 & 11.91 $\pm$0.053 & -219 $\pm$8 & 11.76 $\pm$0.029 & -216 $\pm$11 & 11.34 $\pm$0.079 & -189 $\pm$3 & 11.65 $\pm$0.048 & 17 $\pm$1 & 12.32 $\pm$0.011 & 106 $\pm$2 & 12.37 $\pm$0.01 & 109 $\pm$2 \\
30 & \textless11.46 & - & \textless11.66 & - & \textless12.02 & - & 12.35 $\pm$0.08 & 23 $\pm$1 & 12.59 $\pm$0.036 & 24 $\pm$2 & 12.72 $\pm$0.027 & 25 $\pm$1 \\
31 & \textless11.84 & - & 12.72 $\pm$0.016 & -299 $\pm$16 & 12.2 $\pm$0.052 & -162 $\pm$5 & 12.67 $\pm$0.028 & 118 $\pm$3 & 13.3 $\pm$0.005 & 60 $\pm$7 & 13.32 $\pm$0.005 & 100 $\pm$4 \\
32 & 12.01 $\pm$0.132 & -169 $\pm$12 & 12.05 $\pm$0.045 & -151 $\pm$6 & \textless11.84 & - & \textless11.44 & - & 11.8 $\pm$0.084 & 82 $\pm$6 & \textless11.89 & - \\
33 & \textless11.26 & - & 11.52 $\pm$0.138 & -188 $\pm$8 & 12.29 $\pm$0.023 & -254 $\pm$22 & \textless11.24 & - & 11.84 $\pm$0.061 & 42 $\pm$2 & 12.67 $\pm$0.009 & 96 $\pm$7 \\
34 & \textless11.22 & - & 12.46 $\pm$0.014 & -76 $\pm$-4 & 12.42 $\pm$0.015 & -76 $\pm$-1 & - & - & - & - & - & - \\
35 & \textless11.29 & - & 13.12 $\pm$0.01 & -89 $\pm$-5 & 13.27 $\pm$0.007 & -88 $\pm$-3 & - & - & - & - & - & - \\
36 & *\textless11.36 & *- & *\textless11.26 & *- & *\textless11.68 & *- & *11.49 $\pm$0.122 & *45 $\pm$1 & *12.88 $\pm$0.011 & *45 $\pm$5 & *12.91 $\pm$0.01 & *52 $\pm$4 \\
37 & \textless11.67 & - & 12.38 $\pm$0.025 & -162 $\pm$7 & 12.17 $\pm$0.04 & -177 $\pm$15 & \textless11.32 & - & \textless11.08 & - & \textless11.53 & - \\
38 & \textless11.15 & - & 11.69 $\pm$0.057 & -191 $\pm$8 & \textless11.45 & - & 12.0 $\pm$0.092 & 39 $\pm$1 & 12.86 $\pm$0.01 & 39 $\pm$1 & 12.77 $\pm$0.012 & 40 $\pm$1 \\
39 & \textless11.94 & - & 12.75 $\pm$0.012 & -190 $\pm$7 & 12.43 $\pm$0.024 & -189 $\pm$9 & \textless11.85 & - & 12.05 $\pm$0.04 & 50 $\pm$1 & \textless11.76 & - \\
40 & *\textless11.74 & *- & *12.38 $\pm$0.018 & *-208 $\pm$11 & *12.21 $\pm$0.027 & *-194 $\pm$4 & *\textless12.28 & *- & *12.94 $\pm$0.027 & *22 $\pm$1 & *13.02 $\pm$0.022 & *23 $\pm$2 \\
41 & 12.81 $\pm$0.054 & -274 $\pm$11 & 12.5 $\pm$0.029 & -208 $\pm$6 & 12.22 $\pm$0.055 & -198 $\pm$11 & 12.11 $\pm$0.084 & 69 $\pm$1 & 12.98 $\pm$0.013 & 62 $\pm$3 & 13.07 $\pm$0.01 & 63 $\pm$3 \\
42 & - & - & - & - & - & - & \textless10.98 & - & \textless10.81 & - & 11.55 $\pm$0.04 & 20 $\pm$1 \\
43 & - & - & - & - & - & - & \textless11.37 & - & 11.87 $\pm$0.128 & 87 $\pm$2 & 12.4 $\pm$0.037 & 72 $\pm$5 \\
44 & - & - & - & - & - & - & 12.04 $\pm$0.111 & 69 $\pm$2 & 12.71 $\pm$0.016 & 52 $\pm$2 & 12.92 $\pm$0.01 & 58 $\pm$2 \\
45 & 12.75 $\pm$0.035 & -130 $\pm$2 & 13.43 $\pm$0.006 & -137 $\pm$6 & 13.36 $\pm$0.007 & -110 $\pm$6 & \textless11.18 & - & \textless10.95 & - & \textless11.27 & - \\
46 & *12.56 $\pm$0.041 & *-101 $\pm$3 & *13.13 $\pm$0.007 & *-119 $\pm$4 & *12.66 $\pm$0.02 & *-94 $\pm$2 & *\textless11.44 & *- & *11.99 $\pm$0.044 & *165 $\pm$2 & *\textless11.74 & *- \\
47 & - & - & 11.63 $\pm$0.088 & -4 $\pm$1 & - & - & \textless11.2 & - & 10.99 $\pm$0.151 & 247 $\pm$1 & \textless11.49 & - \\
48 & - & - & 12.22 $\pm$0.064 & -62 $\pm$6 & - & - & \textless11.6 & - & \textless11.11 & - & 12.11 $\pm$0.082 & 5 $\pm$1 \\
49 & \textless12.38 & - & 12.86 $\pm$0.023 & -134 $\pm$7 & 13.04 $\pm$0.015 & -109 $\pm$5 & \textless11.88 & - & \textless11.45 & - & \textless11.72 & - \\
50 & 13.41 $\pm$0.023 & -226 $\pm$6 & 13.14 $\pm$0.015 & -223 $\pm$7 & 12.88 $\pm$0.027 & -187 $\pm$4 & \textless11.77 & - & 11.99 $\pm$0.075 & 108 $\pm$2 & \textless11.95 & - \\
51 & \textless11.58 & - & 12.17 $\pm$0.036 & -37 $\pm$1 & \textless11.68 & - & \textless11.42 & - & 12.23 $\pm$0.034 & 186 $\pm$2 & \textless11.52 & - \\
52 & \textless11.79 & - & \textless11.46 & - & \textless11.74 & - & \textless11.78 & - & \textless11.21 & - & \textless11.71 & - \\
53 & \textless11.72 & - & 12.29 $\pm$0.055 & -89 $\pm$4 & 12.3 $\pm$0.054 & -86 $\pm$4 & \textless11.54 & - & \textless12.04 & - & \textless11.56 & - \\
\hline
\end{tabular}

\end{table*}

%% file: SurfaceDensityCharacteristics.tex
\begin{table*}
\caption{Measured surface mass density and mass flow rate surface densities in different regions of the sky. The values are strictly lower limits as these estimates assume no ionization corrections.}
\label{tab:SMD}
\begin{tabular}{ccc | cc}
\toprule
      & \multicolumn{2}{c|}{Inflow} & \multicolumn{2}{c}{Outflow} \\
\hline
ID & log $\Sigma_{\rm in}$ & log $\dot{\Sigma}_{\rm in}$ & log $\Sigma_{\rm out}$ & log $\dot{\Sigma}_{\rm out}$ \\
- & $\left(M_{\odot}\,\mathrm{kpc}^{-2}\right)$ & $\left(M_{\odot}\,\mathrm{kpc}^{-2}\,\mathrm{yr}^{-1}\right)$ & $\left(M_{\odot}\,\mathrm{kpc}^{-2}\right)$ & $\left(M_{\odot}\,\mathrm{kpc}^{-2}\,\mathrm{yr}^{-1}\right)$\\
\hline
1 & 5.3 $\pm$0.26 & -2.55 $\pm$-0.27 & 2.83 $\pm$0.3 & -5.13 $\pm$0.3 \\
2 & 4.97 $\pm$0.26 & -3.37 $\pm$-0.27 & - & - \\
3 & 5.74 $\pm$0.26 & -3.05 $\pm$-0.26 & 2.34 $\pm$0.32 & -5.27 $\pm$0.32 \\
4 & 5.67 $\pm$0.26 & -2.62 $\pm$-0.26 & 3.97 $\pm$0.27 & -4.22 $\pm$0.28 \\
5 & 4.43 $\pm$0.27 & -4.5 $\pm$-0.27 & 3.73 $\pm$0.28 & -6.04 $\pm$0.28 \\
6 & * & * & * & * \\
7 & 4.9 $\pm$0.26 & -3.16 $\pm$-0.27 & 3.52 $\pm$0.26 & -4.4 $\pm$0.27 \\
8 & 5.4 $\pm$0.26 & -2.68 $\pm$-0.26 & - & - \\
9 & 5.55 $\pm$0.26 & -2.93 $\pm$-0.26 & 3.94 $\pm$0.27 & -5.44 $\pm$0.27 \\
10 & 5.04 $\pm$0.27 & -3.42 $\pm$-0.27 & 3.83 $\pm$0.3 & -5.3 $\pm$0.3 \\
11 & 4.61 $\pm$0.27 & -4.25 $\pm$-0.27 & - & - \\
12 & * & * & * & * \\
13 & 4.82 $\pm$0.26 & -3.15 $\pm$-0.27 & 3.05 $\pm$0.27 & -4.77 $\pm$0.27 \\
14 & 4.92 $\pm$0.27 & -3.07 $\pm$-0.27 & 3.76 $\pm$0.27 & -4.1 $\pm$0.28 \\
15 & 4.74 $\pm$0.28 & -3.44 $\pm$-0.28 & 3.81 $\pm$0.29 & -3.92 $\pm$0.29 \\
16 & 4.4 $\pm$0.28 & -3.61 $\pm$-0.29 & 2.94 $\pm$0.29 & -4.83 $\pm$0.29 \\
17 & 4.6 $\pm$0.26 & -3.58 $\pm$-0.27 & 2.78 $\pm$0.27 & -5.09 $\pm$0.27 \\
18 & 4.44 $\pm$0.27 & -3.49 $\pm$-0.27 & - & - \\
19 & * & * & * & * \\
20 & 4.81 $\pm$0.27 & -3.88 $\pm$-0.27 & - & - \\
21 & 4.27 $\pm$0.27 & -3.54 $\pm$-0.28 & 3.56 $\pm$0.27 & -4.79 $\pm$0.27 \\
22 & 4.79 $\pm$0.27 & -3.01 $\pm$-0.27 & 2.76 $\pm$0.3 & -5.42 $\pm$0.3 \\
23 & 3.9 $\pm$0.28 & -4.73 $\pm$-0.28 & - & - \\
24 & 4.9 $\pm$0.26 & -3.7 $\pm$-0.27 & - & - \\
25 & 4.45 $\pm$0.27 & -3.24 $\pm$-0.27 & 4.29 $\pm$0.27 & -5.56 $\pm$0.27 \\
26 & 4.66 $\pm$0.27 & -3.19 $\pm$-0.27 & 3.22 $\pm$0.27 & -5.02 $\pm$0.27 \\
27 & 4.99 $\pm$0.27 & -2.67 $\pm$-0.27 & 3.18 $\pm$0.28 & -4.94 $\pm$0.28 \\
28 & 4.7 $\pm$0.28 & -3.0 $\pm$-0.28 & - & - \\
29 & 3.7 $\pm$0.27 & -4.04 $\pm$-0.27 & 3.79 $\pm$0.26 & -4.28 $\pm$0.26 \\
30 & - & - & 4.16 $\pm$0.28 & -4.53 $\pm$0.28 \\
31 & 4.33 $\pm$0.27 & -3.31 $\pm$-0.27 & 4.76 $\pm$0.26 & -3.38 $\pm$0.27 \\
32 & 3.86 $\pm$0.29 & -4.0 $\pm$-0.29 & 2.9 $\pm$0.28 & -5.26 $\pm$0.29 \\
33 & 3.85 $\pm$0.3 & -3.83 $\pm$-0.31 & 3.83 $\pm$0.27 & -4.29 $\pm$0.28 \\
34 & 4.28 $\pm$0.27 & -3.79 $\pm$-0.28 & - & - \\
35 & 5.0 $\pm$0.26 & -3.12 $\pm$-0.27 & - & - \\
36 & * & * & * & * \\
37 & 4.08 $\pm$0.28 & -3.76 $\pm$-0.28 & - & - \\
38 & 3.19 $\pm$0.27 & -4.6 $\pm$-0.27 & 4.25 $\pm$0.27 & -4.23 $\pm$0.27 \\
39 & 4.42 $\pm$0.27 & -3.37 $\pm$-0.27 & 3.15 $\pm$0.27 & -5.22 $\pm$0.27 \\
40 & * & * & * & * \\
41 & 4.55 $\pm$0.28 & -3.13 $\pm$-0.28 & 4.46 $\pm$0.27 & -3.81 $\pm$0.27 \\
42 & - & - & 3.91 $\pm$0.27 & -3.8 $\pm$0.26 \\
43 & - & - & 3.61 $\pm$0.3 & -4.58 $\pm$0.3 \\
44 & - & - & 4.27 $\pm$0.27 & -4.05 $\pm$0.27 \\
45 & 5.24 $\pm$0.26 & -2.73 $\pm$-0.27 & - & - \\
46 & * & * & * & * \\
47 & 3.13 $\pm$0.29 & -6.37 $\pm$-0.29 & 2.09 $\pm$0.33 & -5.59 $\pm$0.33 \\
48 & 3.72 $\pm$0.28 & -4.56 $\pm$-0.28 & 3.21 $\pm$0.3 & -6.15 $\pm$0.31 \\
49 & 4.76 $\pm$0.27 & -3.23 $\pm$-0.27 & - & - \\
50 & 5.17 $\pm$0.27 & -2.56 $\pm$-0.27 & 3.09 $\pm$0.28 & -4.95 $\pm$0.28 \\
51 & 3.67 $\pm$0.27 & -4.83 $\pm$-0.27 & 3.33 $\pm$0.27 & -4.47 $\pm$0.27 \\
52 & - & - & - & - \\
53 & 4.1 $\pm$0.28 & -4.03 $\pm$-0.28 & - & - \\
\hline
\end{tabular}
\end{table*}